\documentclass[aps,twocolumn,nobalancelastpage,amsmath,amssymb,a4paper,twosides]{revtex4}

\usepackage{graphics}      

\usepackage{url}           
\usepackage{bm}            
\usepackage[english]{babel}    
\usepackage[latin1]{inputenc}  
\usepackage[dvips]{graphicx} 
\usepackage{feynmp}
\usepackage{mathrsfs}
\def\<{\langle}
\def\>{\rangle}
\def\ddleft{\langle\hspace{-2pt}\langle}
\def\ddright{\rangle\hspace{-2pt}\rangle}

\def\½{\frac{1}{2}}
\def\klaser{k_{{\text{\tiny L}}}}
\def\laser{\omega_{{\text{\tiny L}}}}

\def\mathfb{\mathbf}

\numberwithin{equation}{section}

\begin{document}

\title{Three dimensional theory for light matter interaction}
\author{Martin W. Sørensen} \affiliation{QUANTOP -- Danish quantum
  optics center and the Niels Bohr Institute, University of
  Copenhagen, DK-2100 Copenhagen \O, Denmark}
\author{Anders S. S\o rensen} \affiliation{QUANTOP -- Danish quantum
  optics center and the Niels Bohr Institute, University of
  Copenhagen, DK-2100 Copenhagen \O, Denmark}

\date{\today}

\begin{abstract}
  We present a full quantum mechanical three dimensional theory
  describing an electromagnetic field interacting with an ensemble of
  identical atoms. The theory is constructed such that it describes
  recent experiments on light-matter quantum interfaces, where the
  quantum fluctuations of light are mapped onto the atoms and back
  onto light.  We show that the interaction of the light with the
  atoms may be separated into a mean effect of the ensemble and a
  deviation from the mean. The mean effect of the interaction
  effectively give rise to an index of refraction of the gas. We
  formally change to a dressed state picture, where the light modes
  are solutions to the diffraction problem, and develop a perturbative
  expansion in the fluctuations. The fluctuations are due to quantum
  fluctuations as well as the random positions of the atoms. In this
  perturbative expansion we show how the quantum fluctuations are
  mapped between atoms and light while the random positioning of the
  atoms give rise to decay due to spontaneous emission. Furthermore we
  identify limits, where the full three dimensional theory reduce to
  the one dimensional theory typically used to describe the
  interaction.
\end{abstract}

\maketitle


  \section{Introduction}

  For several applications in quantum information science, such as long
distance quantum communication \cite{briegel}, it is essential to
create an interface linking the photonic states used for transmitting
quantum information to a material state suitable for storing and
processing the information. The generation of the required strong
coherent coupling of light to a single emitter has proven difficult to
achieve in practise, although substantial progress has been made
\cite{kimblecavity,rempe,blinov,yamamoto,imamoglu}. In recent years
optically dense atomic ensembles has emerged as a promising alternative
\cite{juulsgaard1,juulsgaard2,sherson,mabuchi,duan,kuzmichtheory,DLCZ,kuzmich,lukin,kimbleensemble,schmiedmayer,vuletic,lukin-eit,kraus,oxana1,oxana2}.
In this approach one can for instance use classical laser pulses to
engineer a suitable interaction such that an incoming light field is
reversibly stored into the coherence between, e.g., two stable ground
states in the atoms \cite{juulsgaard2}.

Some experiments on atomic ensembles uses atoms that are enclosed inside
a cavity to enhance the coupling \cite{vuletic}. In this situation the
cavity defines a unique mode of the light field and the theoretical
description consists of describing a single optical mode coupled to
the atomic ensembles. Most experiments are, however, performed with
atoms in free space not enclosed in a cavity, and in this situation
the theoretical description is more complicated. Typically this
situation is described in a one dimensional approximation, where one
only considers a single transverse mode and solves a one dimensional
propagation equation for this mode \cite{duan,DLCZ,kuzmichtheory}.

In this paper we explore the range of validity of the one-dimensional
approximation by making a full three dimensional description of the
interaction between light and an atomic ensemble. Our calculations
directly apply to an experimental situations similar to the ones
described in Refs. \cite{juulsgaard1,juulsgaard2,sherson,mabuchi},
where the light is detuned far from the atomic transition, but we
expect the general features of our results to be valid for a much
broader class of problems. 

Some justification for the one-dimensional
description may be found in the literature on superflouressence, e.g.
Refs.  \cite{raymer,mostowski83,mostowski84}. In this context it was
found that the one-dimensional description is valid provided that the
Fresnel number is of order unity $\mathcal{F}\equiv A/\lambda L\approx
1$, where $A$ is the transverse beam area, $\lambda$ is the wavelength
of the light, and $L$ is the length of the ensemble. Based on this
work it has been argued that it is also necessary to have a Fresnel
number of order unity in order for the one-dimensional approximation
to be applicable to the quantum interfaces between light and atomic
ensembles \cite{duan,kuzmichtheory,DLCZ}. It is, however, essential to
realize that the physical situations are very different in the two
cases. The work on superflouressence typically concerns the temporal
distribution of the output light measured by impinging the outgoing light
on a photodetector. Because the photodetector just measures the
incoming flux $I$, this is essentially a multi-mode measurement
\begin{equation}
I\propto\sum_m \hat{a}_m^\dagger\hat{a}_m,
\end{equation}
where the the sum is over all modes $m$ hitting the detector, and each
of these modes are described
by the photon creation (annihilation) operators $\hat{a}_m^\dagger$
$(\hat{a}_m$). In particular the sum here includes all transverse
modes. This is in contrast to the quantum interface work, where one is
interested in the outgoing state of a single light mode, e.g., in
Refs. \cite{juulsgaard1,juulsgaard2,sherson,mabuchi} the measurement
is essentially a homodyne measurement of a single mode, defined by the
field of the strong classical laser. In other experiments 
the outgoing
light is sent through a single mode optical fiber, which filters out
everything except a single transverse mode.  Furthermore the
superflouressence work applies to a nonperterbative situation with a
large optical gain, whereas the quantum interfaces typically operates
in the few excitation regime. The previous analysis is thus not
applicable to the present situation and it is therefore not to be
expected that the condition $\mathcal{F}\sim 1$ is the right condition
for the validity of the one-dimensional approximation. In fact, the
experiments in Refs. \cite{juulsgaard1,juulsgaard2,sherson} are
performed with $\mathcal{F}\sim10^4$, and still give very good
agreement with the one-dimensional description. Here we make a full
three dimensional description of the experiments in Refs.
\cite{juulsgaard1,juulsgaard2,sherson}, and we find that it reduces to
the one-dimensional description in the paraxial approximation provided that $\mathcal F \gg 1$.

In a related work a three dimensional description was also presented
in Ref. \cite{duan02}. Whereas our procedure assumes non-moving atoms,
i.e., cold atoms, that work considered the opposite limit, where the
motion of the atoms wash out any spatial structure of the atomic spin
state. Unlike the situation in Ref. \cite{duan02}, where the motion of
the atoms always lead to certain inefficiencies, the fact that we
consider stationary atoms, allows us to identify certain limits, where
we exactly reproduce the simple result of the one dimensional theory
as discussed in Sec. \ref{sec:strongly-coupl-multi}.

Our theory is developed as a perturbative expansion of the interaction
between light and the atomic ensembles. It is, however, essential to
be very careful about the way this perturbative expansion is
performed. Below we shall present results up to second order in the
interaction between the light and the atoms. We shall use an effective
Hamiltonian, where the excited atomic state has been eliminated, i.e.,
a Hamiltonian of the form
\begin{equation}
  H\sim \sum_{\mathbf k,\mathbf k'} \sum_i g_{\mathbf k,\mathbf k'} u_{\mathbf k}({\bf
    r}_i)u_{\bf k'}^*({\bf r}_i) 
  \hat{a}^\dagger_{\bf k'} \hat{a}_{\bf k},
\end{equation}
where $g_{\mathbf k,\mathbf k'}$ is a coupling constant for the two
modes $\mathbf k$, and
$\mathbf k'$ described by photon creation (annihilation) operators
$\hat{a}_{\mathbf k}^\dagger$ $(\hat{a}_{\mathbf k}$) with mode
functions $u_{\bf k}$, and
${\bf r}_i$ is the position of the $i$th atom. If we take the
mode functions to be simple plane waves with an input field in a
certain mode $\mathbf k_0$ and calculate the intensity in a certain direction
described by $\mathbf k_1$, we find the intensity
\begin{equation}
  I\propto {\left| \sum_{i} {\rm e}^{i {\bf \Delta k}\cdot{\bf r}_i}
    \right|}^2=\sum_{i,j} {\rm e}^{i {\bf \Delta k}\cdot({\bf r}_i-{\bf r}_j)},
\label{eq:simplesum}
\end{equation}
where ${\bf \Delta k}=\mathbf k_1-\mathbf k_0$. The standard way to proceed from here is to
say that the exponential varies rapidly when $i\neq j$ and therefore
neglect all terms except $i=j$ so that one is left with something
proportional to the number of atoms $N_A$, which is known as spontaneous
emission.  For the problem we are interested in here,  we are, however,
mainly concerned with the properties of the light in the forward
direction, where ${\bf \Delta k} \approx 0$. In this case it seems
unjustified to neglect the cross terms which give rise to collective
scattering scaling as $N^2$. Since $N$ is typically a very big number,
the presence of such large $N^2$ contributions may limit the
applicability of perturbation theory.

In order to avoid the problems associated with this collective
scattering, we use a different basis for our perturbative expansion:
instead of starting from the eigenmodes of the propagation equation in
vacuum, we use the solutions to the classical diffraction problem in
the presence of the medium, i.e., we take into account that the atoms
give rise to an index of refraction of the gas, which changes the
propagation of the light. Specifically, we write the Hamiltonian as
\begin{equation}
H= \langle \mathcal{H} \rangle_{{\rm atoms}} +\delta H,
\end{equation}
where $ \langle \mathcal{H} \rangle_{{\rm atoms}} $ is the quantum
mechanical expectation value of the Hamiltonian with respect to the
atomic spin state averaged over the random positions of the atoms.
This averaged Hamiltonian gives rise a continuous quadratic
Hamiltonian in the light field operators similar to a Hamiltonian
describing the interaction with a dielectric medium. When we formally
change to the interaction picture with respect to this averaged
Hamiltonian, we obtain a new set of basis modes. Doing perturbation
theory on these modes, the only effect on the light comes from the
quantum mechanical fluctuations and the fluctuations caused by the
random position of the atoms. These fluctuations are described by the
Hamiltonian $\delta H=H-\langle \mathcal{H} \rangle_{{\rm atoms}}$.
When we average the first order term in the perturbative expansion
with respect to the position of the atoms the resultant expression
describe that the quantum fluctuations of the atoms are mapped onto
the light in analogy with the results derived in a one-dimensional
theory in Ref. \cite{duan}.

If we go to second order in the interaction, our expression will give
terms quadratic in $ \delta H$. In order to take the spatial average of such
terms we need to know the density-density correlation function of the
atoms. Inserting the density-density correlation function for an ideal
gas we no longer find the collective scattering terms described above,
i.e., the collective scattering is essentially the classical
diffraction of the light, which is explicitly taken into account by
our average Hamiltonian, and therefore it does not appear in our
perturbation theory. The spatial average of the second order term does,
however, produce a new term associated with the point particle nature
of the atoms and their random positions. This term is equivalent to
the results obtained by just keeping the $i=j$ terms in Eq.
(\ref{eq:simplesum}), and represents the effect of spontaneous
emission.

Unlike most approaches to the interaction between atoms and light,
which derive coupled equations for the atomic states and the electric
field, our approach considers the electric displacement field ${\bf
  D}$ instead of the electric field. The reason we chose to use the
displacement field is that it is convenient to work with a purely
transverse field, which is the case for the displacement field due to
the macroscopic Maxwell equation $\boldsymbol{\nabla} \cdot {\mathbf
  D}=0$, whereas this is not necessarily the case for the electric
field in a medium.  Formally the two approaches are equivalent and may
be related through a unitary transformation
\cite{claude97:photons_and_atoms}.

The full theory is quite involved. Readers who are mainly interested
in the consequences of our theory for experimental implementations are
therefore advised to skip to Sec. \ref{sec:exper-appl-valid}, where we
discuss such consequences. The sections prior to this mainly focus on
building the theoretical frame using a first-principles strategy.  The
paper is organized as follows: In Sec. \ref{sec:model} we give the
details of the model used to describe the interaction. In Sec.
\ref{sec:equations-motion} we derive a set of equations of motion
describing the system of atoms and light, using Heisenberg's equation
of motion. The wave equation describing the light is expressed in a
form that ideally suits a perturbative treatment. In Sec.
\ref{sec:born-appr-feynm} we express the general solution to the wave
equation in terms of Green's functions and derive the perturbative
expansion of the solution to the wave equations as well as the
equation describing the atoms. This is represented in terms of Feynman
diagrams. In addition we develop the appropriate theoretical tools to
describe point particle effects such as density correlations, and
derive a formal expression for the Green's function. In Sec.
\ref{sec:results} we present our results where we discuss higher order
effects such as spin decay and light scattering. We define operators
that describe photon-measurements, and demonstrate how these are
calculated in the theory.  In Sec.  \ref{sec:exper-appl-valid} we
discuss various limits where the general three dimensional theory
reduce to the usually employed one dimensional model \cite{duan}. We
also describe how a detailed understanding of the spatial modes can be
used to achieve storage and retrieval of information in several
transverse modes of light and atoms simultaneously.  In Sec.
\ref{sec:conclusion} we conclude the work, and in the appendices we
give several details omitted from the main text.

 %
  \section{Model}\label{sec:model}
  The model we consider describes the interaction between an ensemble
  of atoms and an incoming light field. The atomic ensemble is
  considered to be an ideal gas of identical atoms.  The atoms are described
  as non-moving randomly distributed point particles and the
  interaction with the light field is described within the
  dipole-approximation. Each atom is assumed to have a ground level of
  total spin $F$. In addition we assume that the atoms have no other
  stable ground states to which they can decay. See Fig.
  \ref{fig:atom1}. We shall assume that the electric fields are
  sufficiently far-detuned that we may adiabatically eliminate the
  exited states, and work with an effective Hamiltonian involving only
  the ground states. In the following we first discuss the interaction
  between light and a single atom, and then move on to discuss the
  interaction with an ensemble of atoms.

  \subsection{Interaction with single atoms}\label{sec:hamiltonian}
  The aim of this work is to describe the interaction between an
  electromagnetic field and an ensemble of identical atoms.  The
  problem is therefore both to deal with the microscopic behaviour of
  a single atom, and also the collective effect of many atoms. We
  choose here to work in the so called length gauge, where the basic
  interaction is given as the product of the displaced electric field
  and the polarization of the media
  \cite{claude97:photons_and_atoms,footnote1}
  \begin{align}\label{eq:dipole_hamilton}
    \mathcal H_{\text{int}} =- \sum_j^{\text{\scriptsize{Atoms}}}
    \frac{1}{\epsilon_0}\mathbf D(\mathbf r_j,t) \cdot \mathbf P
    (\mathbf r_j,t).
  \end{align}
  Our gauge choice ensures $\boldsymbol{\nabla}\cdot \mathbf D(\mathbf
  r,t)=0$.  We will assume that the fields have a large detuning and
  do not saturate the atomic transition, so that the exited levels may
  be adiabatically eliminated. This procedure is described in Appendix
  \ref{sec:adiab-elim}. The polarization of the atomic ensemble then
  depends linearly on the displaced electric field, that is $\mathbf
  P(\mathbf r,t) = \bar{\bar V}[\hat{\mathbf J}]\mathbf D(\mathbf
  r,t)$. We introduce here the argument $\hat{\mathbf J}$ to indicate
  that the interaction matrix $\bar{\bar V}[\hat{\mathbf J}]$ depends
  on the spin of the atoms.  Next we write the displaced electric
  field as a sum of a positively oscillating part and a negatively
  oscillating part,
    \begin{align}
    {\mathbf D}(\mathbf r,t) = {\mathbf D}^{(+)}(\mathbf r,t) + {\mathbf
      D}^{(-)}(\mathbf r,t).
  \end{align}

    \begin{figure}[t]
    \centering
    \vspace{.2cm}
    \includegraphics[width=0.4\textwidth]{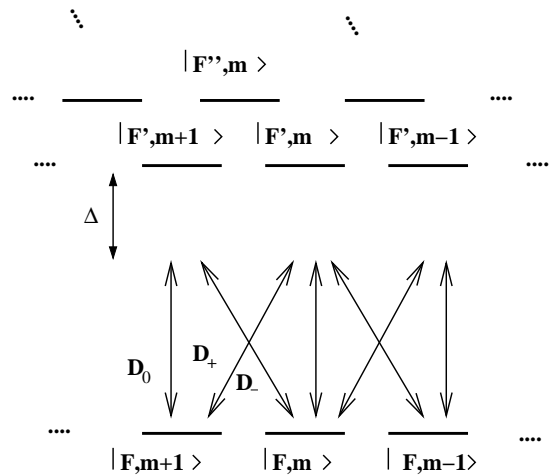}
    \caption{Example of an atomic level structure. The atoms have a
      single ground level with spin F and one or more exited levels.
      The fields have a large detuning $\Delta$ so that the exited
      states may be adiabatically eliminated and we obtain an effective
      ground state Hamiltonian Eq. (\ref{eq:hatm-h_text-=}).}
    \label{fig:atom1}
  \end{figure}

  \noindent In Appendix \ref{sec:adiab-elim} we show that the effective
  interaction Hamiltonian, assuming such linear dependence of the
  polarization on the displaced electric field, reads
  \begin{align}\label{eq:hatm-h_text-=}
    {\mathcal H}_{\text{int}} = -\frac{1}{2 \epsilon_0
    }\sum_j^{\text{Atoms}} \Big(  \Big[ \bar{\bar V}[\hat{\mathbf
      J}_j]^t&\; {\mathbf D}^{(-)}_j \Big] \cdot {\mathbf D}_j^{(+)}
    \notag \\ &+ {\mathbf
      D}_j^{(-)} \cdot \Big[ \bar{\bar V}[\hat{\mathbf J}_j]\; {\mathbf
      D}^{(+)}_j \Big] \Big),
  \end{align}
  where we have also employed the rotating wave approximation.  Here
  the superscript $t$ denotes matrix transposition. 

  Since the Hamiltonian must be rotationally invariant it can only
  contain irreducible tensors of at most rank two.  In the vector
  representation the interaction may thus in general be written as
  \begin{align}\label{eq:barb-vmathbf-j_j=b-1}
    \bar{\bar V}[\hat{\mathbf J}_j]=\beta \Big( c_0 \; \hat{\mathbf J}_j^2
    -ic_1 \; \hat{\mathbf J}_j \times + c_2\;\times \hat{\mathbf
      J}_j \big) \cdot \big( \hat{\mathbf J}_j \times \Big).
  \end{align}
  The meaning of the notation is that when inserted into
  the Hamiltonian the result of, e.g., the last term of the right hand
  side of Eq. (\ref{eq:barb-vmathbf-j_j=b-1}) is
  \begin{align} 
     \beta c_2 \sum_j^{\text{Atoms}} \big( {\mathbf D}^{(-)}(\mathbf
    r_j,t)\times \hat{\mathbf J}_j \big) \cdot \big( \hat{\mathbf J}_j
    \times {\mathbf D}^{(+)}(\mathbf r_j,t) \big).
  \end{align}
  Note that we have here chosen a description which has a simple
  analytical representation, but this means the $c_2$ term is not a
  pure rank two irreducible tensor, but consist of a combination of
  tensors of rank zero, one and two.  In matrix form the interaction
  may be written:
  \begin{widetext}
    \begin{align}\label{eq:barbar-vbarmathbf-j}
      \bar{\bar V}[\bar{\mathbf J}] = \beta \left[
        \begin{array}{lll}
          (c_0 -c_2)\hat{\mathbf J}^2 + c_2 \hat J_x^2 & \quad ic_1\hat J_z
          + c_2\hat J_y\hat J_x & -ic_1\hat J_y + c_2 \hat J_z \hat J_x \\ 
          -ic_1\hat J_z + c_2\hat J_x\hat J_y & (c_0 -c_2)\hat{\mathbf
            J}^2 + c_2 \hat J_y^2 &  \quad ic_1\hat J_x + c_2 \hat J_z \hat J_y   \\ 
          \quad ic_1\hat J_y + c_2\hat J_x\hat J_z &
          -ic_1\hat J_x + c_2 \hat J_y \hat J_z & (c_0 -c_2)\hat{\mathbf
            J}^2 + c_2 \hat J_z^2 
        \end{array} \right].
    \end{align}
  \end{widetext}
  In general the atoms may have several exited levels
  as shown in Fig. \ref{fig:atom1}. The effect of several exited levels can
  be included in the coefficients $c_0$, $c_1$ and $c_2$ that will then
  depend on the detuning.
  For atoms with $F=½$ or for an alkali atom, where the fields
  are detuned by more than the hyperfine structure of the exited
  state, the $c_2$ term disappears \cite{brian_thesis} and the interaction matrix is given
  by
  \begin{align}\label{eq:barb-vmathbf-j_j=b}
    \bar{\bar V}[\hat{\mathbf J}_j]=\beta \Big( c_0 \; \hat{\mathbf J}^2_j
    -ic_1 \; \hat{\mathbf J}_j \times \Big).
  \end{align}
  Here $c_0$ and $c_1$ are constants which depend on the atomic structure as well as the detuning.
  The coupling constant $\beta$ in
  Eq. (\ref{eq:barb-vmathbf-j_j=b}) is given by
  \begin{align}
    \beta=\frac{\pi \gamma }{2 \Delta \klaser^3},
  \end{align}
  where $\gamma$ is the linewidth of the exited level, $\Delta$ the
  detuning of the laser field with respect to the atomic transition,
  and $\klaser$ is the wave vector. With this choice of $\beta$ the
  coefficients $c_0$, $c_1$ and $c_2$ will be of order unity or less.
  Throughout this paper we shall only consider the simple interaction in \eqref{eq:barb-vmathbf-j_j=b}. A discussion of the effect of the $c_2$ term is given in Refs. \cite{oxana1,oxana2} in a one dimensional description.
  
  We will consider a perturbative regime, where the product of the
  atomic density $\rho$ and $\beta$ is small $\beta \rho \ll 1$, and
  make a perturbative expansion in $\beta$. Note, however, that this
  condition does not imply that the total effect of the interaction is
  small. On the contrary, we are most interested in situations, where
  the integrated effect of the interaction significantly alters the
  light beam as it passes through the sample. To take into account
  these collective effects we explicitly include, e.g.,  the diffraction
  of the light caused by the propagation through a medium. To describe
  these effects we discuss in the following section how to quantize
  the field in a medium.
  \subsection{Mode expansion}\label{sec:quant-electr-field}
  To quantize the electromagnetic fields we could: i)
  impose the canonical commutation relations on the vector
  potential and displaced electric field. Or ii) expand the
  electromagnetic fields on an orthonormal set of spatial
  mode-functions $\big\{ \mathbf f_{\mathbf k} \big\} $ conveniently
  chosen to diagonalize the Hamiltonian (in vacuum this is the set of
  plane waves), and then quantizing the mode-amplitudes. Here we will
  use the latter. The
  Hamiltonian describing the electromagnetic field in a medium is given
  by \cite{claude97:photons_and_atoms}
  \begin{align}\label{eq:mathcal-h-=}
    \mathcal H = \frac{1}{2}\int d^3r&\;\Big\{ \frac{\mathbf
      D^2}{\epsilon_0} + \frac{(\boldsymbol{\nabla}\times \mathbf
      A)^2}{\mu_0} \Big\} + \mathcal H_{\text{int}},
  \end{align}
  where $\mathcal H_{\text{int}}$ is given in equation
  (\ref{eq:hatm-h_text-=}).  A careful analysis of how to quantize the
  electromagnetic field in a medium, is given in Ref.
  \cite{glauber91:_quant}, and here we shall only go through the steps
  briefly. 

 By introducing the spin field
  \begin{align}\label{eq:hatmathbf-jmathbf-r-1}
    \hat{\mathbf J}(\mathbf
    r,t)=\sum_j^{\text{Atoms}}\hat{\mathbf J}_j\delta(\mathbf r - \mathbf r_j),
  \end{align}
  the Hamiltonian may be put in an all-integral form.  The main idea
  in our approach is to divide the full Hamiltonian into a spatially averaged
  part, and a point particle part, describing the fluctuations from
  the average caused by the atoms being point particles. For now we only
  consider the spatially averaged part of the theory. We will use calligraphic font to denote that we
  have made a spatial average.  We thus write the spatially
  averaged interaction from equation (\ref{eq:barb-vmathbf-j_j=b}) as
  \begin{align}
    \bar{\bar{\mathcal V}}[\bar{\mathbf J}]=\beta \rho(\mathbf
    r)\Big( c_0 \; \overline{\mathbf J^2} -ic_1 \; \bar{\mathbf
      J}(\mathbf r)\times \Big).
  \end{align}
  Here a bar denotes a single-atom operator, that is $\bar{\mathbf
    J}(\mathbf r)$ is the spin operator of a single atom at position
  $\mathbf r$. We use the bar to distinguish between the spatially
  averaged single-atom spin operator, and the general spin
  field in equation (\ref{eq:hatmathbf-jmathbf-r-1}). The two may be
  related by $\< \hat{\mathbf J}(\mathbf r,t) \>_{\rm s.a.} =
  \rho(\mathbf r)\bar{\mathbf J}(\mathbf r,t)$, where $\< \cdot
  \>_{\rm s.a.}$ denotes spatial average. The function $\rho(\mathbf r)$ denotes
  the average atomic density, which in this model is a continuous scalar field.  

  In the following we will define a mean Hamiltonian, where we have
  taken into account the quantum mechanical average of the spatially
  averaged interaction.  We then write the Hamiltonian as a
  sum of the average Hamiltonian and a point particle Hamiltonian
  \begin{align}
    \mathcal H =& \mathcal H_{0} + \mathcal
    H_{\text{pp}},\label{eq:splitting_ham} \intertext{where} \mathcal
    H_{0} =& \frac{1}{2}\int d^3r \Big\{ \frac{\mathbf
      D(\bar{\bar{\mathcal M}}^t \mathbf D^{(-)}+\bar{\bar{\mathcal
          M}} \mathbf D^{(+)})  }{\epsilon_0} +
    \frac{(\boldsymbol{\nabla}\times \mathbf A)^2}{\mu_0}
    \Big\},\\
    \mathcal H_{\text{pp}}=& -\frac{1}{2 \epsilon_0 }\int d^3r\;
    {\mathbf D} \cdot \Big( \bar{\bar m}[\hat{\mathbf J}]^t\; {\mathbf
      D}^{(-)} + \bar{\bar m}[\hat{\mathbf J}]\; {\mathbf D}^{(+)}
    \Big), \\
    &\bar{\bar{\mathcal M}} = \openone - \bar{\bar{\mathcal
        V}}[{\mathbf J}],
    \intertext{and} &\bar{\bar m}[\hat{\mathbf J}] = \bar{\bar
      V}[\hat{\mathbf J}] - \bar{\bar{\mathcal V}}[{\mathbf J}].
  \end{align}
  Here we simply write $\mathbf J$ (without the hat) to denote that this is now a
  classical field describing the classical expectation of the spin of
  the atoms.  In analogy with Ref. \cite{glauber91:_quant} we
  introduce the mode functions $\big\{ \mathbf f_{\mathbf k} \big\}$
  defined by:
  \begin{subequations}\label{eq:boldsymb-barb-vhatm}
    \begin{align}
      \boldsymbol{\nabla}\times\boldsymbol{\nabla}\times\bar{\bar{\mathcal
          M}} \mathbf f_{\mathbf k}(\mathbf
      r)=&\frac{\omega_{\mathbf k}^2}{c^2} \mathbf f_{\mathbf
        k}(\mathbf
      r),\label{eq:boldsymb-barb-vhatm-1} \\
      \boldsymbol{\nabla}\cdot \mathbf f_{\mathbf k}(\mathbf r) =& 0.
      \label{eq:boldsymb-mathbf-f_ma} 
    \end{align}
  \end{subequations}
  We also define the appropriate inner product on the space spanned by
  these mode functions:
  \begin{align}\label{eq:-boldsymb-r}
    \< \boldsymbol{\phi}(\mathbf r) | \boldsymbol{\psi}(\mathbf r) \>
    = \int d^3r \boldsymbol{\phi}(\mathbf r)^* \cdot \bar{\bar{\mathcal
      M}} \boldsymbol{\psi}(\mathbf r).
  \end{align}
  We will assume that the average interaction term $\bar{\bar{\mathcal
      V}}[{\mathbf J}]$ does not evolve in time, and our appropriate
  mode-functions are therefore time independent vector fields.  One
  can show that the functions $\mathbf f_{\mathbf k}$ span a complete
  orthonormal basis for the space in which we work.  To diagonalize
  the Hamiltonian we expand the vector potential and the displaced
  electric field in these mode functions
  \begin{subequations}\label{eq:begin-mathbf-dmathbf}
    \begin{align}\label{eq:mathbf-dmathbf-r-1}
      \mathbf D(\mathbf r,t) =& -\sum_{\mathbf k} \sqrt{\epsilon_0}\;
      p_{\mathbf k}(t)\mathbf f_{\mathbf k}^*(\mathbf r) \\
      \mathbf A(\mathbf r,t)=& \sum_{\mathbf k} c\sqrt{\mu_0}\;
      q_{\mathbf k}(t)(1-\bar{\bar{\mathcal V}}[{\mathbf
        J}])f_{\mathbf k}(\mathbf r).
    \end{align}
  \end{subequations} 
  The minus sign in Eq. (\ref{eq:mathbf-dmathbf-r-1}) is conventional and
  stems from the relation between the displaced electric field and the
  canonical conjugate field given in terms of the vector potential.
  
  The reality condition on the displaced electric field $\Big[ \; (\mathbf
  D(\mathbf r,t))^{\dagger}=\mathbf D(\mathbf r,t) \; \Big]$ allows us to
  write
  \begin{align}\label{eq:mathbf-dmathbf-r}
    \mathbf D(\mathbf r,t) =& -\sum_{\mathbf k}
    \frac{\sqrt{\epsilon_0}}{2} \left(  p^{\dagger}_{\mathbf k}(t)\mathbf
      f_{\mathbf k}(\mathbf r)  +  p_{\mathbf k}(t)\mathbf f_{\mathbf
        k}^*(\mathbf r) \right).
  \end{align}
  Using the results 
  in Eqs. (\ref{eq:boldsymb-barb-vhatm}) and (\ref{eq:-boldsymb-r}) and
  the expansion in equation (\ref{eq:begin-mathbf-dmathbf}), the
  Hamiltonian attains the desired diagonal form
  \begin{align}\label{eq:mathc-h_0=-frac12}
    \mathcal H_{0}=& \frac{1}{2}\int d^3r \Big\{ \frac{\mathbf D(1 -
      \bar{\bar{\mathcal V}}[{\mathbf J}]) \mathbf D
    }{\epsilon_0} + \frac{(\boldsymbol{\nabla}\times \mathbf A)^2}{\mu_0}
    \Big\} \notag \\ =&
    \frac{1}{2}\sum_{\mathbf k} \Big\{ p^{\dagger}_{\mathbf k}(t)p_{\mathbf k}(t) +
    \omega^2_{\mathbf k} q^{\dagger}_{\mathbf k}(t)q_{\mathbf k}(t) \Big\}.
  \end{align}
  The mode functions $\big\{ \mathbf f_{\mathbf k} \big\}$ are thus
  the spatial basis diagonalizing the spatially averaged Hamiltonian, and
  as we shall see the proper basis describing the diffraction problem.

  The splitting in equation (\ref{eq:splitting_ham}) allows us to
  consider the problem as comprised of two types of properties. The
  effect of single atoms, and the spatially averaged Hamiltonian. The
  effect of the spatially averaged Hamiltonian is well understood in terms
  of the mode-functions defined in equation
  (\ref{eq:boldsymb-barb-vhatm}). The point particle effect we will
  discuss in greater detail when considering the equations of motion
  for the full system. Before deriving these equations of motion we,
  however, briefly need to discuss the commutation relations
  describing the system.
  \subsection{Quantization and Commutation relations}
   Above we expanded the fields in convenient spatial modes.
   The coordinates $p_{\mathbf k}(t)$ and $q_{\mathbf k}(t)$
   are canonically conjugate variables, and we can thus quantize our
   theory by imposing the commutation relations
   \begin{align}
     \label{eq:can_quant_commu_rel}
     \big[ q_{\mathbf k}(t),p_{\mathbf k'}(t) \big]= i \hbar
     \delta_{\mathbf k \mathbf k'}.
   \end{align}
   It will however be convenient to have the commutation relations for
   the fields which we may derive from the mode-amplitude commutation
   relations. It will also be convenient to separate the displaced
   electric field into a positively and a negatively oscillating part
   $\mathbf D = \mathbf D^{(+)} + \mathbf D^{(-)}$, where $\mathbf
   D^{(-)}$ is in accordance with convention chosen so that it only
   contains terms oscillating like $e^{i\omega_{\mathbf k}t}$. Our
   choice of gauge is reflected in the 
   transversality of the mode functions defined in
   Eq. (\ref{eq:boldsymb-barb-vhatm}). We expect this transversality
   condition to be represented in the commutation relations as well.
   With the quantization procedure above one finds the following
   expression for the negative frequency part of the relevant fields
  \begin{subequations}
    \begin{align}
      \hat{\mathbf D}^{(-)}(\mathbf r,t) =& -i\sum_{\mathbf k}
      \sqrt{\frac{\hbar \omega_{\mathbf k}\epsilon_0}{2}} \hat
      a_{\mathbf k}^{\dagger} e^{i\omega_{\mathbf k}t}\mathbf
      f_{\mathbf k}^*(\mathbf r) \label{eq:hatmathbf-d+mathbf-r}\\
      \hat{\mathbf A}^{(-)}(\mathbf r,t) =& \sum_{\mathbf k} c \sqrt{\frac{\hbar
          \mu_0}{2\omega_{\mathbf k}}} \hat a^{\dagger}_{\mathbf k}
      e^{i\omega_{\mathbf k}t} (1-\bar{\bar{\mathcal V}}[{\mathbf
        J}])^{t}\mathbf f^*_{\mathbf k}(\mathbf r).\label{eq:hatmathbf-a+mathbf-r}
    \end{align}
  \end{subequations}
  The positive frequency part may be found by Hermitian conjugation.
  The above result is found from equation (\ref{eq:mathbf-dmathbf-r})
  along with the definitions of creation and annihilation operators
  given by
  \begin{subequations} \label{eq:QP_representation}
    \begin{align}
      q_{\mathbf k}(t)=& \sqrt{\frac{\hbar}{2\omega_{\mathbf k}}} \Big\{
      \hat a_{\mathbf k}(t) + \sum_{\mathbf k'} U^*_{\mathbf k \mathbf
        k'} \hat a_{\mathbf k'}^{\dagger}(t) \Big\}
      \\
      p_{\mathbf k}(t) =& i\sqrt{\frac{\hbar\omega_{\mathbf k}}{2}}
      \Big\{ \hat a_{\mathbf k}^{\dagger}(t) - \sum_{\mathbf k'}
      U_{\mathbf k \mathbf k'} \hat a_{\mathbf k'}(t) \Big\},
    \end{align}
  \end{subequations}
  where the matrix $U_{\mathbf k \mathbf k'}$ is defined as
  \begin{align}
    U_{\mathbf k \mathbf k'} = \int d^3r \bar{\bar{\mathcal M}} \mathbf f_{\mathbf k}(\mathbf r) \cdot \mathbf f_{\mathbf
      k'}(\mathbf r).
  \end{align}
  A detailed discussion of this procedure is found in Ref.
  \cite{glauber91:_quant}.  

From these definitions and the commutation
  relations (\ref{eq:can_quant_commu_rel}) we obtain 
  \begin{align}
    \label{eq:crea_annihi_comm_rel}
    \big[ \hat a_{\mathbf k}(t) , \hat a^{\dagger}_{\mathbf k}(t)
    \big] = \delta_{\mathbf k \mathbf k'}. 
  \end{align}
  Going to the field operators we get
  \begin{align}
    \big[  \hat{\mathbf D}^{(+)}(\mathbf r,t) ,\hat{\mathbf A}^{(+)}(\mathbf
    r',t) \big] =& 0\label{eq:big-hatmathbf-d}  \\
    \big[ \hat{\mathbf D}^{(+)}(\mathbf r,t) ,\hat{\mathbf A}^{(-)}(\mathbf
    r',t) \big] =& \frac{i \hbar}{2} \bar{\bar{\delta}}^T(\mathbf
    r,\mathbf r'),\label{eq:other_commutation_rel}
  \end{align}
  where
  \begin{align} \label{eq:barb-r-mathbf}
   \bar{\bar{\delta}}^T(\mathbf r,\mathbf r') =&
    \sum_{\mathbf k} \mathbf f_{\mathbf k}(\mathbf
    r)\Big[ \bar{\bar{\mathcal M}}^t \mathbf f^*_{\mathbf
      k}(\mathbf r')\Big] .
  \end{align}
  Here $ \bar{\bar{\delta}}^T(\mathbf r,\mathbf r')$ is a generalized
  transverse delta function \cite{glauber91:_quant}. This may be seen
  by considering its action on some transverse vector field
  ($\boldsymbol{\nabla}\cdot \boldsymbol{\psi}(\mathbf r,t)=0$). 
  Since $\big\{ \mathbf
  f_{\mathbf k} \big\}$ is a complete basis on the set of
  transverse fields, we may expand $\boldsymbol{\psi}(\mathbf r,t)$ as
  \begin{align}
    \boldsymbol{\psi}(\mathbf r,t) = \sum_{\mathbf k} \hat C_{\mathbf
      k}(t)\mathbf f_{\mathbf k}(\mathbf r). 
  \end{align}
  If we calculate the effect of the transverse delta-function on a
  transverse field we find 
  \begin{align}
    \int d^3r'\; \bar{\bar{\delta}}^T(\mathbf r,\mathbf r')\cdot \
    \boldsymbol{\psi}(\mathbf r',t) =& \notag \\ \int
    d^3r'\;\sum_{\mathbf k\mathbf k'} \hat C_{\mathbf k}(t) \mathbf
    f_{\mathbf k'}(\mathbf r)\Big[ &\bar{\bar{\mathcal M}}^t \mathbf
    f^*_{\mathbf k'}(\mathbf r') \cdot \mathbf f_{\mathbf k}(\mathbf
    r') \Big] \notag \\ \sum_{\mathbf k\mathbf k'} \hat C_{\mathbf k}(t)
    \mathbf f_{\mathbf k'}(\mathbf r) \delta_{\mathbf k \mathbf k'} =&\;
    \boldsymbol{\psi}(\mathbf r,t),
  \end{align}
  where we have used the orthonormality condition of the basis-functions.
   
  We shall also need the equal-space commutation relations
   \begin{align*}
    \big[ \hat{\mathbf D}^{(+)}(\mathbf r,t) , \hat{\mathbf D}^{(-)}(\mathbf r,t') \big].
  \end{align*}
  A formal expression of this commutation relation can be found from
  Eq. (\ref{eq:hatmathbf-d+mathbf-r}) to be
  \begin{align}\label{eq:big-hatmathbf-d-1}
    \big[ \hat{\mathbf D}^{(+)}(\mathbf r,t) , \hat{\mathbf D}^{(-)}(\mathbf
    r,t') \big] = \frac{\hbar \epsilon_0}{2} \bar{\bar{\eta}}(\mathbf r,t,t'),
  \end{align}
  where
  \begin{align}\label{eq:barbaretamathbf-r-t}
    \bar{\bar{\eta}}(\mathbf r,t,t') = 
    \sum_{\mathbf k}
    \omega_{\mathbf k} \mathbf f_{\mathbf k}(\mathbf r) \mathbf
    f^*_{\mathbf k}(\mathbf r) e^{-i\omega_{\mathbf k}(t-t')}. 
  \end{align}
  In vacuum $\bar{\bar{\eta}}(\mathbf r,t,t')$ is simple to
  evaluate, but for complex systems it is nontrivial to gain
  knowledge of the basis-functions $\big\{ \mathbf f_{\mathbf k}
  \big\}$. In Appendix \ref{sec:calc-infin-short} we calculate
  $\bar{\bar{\eta}}$ using the rotating-wave approximation and the
  local density approximation, where we assume that $\rho (\mathbf r)$
  varies slowly with respect to $\mathbf r$.
  
  %
  \section{Equations of motion}\label{sec:equations-motion}
  In this section we derive the equations of motion for the system,
  and consider their general properties. In the previous section we
  discussed that the theory could be divided into an average part and
  a part representing the deviation from the average. To derive the
  equations of motion we will, however, work with the full Hamiltonian
  and later make the splitting into the average part and the
  deviations from it. The strategy we will use is to first derive the
  quantum mechanical Maxwell equations, and then to combine them into an
  effective wave equation for the field.  

  We will now as an example derive one of the quantum mechanical Maxwell
  equations from Heisenberg's equation of motion:
  \begin{align}\label{eq:fracddth-dmathbf-r}
    \frac{d}{dt}&\hat{\mathbf D}(\mathbf r) = \frac{i}{\hbar} \big[
    \hat{\mathcal H} , \hat{\mathbf D}(\mathbf r) \big] \notag \\
    =& \frac{i}{2\hbar \mu_0} \int d^3r'\; \big[
    (\boldsymbol{\nabla}\times \hat{\mathbf A}(\mathbf r'))^2 ,
    \hat{\mathbf D}(\mathbf r) \big] \notag \\
    =& \frac{i}{2\hbar \mu_0} \int d^3r'\;
    \Big\{(\boldsymbol{\nabla}\times \boldsymbol{\nabla}\times
    \hat{\mathbf A}(\mathbf r') ) \cdot \big[ \hat{\mathbf A}(\mathbf
    r') , \hat{\mathbf D}(\mathbf r) \big] \notag \\ &\hspace{0.5cm} +
    \big[ \hat{\mathbf A}(\mathbf r') , \hat{\mathbf D}(\mathbf r)
    \big] \cdot (\boldsymbol{\nabla}\times \boldsymbol{\nabla}\times
    \hat{\mathbf A}(\mathbf r') ) \Big\}.
  \end{align}
  Here we have used the Hamiltonian given in
  Eq.(\ref{eq:mathcal-h-=}), and the boundary condition that the
  physical fields vanish at infinity.  To shorten the notation we have
  suppressed the explicit time dependence.  
  The commutation relation may be
  found from (\ref{eq:big-hatmathbf-d}) and
  (\ref{eq:other_commutation_rel}) to be
  \begin{align} \label{eq:big-hatm-amathbf}
    \big[ \hat{\mathbf A}(\mathbf r') , \hat{\mathbf D}(\mathbf r)
    \big] = -i\hbar \bar{\bar{\delta}}^T(\mathbf r,\mathbf r').
  \end{align}
  Since the field $\boldsymbol{\nabla}\times \hat{\mathbf A}$ is
  transverse by definition, this gives us the first quantum mechanical
  Maxwell equation.
  \begin{align}\label{eq:fracddth-dmathbf-r-2}
    \frac{d}{dt}\hat{\mathbf D}(\mathbf r) =& \frac{1}{\mu_0}
    \boldsymbol{\nabla}\times \hat{\mathbf B}(\mathbf r),
    \intertext{where} \hat{\mathbf B}(\mathbf r) =&
    \boldsymbol{\nabla}\times \hat{\mathbf A}(\mathbf r) \label{eq:fracddth-dmathbf-r-1}.
  \end{align} 
  Similarly we may derive the Maxwell equation
  $\boldsymbol{\nabla} \times \hat{\mathbf E} = -\partial_t \hat{\mathbf B}$,
  where $\hat{\mathbf E} = -d \hat{\mathbf A}/d
  t=\hat{\mathbf D} - \hat{\mathbf P}$. The
  remaining Maxwell equations $\boldsymbol{\nabla} \cdot \hat{\mathbf B} =0 $
  and $\boldsymbol{\nabla} \cdot \hat{\mathbf D}=0$ follow immediately from
  the definition of $\hat{\mathbf B}$ in Eq. (\ref{eq:fracddth-dmathbf-r-1})
  and from the transversality of $\hat{\mathbf D}$. 

  Because of the nature of the interaction part of the Hamiltonian, it
  is convenient to consider the two frequency components of the
  displaced electric field separately.   The quantum mechanical Maxwell
  equations may be combined into a single wave equation
   \begin{align}\label{eq:big-fracp-t2}
    &\Big( \frac{d^2}{d t^2} + c^2
    \boldsymbol{\nabla}\times \boldsymbol{\nabla}\times \Big)
    \hat{\mathbf D}^{(-)}(\mathbf r,t) = \notag \\ & c^2 \int d^3r\;
    \boldsymbol{\nabla}\times \boldsymbol{\nabla}\times
    \bar{\bar{\delta}}^T(\mathbf r,\mathbf r') \cdot \bar{\bar V}[\hat{
    \mathbf J}]^t \hat{\mathbf D}^{(-)}(\mathbf r',t),
  \end{align}
  where the positive frequency part can be found by Hermitian
  conjugation.  Similarly we may derive equations for the spin of the
  atoms, and for the simple interactions given in Eq.
  (\ref{eq:barb-vmathbf-j_j=b}), one finds
  \begin{align} \label{eq:fracddt-hatm-jmathbf-1}
    \frac{d}{dt} \hat{\mathbf J}(\mathbf r,t) = \frac{i\beta c_1}{\hbar
      \epsilon_0} \hat{\mathbf J}(\mathbf r,t) \times \Big(
    \hat{\mathbf D}^{(-)}(\mathbf r,t) \times \hat{\mathbf D}^{(+)}(\mathbf
    r,t) \Big). 
  \end{align}
  In the remainder of this article we will solve these coupled
  partial differential equations.

  The expression in Eq. (\ref{eq:big-fracp-t2}) is a second order
  differential equation in time. The solution of this equation will in
  general not only depend on the initial value $\mathbf D(\mathbf
  r,t=t_0)$, but also the time derivative $\partial_t \mathbf
  D(\mathbf r,t)|_{t=t_0}$. In deriving our interaction we have, however, already
  used the rotating wave approximation, where we ignore the dynamics on
  a time scale similar to the inverse of the optical frequency. Similarly we shall
  here make a slowly-varying-envelope approximation and write the
  displaced electric field as
    \begin{align}
    \hat{\mathbf D}(\mathbf r,t) = \tilde{\mathbf D}^{(-)}(\mathbf
    r,t)e^{i\laser t} + \tilde{\mathbf D}^{(+)}(\mathbf
    r,t)e^{-i\laser t},
  \end{align}
  where $\tilde{\mathbf D}^{(\pm)}$ are slowly varying in time. If we
  ignore the second derivative of the slowly varying operators 
  ($\partial_t^2 \tilde{\mathbf D}^{(\pm)}(\mathbf r,t)\approx 0$), then  Eq.
  (\ref{eq:big-fracp-t2}) reduces to a first-order differential
  equation in time. 

  Since we are heading towards a perturbation theory in the
  point-particle part of the Hamiltonian (\ref{eq:splitting_ham}), we
  will add and subtract the average part of the source term in Eq.
  (\ref{eq:big-fracp-t2}).  That is we write
   \begin{align}
     \bar{\bar V}[\hat{\mathbf J}] = \bar{\bar V}[\hat{\mathbf J}] -
     \bar{\bar{\mathcal V}}[\mathbf J] + \bar{\bar{\mathcal
         V}}[\mathbf J] \equiv \bar{\bar m}[\hat{\mathbf J}] +
     \bar{\bar{\mathcal V}}[\mathbf J].
   \end{align}
   The idea in this separation is that now $\bar{\bar{\mathcal
       V}}[\mathbf J]$ represents the average effect of the ensemble,
   which may have a big effect, whereas $\bar{\bar{ m}}[\hat{\mathbf
     J}]$ represents the fluctuations around this average. To take
   advantage of this we first consider the average term
     \begin{align}
    \int d^3r\; \boldsymbol{\nabla}\times \boldsymbol{\nabla}\times
    \bar{\bar{\delta}}^T(\mathbf r,\mathbf r') \cdot \bar{\bar{\mathcal
      V}}[{ \mathbf J}]^t \hat{\mathbf D}^{(-)}(\mathbf r',t).
  \end{align} 
  This term is continuous and we may use partial integration twice.
  Using the expression for the general transverse delta-function one
  finds
  \begin{align}
    \int d^3r\; \boldsymbol{\nabla}&\times \boldsymbol{\nabla}\times
    \bar{\bar{\delta}}^T(\mathbf r,\mathbf r') \cdot \bar{\bar{\mathcal
        V}}[{ \mathbf J}]^t \hat{\mathbf D}^{(-)}(\mathbf r',t) \notag \\
    &= \boldsymbol{\nabla}\times \boldsymbol{\nabla}\times \bar{\bar{\mathcal
        V}}[{ \mathbf J}]^t \hat{\mathbf D}^{(-)}(\mathbf r,t).
  \end{align}
  This term we will move to the left hand side of Eq.
  (\ref{eq:big-fracp-t2}), and we are left with a diffusion equation
  involving only the fluctuations as a source term on the right hand
  side
     \begin{align}
    \Big( 2i\laser\frac{d}{d t} -&\laser^2 + c^2
    \boldsymbol{\nabla}\times \boldsymbol{\nabla}\times
    \bar{\bar{\mathcal M}}^t  \Big)  \tilde{\mathbf
      D}^{(-)}(\mathbf r,t) \notag \\ =&\;c^2\int d^3r\;
    \boldsymbol{\nabla}\times \boldsymbol{\nabla}\times
    \bar{\bar{\delta}}^T(\mathbf r,\mathbf r') \cdot \bar{\bar
      m}[\hat{ \mathbf J}]^t \tilde{\mathbf D}^{(-)}(\mathbf r',t).
    \label{eq:-=c2int-d3r}  
  \end{align}
  If we put the right hand side of this equation to zero, i.e., ignore the
  fluctuations, this equation describes the propagation and
  diffraction of the field in a medium.  
  For instance if we take the simplest case where the medium is
  isotropic so that the matrix $\bar{\bar{\mathcal{V}}}[\mathbf J]$ is
  just a scalar, this equation describes the propagation through a
  medium with an index of refraction given by
  $n=1/\sqrt{1-{\bar{\bar{\mathcal{V}}}[\mathbf J]}}$, see
    Ref. \cite{glauber91:_quant}.
  
  \section{General solution and Feynman diagrams}\label{sec:born-appr-feynm}
  In this section we discuss the solution of Eq.
  (\ref{eq:-=c2int-d3r}) in terms of its Green's function.  Let us for
  convenience define the differential operator
  \begin{align}\label{eq:mathcal-d-=}
    \mathcal D = 2i\laser\frac{d}{d t} - \laser^2 + c^2
    \boldsymbol{\nabla}\times \boldsymbol{\nabla}\times \bar{\bar{\mathcal
      M}}^t(\mathbf r).
  \end{align}
  We then define the Green's function by
  \begin{align}\label{eq:mathc-dbarb-gmathbf}
    \mathcal D\bar{\bar G}^{(-)}(\mathbf r,t|\mathfb r_0,t_0) =
    \bar{\bar{\delta}}^{T}(\mathbf r,\mathbf r_0)\delta(t-t_0).
  \end{align}
  The right hand side of this equation describes an identity
  functional on the inner product space we are working in.  We want
  the Green's function to describe an evolution of the system forward
  in time. We therefore define a cut-off on the Green's function in
  time
  \begin{align}
    \bar{\bar G}^{(-)}(\mathbf r,t|\mathbf r_0,t_0)=0 \quad \text{for} \quad
    t<t_0.
  \end{align}
  The general solution to Eq. (\ref{eq:-=c2int-d3r}) in terms of
  Green's functions is discussed in detail in Appendix
  \ref{sec:recipr-equat-greens}, and reads
      \begin{widetext}
    \begin{align}\label{eq:tild-d+mathbf-r}
      \tilde{\mathbf D}^{(-)}(\mathbf r,t) =&2i\laser \int d^3 r'
      \;\bar{\bar{\mathcal M}}^t(\mathbf r') \bar{\bar
        G}^{(-)}(\mathbf r,t|\mathbf r',t_0) \cdot \tilde{\mathbf
        D}^{(-)}(\mathbf r',t_0) \notag \\ &+ c^2\iint_{t_0}^{t^+}
      d^3r'dt'\;\bar{\bar{\mathcal M}}^t(\mathbf r') \bar{\bar
        G}^{(-)}(\mathbf r,t|\mathbf r',t') \cdot \int
      d^3r''\; \boldsymbol{\nabla}'\times \boldsymbol{\nabla}'\times
      \bar{\bar{\delta}}^T(\mathbf r',\mathbf r'') \cdot \bar{\bar
        m}[\hat{ \mathbf J}]^t \tilde{\mathbf D}^{(-)}(\mathbf
      r'',t').
    \end{align}
  \end{widetext}
  The upper limit is understood to be $t^+ =\lim_{\varepsilon
    \rightarrow 0} [t+\varepsilon]$. Before continuing a few comments
  are in order.  Here we have used the boundary conditions, that all
  fields vanish at infinity, i.e., we imagine that at time $t=0$ we
  have generated an optical pulse inside the volume we are describing,
  which travels toward the atomic medium. Alternatively we could have
  described the incomming field by a boundary term.
  The positive frequency part may be found by Hermitian conjugation.

  Let us now consider the last term of
  Eq. (\ref{eq:tild-d+mathbf-r}). We notice that the involved fields
  are all continuous and differentiable with respect to the primed
  spatial coordinates. Using partial integration twice and introducing
  the propagator defined by
  \begin{align}\label{eq:barbar-p+mathbf-r}
    \bar{\bar P}^{(-)}(\mathbf r,t|\mathbf r',t')=
    \boldsymbol{\nabla}'\times \boldsymbol{\nabla}'\times
    \bar{\bar{\mathcal M}}^t(\mathbf r') \bar{\bar G}^{(-)}(\mathbf
    r,t|\mathbf r',t')
  \end{align}
  the last term of Eq. (\ref{eq:tild-d+mathbf-r}) may be written as
  \begin{align}\label{eq:c2iint_t_0t-d3rdt-d3}
    c^2\iint_{t_0}^t d^3r'dt'\int d^3r''\; \bar{\bar P}^{(-)}(\mathbf
    r,t|\mathbf r',t') \cdot &\bar{\bar{\delta}}^T(\mathbf r',\mathbf
    r'') \notag \\ &\cdot \bar{\bar m}[\hat{ \mathbf J}]^t
    \tilde{\mathbf D}^{(-)}(\mathbf r'',t').
  \end{align}
  Due to the cross product in Eq. (\ref{eq:barbar-p+mathbf-r}) the
  propagator is transverse with respect to primed coordinates and the
  transverse delta function in (\ref{eq:c2iint_t_0t-d3rdt-d3}) may be
  integrated out, giving
  \begin{align}
    c^2\iint_{t_0}^t d^3r'dt'\; \bar{\bar P}^{(-)}(\mathbf r,t|\mathbf
    r',t') \cdot \bar{\bar m}[\hat{ \mathbf J}]^t \tilde{\mathbf
      D}^{(-)}(\mathbf r',t').
  \end{align}

  The first term of the right hand side of equation
  (\ref{eq:tild-d+mathbf-r}) we will denote as $\tilde{\mathbf
    D}_0^{(-)}(\mathbf r,t)$
  \begin{align}\label{eq:2ilaser-int-d3}
     \tilde{\mathbf D}_0^{(-)}&(\mathbf
    r,t) = \notag \\  &2i\laser \int d^3 r' \;\bar{\bar M}^t(\mathbf r') \bar{\bar
      G}^{(-)}(\mathbf r,t|\mathbf r',t_0) \cdot \tilde{\mathbf
      D}^{(-)}(\mathbf r',t_0). 
  \end{align}
  If there were no deviation from the mean, i.e. $\bar{\bar
    m}[\hat{\mathbf J}]=0$, the solution would simply be
  $\tilde{\mathbf D}^{(-)}(\mathbf r,t) = \tilde{\mathbf
    D}_0^{(-)}(\mathbf r,t)$.  $\tilde{\mathbf D}_0^{(-)}(\mathbf r,t)
  $ thus denotes the solution to the diffraction problem, where the
  atomic medium is treated as a continuous medium with a diffraction
  matrix $\bar{\bar{\mathcal M}}$. 
  
  \bigskip 

  \subsection{Perturbative expansion}

  Below we shall develop a perturbative expansion in the deviation
  from the mean due to quantum fluctuations and from the fact that the
  medium is not continuous but consists of a large number of point
  particles.  The starting point for the perturbative expansion will
  be the \emph{field equation}
  \begin{align}\label{eq:tild-d+mathbf-r-1}
    \tilde{\mathbf D}^{(-)}&(\mathbf r,t) =\tilde{\mathbf
      D}_0^{(-)}(\mathbf r,t) \notag \\ &+ c^2\iint_{t_0}^t d^3r'dt'\;
    \bar{\bar P}^{(-)}(\mathbf r,t|\mathbf r',t') \cdot \bar{\bar
      m}[\hat{ \mathbf J}]^t \tilde{\mathbf D}^{(-)}(\mathbf r',t').
  \end{align}
  In addition to this we shall also need the solution to the equations
  of motion for the spin (\ref{eq:fracddt-hatm-jmathbf-1}), which may
  be formally solved to give the \emph{spin equation}
  \begin{align}\label{eq:hatmathbf-jmathbf-r}
    \hat{\mathbf J}(\mathbf r,&t) = \hat{\mathbf J}(\mathbf
    r,t_0)\notag \\ &+ \frac{i\beta c_1}{\hbar \epsilon_0}
    \int_{t_0}^t dt'\; \hat{\mathbf J}(\mathbf r,t') \times \Big(
    \tilde{\mathbf D}^{(-)}(\mathbf r,t') \times \tilde{\mathbf
      D}^{(+)}(\mathbf r,t') \Big).
  \end{align}
  These are the equations we wish to treat using the Born
  approximation, where we make an expansion in the interaction
  parameter $\beta$. ( In Eq.  (\ref{eq:tild-d+mathbf-r-1}) the
  interaction $\bar{\bar m}[\hat{\mathbf J}]^t$ is proportional to
  the expansion parameter $\beta$.)

  In terms
  of notation this expansion gets extremely cumbersome.  It is
  therefore convenient to introduce Feynman diagrams to represent the
  various terms of the expansion. We will be dealing with two types of
  interactions: the one given in Eq.  (\ref{eq:tild-d+mathbf-r-1})
  which we
  will represent with a shaded circle, and the one given in Eq.
  (\ref{eq:hatmathbf-jmathbf-r}) which we will represent with a shaded
  triangle. The field equation, we diagrammatically represent as
\begin{widetext}
\begin{fmffile}{feynman_d_1}
  \fmfcmd{%
    style_def wiggly_arrow expr p =
    cdraw (wiggly p);
    shrink (1);
       cfill (arrow p);
     endshrink;
     enddef;}

 \fmfcmd{%
    style_def DwigA expr p =
    draw_double (wiggly p);
    shrink (1.5);
       cfill (arrow p);
     endshrink;
     enddef;}
   \begin{align}
  \begin{fmfgraph*}(60,30) \fmfleft{i} \fmfright{o}
         \fmflabel{$t$}{o}
    \fmf{DwigA}{i,o} 
    \fmfdot{o}
    \end{fmfgraph*} 
    \hspace{.6cm} &\raisebox{0.4cm}{=} \hspace{.2cm}
       \begin{fmfgraph*}(60,30) \fmfleft{i} \fmfright{o}
         \fmflabel{$t$}{o}
    \fmf{wiggly_arrow}{i,o} 
    \fmfdot{o}
  \end{fmfgraph*} \hspace{.6cm} \raisebox{0.4cm}{+} \hspace{.2cm}
    \begin{fmfgraph*}(70,30) \fmfleftn{i}{2} \fmfright{o}
        \fmflabel{$t$}{o}
    \fmf{DwigA}{i2,v} \fmf{dbl_plain_arrow}{i1,v}
    \fmf{wiggly_arrow}{v,o}
    \fmffreeze
    \fmfshift{0.2w,0h}{v}
    \fmfdot{o} \fmfblob{.2w}{v}
    \end{fmfgraph*}\hspace{.5cm}, 
\intertext{and the spin equation is represented as}
   \begin{fmfgraph*}(60,30) \fmfleft{i} \fmfright{o}
         \fmflabel{$t$}{o}
    \fmf{dbl_plain_arrow}{i,o} 
    \fmfdot{o}
    \end{fmfgraph*} 
    \hspace{.6cm} &\raisebox{.4cm}{=} \hspace{.1cm}
       \begin{fmfgraph*}(60,30) \fmfleft{i} \fmfright{o}
         \fmflabel{$t$}{o}
    \fmf{fermion}{i,o} 
    \fmfdot{o}
  \end{fmfgraph*} \hspace{.6cm} \raisebox{.4cm}{+} \hspace{.1cm}
  \begin{fmfgraph*}(70,30) \fmfleftn{i}{3} \fmfright{o}
        \fmflabel{$t$}{o}
        \fmfpolyn{triagram,tension=.7,filled=shaded}{v}{3}
    \fmf{DwigA,tension=.5,left=.5}{i3,v1} 
    \fmf{DwigA,tension=.5,left=.5}{v1,i2}
    \fmf{dbl_plain_arrow}{i1,v2}
    \fmf{fermion}{v3,o}
    \fmffreeze
    \fmfshift{0.2w,0h}{v1,v2,v3}
    \fmfshift{.09w,.1h}{i2}
    \fmfdot{o} 
  \end{fmfgraph*}
\hspace{.5cm}.
\end{align}
      \end{fmffile}
    \end{widetext}
  The orientation of the diagram is such that time is going from left
  to right, and the evaluation at time $t$ is marked by a dot.
  Spin propagation is represented by a line with an arrow pointing in the
  positive-time direction. A wiggly line represents propagation of the
  displaced electric field. The arrow denotes whether the line
  represent the photon-generating part of the field, $\tilde{\mathbf
    D}^{(-)}(\mathbf r,t)$, where the arrow points forward in time, or
  the photon-annihilating part of the field, where the arrow points
  backward in time.  The full solution to the spin $\hat{\mathbf
    J}(t)$ is denoted with a double straight line, and the full
  solution to the displaced electric field is denoted with a double
  wiggly line. 

  The field equation and the spin equation can be represented as a
  perturbation series, and in the following we shall discuss the
  effect of the terms in this perturbation series. An important
  feature of our system is the random distribution of the atoms in the
  ensemble.  The equations that we have derived so far apply to each
  realization of the atomic distribution $\big\{ \mathbf r_1,\mathbf
  r_2, \ldots ,\mathbf r_N \big\}$.  However since we have no control
  of the position of the atoms we will have to make a spatial average
  of our equations, that is of the terms in the perturbation series.
  To do this we need to know the density correlations of the gas.

  \subsection{Density correlations. }\label{sec:dens-corr}

  We assume that we are dealing with an ideal gas, i.e., we assume
  that the distribution of the atoms is completely random but has a
  distribution given by the possible spatially varying density
  $\rho(\mathbf r)$, and we assume that there are no correlations
  between the positions of different atoms.  The correlation function
  for the density distribution $\rho(\mathbf r)=\sum_j \delta(\mathbf
  r - \mathbf r_j)$ is thus
  \begin{align}
    \< \rho(\mathbf r) \rho(\mathbf r')\>_{\text{s.a.}} =&\; \< \sum_{jl} \delta
    (\mathbf r - \mathbf r_j) \delta
    (\mathbf r' - \mathbf r_l) \>_{\text{s.a.}} \notag \\
     =&\sum_{j\neq l} \< \delta (\mathbf r - \mathbf r_j) \delta
    (\mathbf r' - \mathbf r_l) \>_{\text{s.a.}} \notag \\ &
    \hspace{1.5cm}+ \sum_j \delta (\mathbf r - \mathbf
    r') \< \delta ( \mathbf r - \mathbf r_j) \>_{\text{s.a.}} \notag \\
    =& \;\< \rho (\mathbf r )\>_{\text{s.a.}} \< \rho (\mathbf r') \>_{\text{s.a.}} + \delta
    (\mathbf r -\mathbf r') \< \rho (\mathbf r) \>_{\text{s.a.}}. 
\label{eq:densitydensity}
  \end{align}
  Here $\<\cdot \>_{\text{s.a.}}$ denotes spatial averaging. In
  the last step we used that the distribution is independent for different
  atoms, and we ignored the small difference between $N_A^2$ and $N_A(N_A-1)$,
  where $N_A$ is the number of atoms. We have also neglected the
  effect that two different atoms can not be found at the same point
  in space. While this may seem insignificant for a low density gas,
  we show in Appendix \ref{sec:lorenz-lorenz-corr} that including this
  effect to all orders in the perturbation series gives the Lorentz-Lorenz
  correction to the index of refraction. 

Below we shall also use the
  correlation functions for the spin. Similar to the calculation above
  we find
    \begin{align} \label{eq:spin_correalation} \< \hat J_n(\mathbf r)
    \hat J_m (\mathbf r') \>_{\text{s.a.}} = \rho(\mathbf r)&\rho(\mathbf r')\bar
    J_n(\mathbf r) \bar J_m(\mathbf r') \notag \\ &+ \rho (\mathbf r)
    \delta(\mathbf r -\mathbf r') \bar J_n(\mathbf r)\bar J_m(\mathbf
    r),
  \end{align}
  where the index $n,m$ refer to the spatial components of the
  operators.  To shorten notation we have written $\rho(\mathbf r)$
  instead of $\<\rho(\mathbf r)\>_{\text{s.a.}}$. As discussed
  previously the bar denotes a single atom operator. 
  We will preserve the quantum mechanical behavior of the operators by
  not taking the quantum mechanical mean.  The first term on the right
  hand side of Eq.  (\ref{eq:spin_correalation}) arises from the
  contribution from different atoms (signified by the prime on the
  second spin operator). In the second term on the other hand the two
  operators refer to the same atom, and the operator product should be
  evaluated for a single atom.  For example for a spin-$½$ system, we have
  the following relation between products of spin operators on single
  atoms
  \begin{align} \label{eq:spin_one_half}
    \bar J_n(\mathbf r)\bar J_m(\mathbf r) =
    \frac{i}{2}\varepsilon_{nml}\bar J_l(\mathbf r).
  \end{align}
  The generalization to even higher-order density correlations is
  straight-forward. 

  These considerations become important when we calculate the spatial
  average of the second-order terms of the perturbation series. Let us
  as an example consider the second order term of the spin equation
  representing a photon first interacting with one atom and then later
  with the atom in consideration.

\begin{fmffile}{feynman_d_1b}
  \fmfcmd{%
    style_def wiggly_arrow expr p =
    cdraw (wiggly p);
    shrink (1);
       cfill (arrow p);
     endshrink;
     enddef;}

 \fmfcmd{%
    style_def DwigA expr p =
    draw_double (wiggly p);
    shrink (1.5);
       cfill (arrow p);
     endshrink;
     enddef;}
   \begin{align}\label{eq:rais-.5cmb-50}
\raisebox{-.5cm}{\begin{fmfgraph*}(60,50) \fmfleftn{i}{4} \fmfright{o}
        \fmflabel{$t$}{o}
        \fmfpolyn{triagram,tension=.8,filled=shaded}{v}{3}
    \fmf{wiggly_arrow,tension=.5,left=.5}{i3,k,v1} 
    \fmf{wiggly_arrow,tension=.5,left=.5}{v1,i2}
    \fmf{fermion}{i1,v2}
    \fmf{fermion}{v3,o}
    \fmf{fermion}{i4,k}
    \fmffreeze
    \fmfshift{0.25w,0h}{v1,v2,v3}
    \fmfshift{.15w,0h}{k}
    \fmfshift{-.1w,0h}{i4}
    \fmfshift{0w,.06h}{i2}
    \fmfshift{0w,-.06h}{i3}
    \fmfdot{o} \fmfblob{.1w}{k}
    \end{fmfgraph*}}
 \qquad \overrightarrow{\;\text{s.a.}\;}\;
 \begin{array}{c}
   \begin{fmfgraph*}(60,50) \fmfleftn{i}{4} \fmfright{o}
        \fmflabel{$t$}{o}
        \fmfpolyn{triagram,tension=.8,filled=30}{v}{3}
    \fmf{wiggly_arrow,tension=.5,left=.5}{i3,k,v1} 
    \fmf{wiggly_arrow,tension=.5,left=.5}{v1,i2}
    \fmf{fermion}{i1,v2}
    \fmf{fermion}{v3,o}
    \fmf{fermion}{i4,k}
    \fmffreeze
    \fmfshift{0.25w,0h}{v1,v2,v3}
    \fmfshift{.15w,0h}{k}
    \fmfshift{-.1w,0h}{i4}
    \fmfshift{0w,.06h}{i2}
    \fmfshift{0w,-.06h}{i3}
    \fmfdot{o} \fmfv{decor.shape=circle,decor.size=.1w,decor.filled=30}{k}
    \end{fmfgraph*} \vspace{.4cm} \\
     \begin{fmfgraph*}(70,30) \fmfleftn{i}{3} \fmfright{o}
        \fmflabel{$t$}{o}
        \fmfpolyn{pentagram,pull=.6,tension=0.9,filled=gray50}{v}{5}
    \fmf{wiggly_arrow,tension=.5,left=.5}{i3,v5} 
    \fmf{wiggly_arrow,tension=.5,left=.5}{v5,i2}
    \fmf{photon,tension=.4}{v4,v4}
    \fmf{fermion}{i1,v2}
    \fmf{fermion}{v3,o}
    \fmffreeze
    \fmfshift{0.2w,0h}{v1,v2,v3,v4,v5}
    \fmfshift{0.06w,.17h}{i2}
    \fmfdot{o} 
    \end{fmfgraph*}
 \end{array}
    \end{align}
  \end{fmffile}
  
  \noindent When taking spatial average this term generates two terms
  in the perturbative expansion as indicated with the arrow in Eq.
  (\ref{eq:rais-.5cmb-50}). The first term involving the spin of two
  different atoms we will refer to as a coherent interaction, which we
  will discuss later. The second term involving the delta
  function corresponds to  the incoherent interaction (for reasons which will become clear below).  We include
  this situation in the diagrammatic notation by introducing a 
  hatched star and a loop signifying the infinitely
  short propagation stemming from the delta-function term of the
  correlation function Eq. (\ref{eq:spin_correalation}), i.e. 
    \begin{align}\label{eq:int-d3r-barbar}
    \int d^3r \bar{\bar P}^{\pm}(\mathbf r,t|\mathbf r',t') \cdot
    \boldsymbol{\psi}(\mathbf r',t') &\delta(\mathbf r- \mathbf r') =
    \notag \\ & \bar{\bar P}^{\pm} (\mathbf r,t|\mathbf r,t') \cdot
    \boldsymbol{\psi}(\mathbf r,t').
  \end{align}
  The loop is placed on the top of the star when it comes from the
  positively oscillating propagator $\bar{\bar P}^{(-)}$, and in the
  bottom of the star when we refer to the negatively oscillating
  propagator $\bar{\bar P}^{(+)}$.  A star scales with the expansion
  coefficient $\beta$ squared since it involves two interactions.  In
  the next section we will calculate the infinitely short propagator
  appearing in these expressions in the local density approximation.


  \subsection{Green's function and propagator}\label{sec:greens-funct-prop}
  In this section we first derive a formal expression for the Green's
  function. Within our inner
  product space the Green's function is defined by
  \eqref{eq:mathcal-d-=} and \eqref{eq:mathc-dbarb-gmathbf}.
  Expanding our Green's function in the
  basis $\mathbf f^*_{\mathbf k}(\mathbf r)$ we find the representation
  \begin{align}
    \bar{\bar G}^{(-)}(\mathbf r,t|\mathbf r',t') = \sum_{\mathbf k}
    \mathbf f^*_{\mathbf k}(\mathbf r) \mathbf f_{\mathbf k}(\mathbf
    r') g_{\mathbf k}^{(-)}(t,t').
  \end{align}
  We have here expanded on the complex conjugated set $\mathbf f_{\mathbf
    k}^*(\mathbf r)$ to match the expansion of the displaced electric
  field in Eq. (\ref{eq:hatmathbf-d+mathbf-r}).  The transverse
  delta-function has the representation
  \begin{align}
    \bar{\bar{\delta}}^T(\mathbf r,\mathbf r') = \sum_{\mathbf k}
    \mathbf f^*_{\mathbf k}(\mathbf r) \mathbf f_{\mathbf k}(\mathbf
    r')
  \end{align}
  where we are now working in the inner-product space with inner
  product defined in Eq. (\ref{eq:-boldsymb-r}).
  The scalar function $g^{(-)}_{\mathbf k}(t,t')$ is defined by
  \begin{align}\label{eq:big-fracp-t-1}
    \Big( 2i\laser \frac{d}{d t} - \laser^2
    +\omega_{\mathbf k}^2 \Big) g^{(-)}_{\mathbf k}(t,t') = \delta(t-t'),
  \end{align} 
  along with the condition that the function $g_{\mathbf k}(t,t')$
  vanish for $t<t'$.  We will consider the following form of the scalar function,
  where we explicitly write this cut-off in terms of a step function
  \begin{align}
    g^{(-)}_{\mathbf k}(t,t') =Ce^{i\gamma_{\mathbf k}( t -t')} \Theta (t-t').
  \end{align}
  The coefficients $\gamma_{\mathbf k}$ and $C$ is found by inserting this result
  into equation (\ref{eq:big-fracp-t-1}).   \begin{subequations}
    \begin{align}
      \gamma_{\mathbf k} =& \frac{\omega_{\mathbf k}^2 - 
        \laser^2 }{2\laser} \approx \omega_{\mathbf k} - \laser \\
      C =& \frac{-i}{2\laser}.
    \end{align}
  \end{subequations}
  The Green's function is thus given by
  \begin{align}\label{eq:barbar-g+-mathbf}
    \bar{\bar G}^{(-)} (\mathbf r,t|\mathbf r'\hspace{-2pt} ,
    \hspace{-1pt}t') =& -\hspace{-1pt} i \hspace{-1pt}\sum_{\mathbf
      k} \mathbf f^*_{\mathbf k}(\mathbf r) \mathbf f_{\mathbf
      k}(\mathbf r') \frac{e^{i(\omega_{\mathbf k} - \laser)(t-t')}}{2
      \laser}\Theta(t\hspace{-2pt}-\hspace{-1pt}t').
  \end{align}
  
  Next we will look at the infinitely short propagator in Eq.
  (\ref{eq:int-d3r-barbar}).  Using the Green's function given in
  equation (\ref{eq:barbar-g+-mathbf}) along with definition
  (\ref{eq:boldsymb-barb-vhatm-1}) and (\ref{eq:barbar-p+mathbf-r})
  the propagator may be written as
  \begin{align}\label{eq:barbar-p+mathbf-r-2}
    \bar{\bar P}^{(-)}(&\mathbf r,t|\mathbf r,t') =\frac{-i}{2\laser c^2}
    \sum_{\mathbf k} \omega_{\mathbf k}^2 \mathbf f^*_{\mathbf
      k}(\mathbf r) \mathbf f_{\mathbf k}(\mathbf r)
    e^{i(\omega_{\mathbf k} - \laser)(t-t')},
  \end{align}
  where we have omitted the step function since it automatically gives
  unity for the integration limits we are using here.  We will now relate
  this infinitely short propagator to some already known parameter. If we go
  back and consider the general result for the equal-space commutator,
  this may in terms of the basis-functions $\big\{ \mathbf f_{\mathbf
    k}\big\} $ be written as:
  \begin{align}
    \big[ \tilde{\mathbf D}^{(-)}(\mathbf r,t) ; \tilde{\mathbf
      D}^{(+)}(\mathbf r,t') \big] =& \notag \\ -\frac{\hbar \epsilon_0}{2}&
    \sum_{\mathbf k} \omega_{\mathbf k} \mathbf f^*_{\mathbf
      k}(\mathbf r) \mathbf f_{\mathbf k}(\mathbf r)
    e^{i(\omega_{\mathbf k} - \laser)(t-t')}.
  \end{align}
  Comparing with \eqref{eq:barbar-p+mathbf-r-2} we immediately get a
  formal relationship between this commutator and the infinitely short
  propagator
  \begin{align}\label{eq:big-fracp-t-3}
    \Big( \frac{d}{d t'} - i\laser \Big) \big[
    \tilde{\mathbf D}^{(-)}(\mathbf r,t) ; \tilde{\mathbf D}^{(+)}&(\mathbf
    r,t') \big] \notag \\ &= - \hbar \epsilon_0 \laser c^2 \bar{\bar P}^{(-)}(\mathbf
    r,t|\mathbf r,t').
  \end{align}
  Using Eq. (\ref{eq:big-hatmathbf-d-1}) this relation can also be
  written as
  \begin{align}\label{eq:barbar-p+mathbf-r-1}
    \bar{\bar P}^{(-)}(\mathbf r,t|\mathbf r,t') = \frac{1}{2c^2}\Big(
    \frac{d}{d t'} - i\laser \Big) \bar{\bar{
        \eta}}^{*t}(\mathbf r,t,t').
  \end{align}

  To illustrate how the indefinitely short propagator enters into the
  equations we will  again consider the second order term in the spin
  equation represented in Eq. (\ref{eq:rais-.5cmb-50}). The term prior
  to spatial average is given as 
  \begin{align}
    \frac{i\beta c_1 c^2}{\hbar \epsilon_0} \int_{t_0}^t dt'
    \hat{\mathbf J} \times \Big[ \iint_{t_0}^{t'} & dt''d^3r' \Big\{ \bar{\bar
      P}^{(-)}(\mathbf r,t'|\mathbf r',t'')\cdot \notag \\ & \bar{\bar m}[\hat{\mathbf
      J}]^t \tilde{\mathbf D}_0^{(-)}(\mathbf r',t'')\Big\}  \times
    \tilde{\mathbf D}_0^{(+)}(\mathbf r,t') \Big].
  \end{align}
  After spatial average we get two terms, representing the coherent
  and the incoherent interaction. The incoherent
  interaction may then be written as 
    \begin{align}
    \frac{ i \beta c_1}{2 \hbar \epsilon_0}\int_{t_0}^t dt'\:&\int_{t_0}^{t'}dt''\:
    \bar{\mathbf J}(\mathbf r) \times \Big[ \Big( \frac{\partial}{\partial t''}-
    i\laser \Big) \notag \\  & \Big[\bar{\bar{ \eta}}^{*t}(\mathbf r,t',t'')
    \bar{\bar{\mathcal V}}^t[\bar{\mathbf J}]
    \tilde{\boldsymbol{\mathcal D}}^{(-)}_0(\mathbf r,t'')\Big] 
    \times \tilde{\boldsymbol{\mathcal D}}^{(+)}_0(\mathbf r,t') \Big].
  \end{align}
  To simplify notation, we have signified spatial averaging with
  calligraphic letters, e.g. $\< \mathbf D(\mathbf r,t)
  \>_{\text{s.a.}} \equiv \boldsymbol{\mathcal D}(\mathbf r,t)$. This
  convention will be used in the remainder of this article.

  We have now developed all the necessary theoretical tools to
  describe the system. In the next section we shall use these tools to
  discuss a pertubative expansion of the evolution of the system.

  %

  \section{Time evolution}\label{sec:results}
  This section is divided into three parts. In the first part we
  examine the general behaviour of the atomic spin in the presence of
  a light field. The aim is to understand the effect of the loops
  introduced in the Feynman diagrams. In the second part we consider
  the light field and we show how the theory introduce a decay of the
  field strength of the light as it interacts with the atoms. Again
  this is connected to the loops introduced in the Feynman diagrams.
  Finally we will introduce and discuss Stokes operators, which are
  the appropriate operators for describing the experiments in
  Ref. \cite{juulsgaard1,juulsgaard2,sherson}. 

  
  \subsection{Evolution of the spin } \label{sec:evolution-spin}
  In this section we will consider the spin equation in detail for
  the simple interaction (\ref{eq:hatmathbf-jmathbf-r}). We will begin
  our analysis by considering the first order term in the perturbative
  expansion of the solution to the spin equation, formally given by
  the diagram

\begin{fmffile}{feynman_d_5aa}
  \fmfcmd{%
    style_def wiggly_arrow expr p =
    cdraw (wiggly p);
    shrink (1);
       cfill (arrow p);
     endshrink;
     enddef;}

 \fmfcmd{%
    style_def DwigA expr p =
    draw_double (wiggly p);
    shrink (1.5);
       cfill (arrow p);
     endshrink;
     enddef;}
\begin{equation}    \label{eq:beginc-beginfmff-fmf}
   \begin{fmfgraph*}(70,30) \fmfleftn{i}{3} \fmfright{o}
        \fmflabel{$t$}{o}
        \fmfpolyn{triagram,tension=.7,filled=shaded}{v}{3}
    \fmf{wiggly_arrow,tension=.5,left=.5}{i3,v1} 
    \fmf{wiggly_arrow,tension=.5,left=.5}{v1,i2}
    \fmf{fermion}{i1,v2}
    \fmf{fermion}{v3,o}
    \fmffreeze
    \fmfshift{0.2w,0h}{v1,v2,v3}
    \fmfshift{.09w,.1h}{i2}
    \fmfdot{o} 
  \end{fmfgraph*}\quad .
     \end{equation}
   \end{fmffile}
    
  This term gives no extra contributions when doing the spatial averaging,
  and we readily write down the expression describing this term
  \begin{align}
    \label{eq:first_order_spin_eq}
    \frac{i\beta
        c_1\rho(\mathbf r)}{\hbar \epsilon_0} \int_{t_0}^t dt'\;
      \bar{\mathbf J}(\mathbf r,t_0) \times
      \Big( \tilde{\boldsymbol{\mathcal D}}_0^{(-)}(\mathbf r,t') \times
      \tilde{\boldsymbol{\mathcal D}}_0^{(+)}(\mathbf r,t') \Big).
  \end{align}

  We now continue with  the second order terms represented by the
  following Feynman diagrams
  \begin{fmffile}{feynman_d_5}
         \fmfcmd{%
    style_def wiggly_arrow expr p =
    cdraw (wiggly p);
    shrink (1);
       cfill (arrow p);
     endshrink;
     enddef;}

 \fmfcmd{%
    style_def DwigA expr p =
    draw_double (wiggly p);
    shrink (1.5);
       cfill (arrow p);
     endshrink;
     enddef;}
   \begin{align}&
     \begin{fmfgraph*}(60,50) \fmfleftn{i}{4} \fmfright{o}
        \fmflabel{$t$}{o}
        \fmfpolyn{triagram,tension=.8,filled=shaded}{v}{3}
    \fmf{wiggly_arrow,tension=.5,left=.5}{i3,k,v1} 
    \fmf{wiggly_arrow,tension=.5,left=.5}{v1,i2}
    \fmf{fermion}{i1,v2}
    \fmf{fermion}{v3,o}
    \fmf{fermion}{i4,k}
    \fmffreeze
    \fmfshift{0.25w,0h}{v1,v2,v3}
    \fmfshift{.15w,0h}{k}
    \fmfshift{-.1w,0h}{i4}
    \fmfshift{0w,.06h}{i2}
    \fmfshift{0w,-.06h}{i3}
    \fmfdot{o} \fmfblob{.1w}{k}
    \end{fmfgraph*} \qquad \raisebox{.5cm}{+}
    \qquad 
    \begin{fmfgraph*}(60,50) \fmfleftn{i}{4} \fmfright{o}
        \fmflabel{$t$}{o}
        \fmfpolyn{triagram,tension=.8,filled=shaded}{v}{3}
    \fmf{wiggly_arrow,tension=.5,left=.5}{i4,v1} 
    \fmf{wiggly_arrow,tension=.5,left=.5}{v1,k,i3}
    \fmf{fermion}{i1,v2}
    \fmf{fermion}{v3,o}
    \fmf{fermion}{i2,k}
    \fmffreeze
    \fmfshift{0.25w,0h}{v1,v2,v3}
    \fmfshift{.15w,0h}{k}
    \fmfshift{-.1w,0h}{i3}
    \fmfshift{0w,-.1h}{i2}
    \fmfshift{-.18w,-.11h}{i4}
    \fmfdot{o} \fmfblob{.1w}{k}
    \end{fmfgraph*} 
    \qquad \notag \\
    &\raisebox{.5cm}{+}\qquad 
     \begin{fmfgraph*}(60,50) 
       \fmfleftn{i}{5} \fmfright{o}
        \fmflabel{$t$}{o}
        \fmfpolyn{triagram,tension=.8,filled=shaded}{v}{3}
    \fmf{wiggly_arrow,tension=.5,left=.5}{i5,v1} 
    \fmf{wiggly_arrow,tension=.5,left=.5}{v1,i4}
    \fmfpolyn{triagram,tension=.7,filled=shaded}{k}{3}
    \fmf{fermion}{k1,v2}
    \fmf{fermion}{v3,o}
    \fmf{wiggly_arrow,tension=.5,left=.5}{i3,k2} 
    \fmf{wiggly_arrow,tension=.5,left=.5}{k2,i2}
    \fmf{fermion}{i1,k3}
    \fmffreeze
    \fmfshift{0.2w,0h}{v1,v2,v3}
    \fmfshift{.2w,-.1h}{k1,k2,k3}
    \fmfshift{-.01w,.1h}{i2}
    \fmfdot{o} 
    \end{fmfgraph*}\quad . \label{fig:feynman-diagram-spin}
  \end{align}
  \end{fmffile}
  When taking spatial average of these terms, we have argued that the
  first two diagrams will give an additional set of Feynman diagrams
  containing loops and stars.  It still remains to consider the last
  diagram of Fig \eqref{fig:feynman-diagram-spin}, representing two
  photons interacting with the same atom at time $t$ and $t'$. In
  this diagram it is necessary to pay special attention to the case
  where the two interactions happen at the same time $t=t'$. The
  contribution of this term is proportional to $\boldsymbol{\mathcal
    D}^{(-)}(t'')\boldsymbol{\mathcal
    D}^{(+)}(t'')\boldsymbol{\mathcal D}^{(-)}(t')\boldsymbol{\mathcal
    D}^{(+)}(t')$ which is not normal-ordered, and it will be
  convenient to separate it into normal-ordered terms. When commuting
  $\boldsymbol{\mathcal D}^{(-)}(\mathbf r,t'')$ and
  $\boldsymbol{\mathcal D}^{(+)}(\mathbf r,t')$ we once again get an
  infinitely short propagator c.f. (\ref{eq:big-hatmathbf-d-1}). This
  extra term we will denote by a filled star with a loop. This
  commutator term will produce an interaction which is linear in the
  field intensity (involves $\boldsymbol{\mathcal
    D}^{(-)}\boldsymbol{\mathcal D}^{(+)}$) whereas the normally
  ordered term ($\boldsymbol{\mathcal D}^{(-)}\boldsymbol{\mathcal
    D}^{(-)}\boldsymbol{\mathcal D}^{(+)}\boldsymbol{\mathcal
    D}^{(+)}$) will be quadratic in the intensity. Ignoring for now
  this quadratic term as well as the coherent interactions, the second
  order diagrams for the spin equation after spatial average reads
   \begin{widetext}
          \begin{fmffile}{feynman_d_6a}
\fmfcmd{%
    style_def wiggly_arrow expr p =
    cdraw (wiggly p);
    shrink (1);
       cfill (arrow p);
     endshrink;
     enddef;}

 \fmfcmd{%
    style_def DwigA expr p =
    draw_double (wiggly p);
    shrink (1.5);
       cfill (arrow p);
     endshrink;
     enddef;}

   \begin{align}\label{eq:beginfmfgr-30-fmfl}
    &\begin{fmfgraph*}(70,30) \fmfleftn{i}{3} \fmfright{o}
        \fmflabel{$t$}{o}
        \fmfpolyn{pentagram,pull=.6,tension=0.9,filled=gray50}{v}{5}
    \fmf{wiggly_arrow,tension=.5,left=.5}{i3,v5} 
    \fmf{wiggly_arrow,tension=.5,left=.5}{v5,i2}
    \fmf{photon,tension=.4}{v4,v4}
    \fmf{fermion}{i1,v2}
    \fmf{fermion}{v3,o}
    \fmffreeze
    \fmfshift{0.2w,0h}{v1,v2,v3,v4,v5}
    \fmfshift{0.06w,.17h}{i2}
    \fmfdot{o} 
    \end{fmfgraph*}
    \qquad \raisebox{0.4cm}{+} \quad 
    \begin{fmfgraph*}(60,40) \fmfleftn{i}{3} \fmfright{o}
        \fmflabel{$t$}{o}
        \fmfpolyn{pentagram,pull=.6,tension=0.9,filled=gray50}{v}{5}
    \fmf{wiggly_arrow,tension=.5,left=.5}{i3,v5} 
    \fmf{wiggly_arrow,tension=.7,left=.3}{v5,i2}
    \fmf{photon,tension=.4}{v1,v1}
    \fmf{fermion}{i1,v2}
    \fmf{fermion}{v3,o}
    \fmffreeze
    \fmfshift{0.2w,0h}{v1,v2,v3,v4,v5}
    \fmfshift{0.06w,.25h}{i2}
    \fmfshift{0.15w,-.27h}{i1}
    \fmfdot{o} 
    \end{fmfgraph*}
    \qquad \raisebox{0.4cm}{+} \quad
     \begin{fmfgraph*}(70,30) \fmfleftn{i}{3} \fmfright{o}
        \fmflabel{$t$}{o}
        \fmfpolyn{pentagram,pull=.6,tension=0.9,filled=full}{v}{5}
    \fmf{wiggly_arrow,tension=.5,left=.5}{i3,v5} 
    \fmf{wiggly_arrow,tension=.5,left=.5}{v5,i2}
    \fmf{fermion,tension=.5}{v2,v2}
    \fmf{fermion}{i1,v2}
    \fmf{fermion}{v3,o}
    \fmffreeze
    \fmfshift{0.2w,0h}{v1,v2,v3,v4,v5}
    \fmfshift{0.06w,.17h}{i2}
    \fmfdot{o} 
  \end{fmfgraph*} \quad ,
 \intertext{which can also be written as}
 &\:\: \begin{fmfgraph*}(30,40) \fmfleftn{i}{3}
      \fmf{wiggly_arrow,tension=.5,left=.5}{i3,v1} 
      \fmf{wiggly_arrow,tension=.7,left=.3}{v1,i2}
      \fmf{fermion}{i1,v2}
      \fmffreeze
      \fmfshift{1w,0h}{v1}
      \fmfshift{1w,.2h}{v2}
      \fmfshift{0.06w,.25h}{i2}
      \fmfshift{0.15w,-.27h}{i1}
    \end{fmfgraph*}
    \quad ^{\text{\large{ $ \times \Bigg\{ $ }}}
    \begin{fmfgraph*}(30,40) \fmfleft{i}\fmfright{o}
      \fmf{phantom}{i,v1}
      \fmf{phantom}{v3,o}
      \fmfpolyn{pentagram,pull=.6,tension=.2,filled=gray50}{v}{5}
      \fmf{photon,tension=.3}{v4,v4}
    \end{fmfgraph*}
    \: \raisebox{.6cm}{+}
     \begin{fmfgraph*}(30,40) \fmfleft{i}\fmfright{o}
      \fmf{phantom}{i,v5}
      \fmf{phantom}{v3,o}
      \fmfpolyn{pentagram,pull=.6,tension=.2,filled=gray50}{v}{5}
      \fmf{photon,tension=.3}{v1,v1}
    \end{fmfgraph*}
    \: \raisebox{.6cm}{+}
     \begin{fmfgraph*}(30,40) \fmfleft{i}\fmfright{o}
      \fmf{phantom}{i,v5}
      \fmf{phantom}{v3,o}
      \fmfpolyn{pentagram,pull=.6,tension=.2,filled=full}{v}{5}
      \fmf{fermion,tension=.3}{v2,v2}
    \end{fmfgraph*}
    \qquad ^{\text{\large{ $  \Bigg\} \times $ }}} 
     \begin{fmfgraph*}(30,40) \fmfleft{i}\fmfright{o}
      \fmf{fermion}{i,o}
      \fmfdot{o}
      \fmflabel{$t$}{o}
    \end{fmfgraph*}\qquad .\label{eq:beginfmfgr-30-fmfl-1}
    \end{align}
  \end{fmffile}
  \end{widetext}
  The interpretation of the diagrams is given below.

  To simplify the expression we will make the slowly varying envelope
  approximation which simplifies Eq. (\ref{eq:barbar-p+mathbf-r-1}) to
  \begin{align}
    \bar{\bar P}^{(-)}(\mathbf r,t|\mathbf r,t') \approx \frac{-i \laser}{2c^2}
    \bar{\bar{ \eta}}^{*t}(\mathbf r,t,t').
  \end{align}
  Secondly we shall evaluate $\eta$ in a local density approximation,
  where we assume that $\eta(\mathbf r,t,t')$ is the same as if we
  were in an infinite medium with a constant density $\rho(\mathbf r)$
  and spin density $\mathbf J(\mathbf r)$. By
  doing this we ignore the reflection of the field on the surface of
  the ensemble or other inhomogeneities.  The infinitely short
  propagator which expresses the amplitude for the field to be found
  at the same position at some later time, therefore becomes a
  delta-function in time. This approximation is valid provided that the diffraction matrix
  $\bar{\bar{\mathcal M}}(\mathbf r)$ varies slowly on the scale of the
  wavelength of the light.  Furthermore $\bar{\bar{\eta}}(\mathbf
  r,t,t')$ also contain the Lamb shift which we ignore for simplicity.
  A detailed calculation of $\bar{\bar{\eta}}$ is presented in
  Appendix \ref{sec:calc-infin-short}, where we find
  \begin{align}
    \bar{\bar P}^{(-)}(\mathbf r,t|\mathbf r,t')=& \frac{-i \delta (t-t')}{c^2} \left[ 
     \begin{array}{ccc}
       \varrho_{||}(\mathbf r) & 0 & 0 \\
       0& \varrho_{\perp}(\mathbf r) & - i \varrho_{\Gamma}(\mathbf r) \\ 
       0& i \varrho_{\Gamma}(\mathbf r) & \varrho_{\perp}(\mathbf r)
     \end{array} \right] \notag \\ 
   \equiv & \frac{-i \delta (t-t')}{c^2} \bar{\bar
     A}^{(-)}(\mathbf r),
  \end{align}
  where the coefficients $\varrho_{||}$, $\varrho_{\perp}$ and
  $\varrho_{\Gamma}$ may be found in Eq.
  (\ref{eq:beginalign-varrho_-=}).  Here the result is given in an
  Euclidean basis, where $\mathbf J$ is assumed to be along the the
  $x$-axis. The result may also be expressed in a coordinate-independent
  form as
  \begin{align}\label{eq:barbar-p+mathbf-r-3}
    \bar{\bar P}^{(-)}(\mathbf r,t|\mathbf r,t')= \frac{-i \delta
      (t-t')}{c^2} \Big\{  \varrho&_{\perp}(\mathbf r) -i \gamma (\mathbf r)
    \hat{\mathbf j} \times
     \notag \\ &  + \big[ \varrho_{||}(\mathbf r) - \varrho_{\perp}(\mathbf r) \big]
    \hat{\mathbf j}\;( \;\hat{\mathbf j} \cdot \Big\} ,
  \end{align}
  where $\hat{\mathbf j}$ is a unit vector parallel to $\mathbf J$.
  This infinitely short propagator is inserted into the second-order
  terms in the spin equation.  The second-order incoherent interaction
  given in Eq. (\ref{eq:beginfmfgr-30-fmfl-1}) then reads
  \begin{widetext}
    \begin{align}\label{eq:fracb-c_12-rhom}
      & \frac{\beta^2  }{ \hbar
        \epsilon_0} \int_{t_0}^t dt'\Bigg\{ c_1 c_0 \bar{\mathbf J}^2 \Big[ \bar{\bar
        A}^{(-)}\boldsymbol{\mathcal D}^{(-)}_0 ( \bar{\mathbf J} \cdot
      \boldsymbol{\mathcal D}^{(+)}_0 ) - \boldsymbol{\mathcal D}^{(-)}_0 (
      \bar{\mathbf J} \cdot \bar{\bar A}^{(+)} \boldsymbol{\mathcal
        D}^{(+)}_0) + H.c. \Big] \notag \\ &
      + \frac{c_1^2}{2} \Big[   \bar{\bar A}^{(+)}\boldsymbol{\mathcal
        D}^{(-)}_0 (\bar{\mathbf J} \cdot
      \boldsymbol{\mathcal D}^{(+)}_0 ) - (\boldsymbol{\mathcal D}^{(-)}_0 \cdot
      \boldsymbol{\mathcal D}^{(+)}_0 )\bar{\bar A}^{(-)}\bar{\mathbf J}
      - \rm{Tr}[\bar{\bar A}^{(-)}] \boldsymbol{\mathcal D}^{(-)}_0( \bar{\mathbf J}
      \cdot \boldsymbol{\mathcal D}^{(+)}_0) + \boldsymbol{\mathcal D}^{(-)}_0
      ( \bar{\mathbf J} \cdot \bar{\bar A}^{(-)} \boldsymbol{\mathcal
        D}^{(+)}_0)  + H.c. \Big] \Bigg\},
    \end{align}
  \end{widetext}
  where we have suppressed the space and time dependencies. 

  In the simple case, where the matrix $\bar{\bar A}^{(\pm)}$ is
  proportional to the identity matrix,($\varrho_{\Gamma}\approx 0$,
  $\varrho_{||}\approx\varrho_{\perp}=\varrho$), which is the case to
  lowest order, the terms proportional to $c_1c_0$ cancels and the
  expression reduces to
  \begin{align}\label{eq:fracddt-barmathbf-j}
   -\frac{\beta^2
      c_1^2\varrho}{2\hbar \epsilon_0}\int_{t_0}^t\hspace{-4pt}
    dt'\Big[ ( \boldsymbol{\mathcal
      D}^{(-)}_0 \cdot \boldsymbol{\mathcal D}^{(+)}_0 )\bar{\mathbf J} +
    \boldsymbol{\mathcal D}^{(-)}_0( \bar{\mathbf J} \cdot
    \boldsymbol{\mathcal D}^{(+)}_0)+ H.c. \Big].
  \end{align}
  This term scale with the power of
  the incident light, and linearly polarized light will affect the spin
  component parallel to the field with twice the rate than the
  perpendicular spin components. To see this we may introduce a decay-rate
  $\Gamma_{\boldsymbol{\mathcal D}}$, and writing expression
  (\ref{eq:fracddt-barmathbf-j}) on a differential form, we thus see
  that the term indeed describes a decay of the spin-components.
  \begin{subequations}
    \begin{align}
      \label{eq:spinlowest-order-possible}
      \partial_t \bar J_x =& -2\Gamma_{\boldsymbol{\mathcal D}}\bar
      J_x
      \\
      \partial_t\bar J_y =& -\Gamma_{\boldsymbol{\mathcal D}}\bar
      J_y
      \\
      \partial_t\bar J_z =& -\Gamma_{\boldsymbol{\mathcal D}}\bar
      J_z \intertext{where} \Gamma_{\boldsymbol{\mathcal D}} =&
      \frac{\beta^2 c_1^2\varrho}{\hbar \epsilon_0}\<
      \boldsymbol{\mathcal D}^{(-)}_{0,x} \boldsymbol{\mathcal
        D}^{(+)}_{0,x} \> \notag,
    \end{align}
  \end{subequations}
  and where we have assumed that the light is linearly polarized in the
  $x$-direction. 
  
  Let us now turn to the coherent part of the interaction represented
  by the Feynman diagrams in Eq. (\ref{fig:feynman-diagram-spin}). The
  first two terms containing a dot
  are by construction very small, and will vanish when taking the
  quantum mechanical average, as discussed in Sec.
  \ref{sec:dens-corr}. The only important second-order coherent
  interaction is therefore the following Feynman diagram for
  normal-ordered fields.
       \begin{fmffile}{feynman_d_5a}
\fmfcmd{%
    style_def wiggly_arrow expr p =
    cdraw (wiggly p);
    shrink (1);
       cfill (arrow p);
     endshrink;
     enddef;}

 \fmfcmd{%
    style_def DwigA expr p =
    draw_double (wiggly p);
    shrink (1.5);
       cfill (arrow p);
     endshrink;
     enddef;}
   \begin{align}\label{eq:rais-:-}
   \raisebox{.6cm}{{\large{\bf :}} } 
      \begin{fmfgraph*}(60,50) \fmfleftn{i}{5} \fmfright{o}
        \fmflabel{$t$}{o}
        \fmfpolyn{triagram,tension=.8,filled=30}{v}{3}
    \fmf{wiggly_arrow,tension=.5,left=.5}{i5,v1} 
    \fmf{wiggly_arrow,tension=.5,left=.5}{v1,i4}
    \fmfpolyn{triagram,tension=.7,filled=30}{k}{3}
    \fmf{fermion}{k1,v2}
    \fmf{fermion}{v3,o}
    \fmf{wiggly_arrow,tension=.5,left=.5}{i3,k2} 
    \fmf{wiggly_arrow,tension=.5,left=.5}{k2,i2}
    \fmf{fermion}{i1,k3}
    \fmffreeze
    \fmfshift{0.2w,0h}{v1,v2,v3}
    \fmfshift{.2w,-.1h}{k1,k2,k3}
    \fmfshift{-.01w,.1h}{i2}
    \fmfdot{o} 
    \end{fmfgraph*} \hspace{.6cm}\raisebox{.6cm}{{\large{\bf :}} }
     \end{align}
   \end{fmffile}
   \noindent Suppressing the spatial dependence of the displaced
   electric field, this normal-ordered coherent interaction is given in
   vector representation by
  \begin{widetext}
      \begin{align}\label{eq:fracb-c_12-rhom-1}
        -\frac{\beta^2 c_1^2 }{\hbar^2 \epsilon_0^2} \int_{t_0}^t
        dt'\;& \Bigg\{  \tilde{\boldsymbol{\mathcal D}}_0^{(-)}(t')
        \Big(  \tilde{\boldsymbol{\mathcal D}}_0^{(-)}(t)\cdot
        \tilde{\boldsymbol{\mathcal D}}_0^{(+)}(t') \Big) \Big(
        \bar{\mathbf J}_0 \cdot  \tilde{\boldsymbol{\mathcal
            D}}_0^{(+)}(t) \Big) - \Big(  \tilde{\boldsymbol{\mathcal
            D}}_0^{(-)}(t) \cdot  \tilde{\boldsymbol{\mathcal
            D}}_0^{(-)}(t') \Big) \Big( \bar{\mathbf J}_0 \cdot
        \tilde{\boldsymbol{\mathcal D}}_0^{(+)}(t) \Big)
        \tilde{\boldsymbol{\mathcal D}}_0^{(+)}(t') + H.c. \Bigg\} .
      \end{align}
  \end{widetext}
  In the case of linearly polarized light, say
  $\tilde{\boldsymbol{\mathcal D}}_0^{(-)}\;||\;\mathbf e_x$ this term
  vanishes, but this is in general not the case. In
  Sec. \ref{sec:exper-appl-valid} we examine the term in some
  simplified system.

  \subsection{Evolution of the light} \label{sec:evolution-light}

  The treatment of the displaced electric field is similar to the
  spin, but there are a few important differences. Let us consider
  the negative-frequency part of the field, and write the expansion of
  the displaced electric field ignoring for now the evolution of
  the spin
    \begin{fmffile}{feynman_d_7_f}
  
\fmfcmd{%
    style_def wiggly_arrow expr p =
    cdraw (wiggly p);
    shrink (1);
       cfill (arrow p);
     endshrink;
     enddef;}

 \fmfcmd{%
    style_def DwigA expr p =
    draw_double (wiggly p);
    shrink (1.5);
       cfill (arrow p);
     endshrink;
     enddef;}

   \begin{align}&
     \begin{fmfgraph*}(30,40)
       \fmfleft{i} \fmfright{o}
       \fmflabel{t}{o}
       \fmf{DwigA,tension=.1,left=.2}{i,o}
       \fmfdot{o}
     \end{fmfgraph*} \hspace{.7cm}
     \raisebox{.6cm}{\text{$\approx$}}
  \begin{fmfgraph*}(50,40) 
   \fmfleftn{i}{2} \fmfright{o}
   \fmflabel{$t$}{o} 
   \fmf{wiggly_arrow,tension=.1,left=.2}{i2,v1}
   \fmf{wiggly_arrow,tension=.1,left=.2}{v1,o}
   \fmf{fermion}{i1,v1}
   \fmffreeze
   \fmfshift{.5w,.7h}{v1}
   \fmfshift{0w,.5h}{i1}
   \fmfblob{.15w}{v1}
   \fmfdot{o}
 \end{fmfgraph*}
 \hspace{.7cm} \raisebox{.7cm}{+} \hspace{.2cm}   
 \begin{fmfgraph*}(70,40) 
   \fmfleftn{i}{4} \fmfright{o}
   \fmflabel{$t$}{o} 
   \fmf{wiggly_arrow,tension=.1,left=.2}{i4,v1}
   \fmf{wiggly_arrow,tension=.1,left=.2}{v1,v2}
   \fmf{wiggly_arrow,tension=.1,left=.2}{v2,o}
   \fmf{fermion}{i3,v1}
   \fmf{fermion}{i2,v2}
   \fmffreeze
   \fmfshift{.4w,.3h}{v1}
   \fmfshift{.65w,.5h}{v2}
   \fmfblob{.1w}{v1}
   \fmfblob{.1w}{v2}
   \fmfdot{o}
 \end{fmfgraph*} \notag \\ &
 \raisebox{.7cm}{+}\hspace{.1cm}
 \begin{fmfgraph*}(70,40)
   \fmfleftn{i}{4} \fmfright{o}
   \fmflabel{$t$}{o}
   \fmf{wiggly_arrow,tension=.1,left=.2}{i4,v1}
   \fmf{wiggly_arrow,tension=.1,left=.2}{v1,v2}
   \fmf{wiggly_arrow,tension=.1,left=.2}{v2,v3}
   \fmf{wiggly_arrow,tension=.1,left=.2}{v3,o}
   \fmf{fermion}{i3,v1}
   \fmf{fermion}{i2,v2}
   \fmf{fermion}{i1,v3}
   \fmffreeze
   \fmfshift{.4w,.3h}{v1}
   \fmfshift{.6w,.5h}{v2}
   \fmfshift{.6w,.5h}{v3}
   \fmfshift{0w,-.3h}{o}
   \fmfblob{.08w}{v1}
   \fmfblob{.08w}{v2}
   \fmfblob{.08w}{v3}
   \fmfdot{o}
 \end{fmfgraph*}
 \hspace{.5cm} \raisebox{.7cm}{+} \hspace{.2cm}
 \begin{fmfgraph*}(80,40)
   \fmfleftn{i}{5} \fmfright{o}
   \fmflabel{$t$}{o}
   \fmf{wiggly_arrow,tension=.1,left=.2}{i5,v1}
   \fmf{wiggly_arrow,tension=.1,left=.2}{v1,v2}
   \fmf{wiggly_arrow,tension=.1,left=.2}{v2,v3}
   \fmf{wiggly_arrow,tension=.1,left=.2}{v3,v4}
   \fmf{wiggly_arrow,tension=.2,left=.2}{v4,o}
   \fmf{fermion}{i4,v1}
   \fmf{fermion}{i3,v2}
   \fmf{fermion}{i2,v3}
   \fmf{fermion}{i1,v4}
   \fmffreeze
   \fmfshift{.4w,.3h}{v1}
   \fmfshift{.6w,.5h}{v2}
   \fmfshift{.7w,.5h}{v3}
   \fmfshift{.7w,.4h}{v4}
   \fmfshift{0w,-.3h}{o}
   \fmfblob{.07w}{v1}
   \fmfblob{.07w}{v2}
   \fmfblob{.07w}{v3}
   \fmfblob{.07w}{v4}
   \fmfdot{o}
 \end{fmfgraph*} \hspace{.7cm}\raisebox{.7cm}{\text{$+ \cdots$}} \quad \raisebox{-.0cm}{.}
 \label{eq:-rais-beginfmfgr}
      \end{align}
  \end{fmffile}
  \noindent When we take spatial average of diagrams like these, we
  introduce delta-function correlations between vertex points. So far
  we have treated the atoms in the ideal gas approximation, where we
  ignore any correlation in the position of the atoms but in
  reality we should include a short-range correlation functions
  describing that two different atoms cannot be at the same position.
  In Appendix \ref{sec:lorenz-lorenz-corr} we show that including this leads to
  the Lorentz-Lorenz or Clausius-Mossotti relation. In the following
  we will only discuss loops, where two consecutive vertex points are
  evaluated for the same atom. Since we have subtracted the quantum
  mechanical average from the vertex, no first-order vertex will give
  a contribution to the evolution of the light, and therefore these
  second-order loop diagrams are the most important effects apart from
  the diffraction effects included in the mode-functions $\{ \mathbf
  f_{\mathbf q}\}$. Later in
  section \ref{sec:appl-spin-meas} we shall discuss the operator nature of
  the light field and then we keep the first-order vertex in the
  calculations. In the current approximation Eq.
  \eqref{eq:-rais-beginfmfgr} reduces to
  \begin{fmffile}{feynman_d_7_g}
  
    \fmfcmd{%
      style_def wiggly_arrow expr p =
      cdraw (wiggly p);
      shrink (1);
      cfill (arrow p);
      endshrink;
      enddef;}

    \fmfcmd{%
      style_def DwigA expr p =
      draw_double (wiggly p);
      shrink (1.5);
      cfill (arrow p);
      endshrink;
      enddef;}
    
    \begin{align}
       \begin{fmfgraph*}(30,35)
       \fmfleft{i} \fmfright{o}
       \fmflabel{t}{o}
       \fmf{DwigA,tension=.1,left=.2}{i,o}
       \fmfdot{o}
     \end{fmfgraph*} \hspace{.5cm}
     \raisebox{.6cm}{\text{$\approx$}}
      \begin{fmfgraph*}(60,40) 
   \fmfleftn{i}{2} \fmfright{o}
   \fmflabel{$t$}{o} 
   \fmf{wiggly_arrow,tension=.1,left=.2}{i2,v1}
   \fmf{wiggly_arrow,tension=.1,left=.2}{v1,o}
   \fmf{photon,tension=.8,left=-90}{v1,v1}
   \fmf{fermion}{i1,v1}
   \fmffreeze
   \fmfshift{.4w,.6h}{v1}
   \fmfshift{0w,.5h}{i1}
   \fmfv{decor.shape=pentagon,decor.filled=hatched,decor.size=.15w}{v1}
      \fmfdot{o}
 \end{fmfgraph*}
 \hspace{.5cm} \raisebox{.7cm}{+} \hspace{.2cm}
 \begin{fmfgraph*}(60,35) 
   \fmfleftn{i}{4} \fmfright{o}
   \fmflabel{$t$}{o} 
   \fmf{wiggly_arrow,tension=.1,left=.2}{i4,v1}
   \fmf{wiggly_arrow,tension=.1,left=.2}{v1,v2}
   \fmf{wiggly_arrow,tension=.1,left=.2}{v2,o}
   \fmf{photon,tension=.7,left=-90}{v1,v1}
   \fmf{photon,tension=.9,left=-90}{v2,v2}
   \fmf{fermion}{i3,v1}
   \fmf{fermion}{i2,v2}
   \fmffreeze
   \fmfshift{.4w,.3h}{v1}
   \fmfshift{.7w,.5h}{v2}
   \fmfv{decor.shape=pentagon,decor.filled=hatched,decor.size=.15w}{v1}
   \fmfv{decor.shape=pentagon,decor.filled=hatched,decor.size=.15w}{v2}
   \fmfdot{o}
 \end{fmfgraph*}\hspace{.5cm} \hspace{.1cm}\raisebox{.7cm}{\text{$+
     \cdots$}} \label{eq:beginfmfgr-35-fmfl}  \raisebox{.5cm}{.}
    \end{align}
  \end{fmffile}
  We have here introduced an interaction denoted by a hatched pentagon which
  scales with $\beta^2 \rho \klaser^3$, and describes two $
  \begin{fmffile}{inseted_dot}
    \begin{fmfgraph*}(7,7)
      \fmfleft{i} \fmfright{o}
      \fmf{phantom}{i,v}
      \fmf{phantom}{v,o}
      \fmfblob{1w}{v}
    \end{fmfgraph*}
  \end{fmffile}
  $ connected by the infinitely short propagator. Using the results
  for the infinitely short propagator, and taking quantum mechanical
  average this interaction reads on matrix form
    \begin{fmffile}{feynman_d_7_h}
  
\fmfcmd{%
    style_def wiggly_arrow expr p =
    cdraw (wiggly p);
    shrink (1);
       cfill (arrow p);
     endshrink;
     enddef;}

 \fmfcmd{%
    style_def DwigA expr p =
    draw_double (wiggly p);
    shrink (1.5);
       cfill (arrow p);
     endshrink;
     enddef;}

   \begin{align}
   \raisebox{-.5cm}{   \begin{fmfgraph}(20,20)
   \fmfleft{i} \fmfright{o}
   \fmf{phantom}{i,v}
   \fmf{phantom}{v,o}
   \fmf{photon,tension=.3,left=-90}{v,v}
   \fmfv{decor.shape=pentagon,decor.filled=hatched,decor.size=.6w}{v}
 \end{fmfgraph}} \hspace{.0cm}
          =  i\beta^2\rho(\mathbf r)\left[ 
      \begin{array}[c]{ccc}
         \Gamma_{||}(\mathbf r) & 0 & 0 \\
              0 & \Gamma_{\perp,1}(\mathbf r) & i\Gamma_{\Gamma} (\mathbf r) \\ 
              0 & -i \Gamma_{\Gamma}(\mathbf r) & \Gamma_{\perp,2}(\mathbf r)
      \end{array} \right]  \equiv i \bar{\bar{\mathcal M}}'^t(\mathbf r),
      \end{align}
  \end{fmffile}
  where the coefficients entering the matrix are given by
  \begin{subequations}
    \begin{align}
      \label{eq:lightdecay_matrix}
      \Gamma_{||}(\mathbf r) =& c_0^2 \mathbf J^4\varrho_{||} + c_1^2
      \varrho_{\perp}( J_z^2 + J_y^2), \\
      \Gamma_{\perp,1}(\mathbf r) =& c_0^2 \mathbf J^4\varrho_{\perp}
      + 2c_0c_1\varrho_{\Gamma}\mathbf J^2J_x + c_1^2(
      \varrho_{||} J_z^2 +   \varrho_{\perp}J_x^2), \\
      \Gamma_{\perp,2}(\mathbf r) =& c_0^2 \mathbf J^4\varrho_{\perp}
      + 2c_0c_1\varrho_{\Gamma}\mathbf J^2J_x + c_1^2(
      \varrho_{||} J_y^2 +   \varrho_{\perp}J_x^2), \\
      \Gamma_{\Gamma} (\mathbf r) =& \varrho_{\perp}2c_1c_0\mathbf
      J^2J_x -\varrho_{||}\frac{c_1^2}{2}J_x+
      \varrho_{\Gamma}(c_0^2\mathbf J^2+c_1^2J_x^2).
    \end{align}
  \end{subequations}
  We have here suppressed the spatial dependence to shorten notation.  The
  series in Eq. \eqref{eq:beginfmfgr-35-fmfl} can be included in the
  differential equation describing the displaced electric field,
    \begin{align}
      \Big( 2i\laser\frac{d}{d t} -&\laser^2 + c^2
      \boldsymbol{\nabla}\times \boldsymbol{\nabla}\times \big[
      \bar{\bar{\mathcal M}}^t(\mathbf r) + i \bar{\bar{\mathcal
          M}}'^t(\mathbf r) \big] \Big) \tilde{\mathbf
        D}^{(-)}(\mathbf r,t) \notag \\ =&\;c^2\int d^3r\;
      \boldsymbol{\nabla}\times \boldsymbol{\nabla}\times
      \bar{\bar{\delta}}^T(\mathbf r,\mathbf r') \cdot \bar{\bar
        m}[\hat{ \mathbf J}]^t_{\text{mod}} \tilde{\mathbf
        D}^{(-)}(\mathbf r',t),
    \label{eq:decay_of_light-equation}
  \end{align}
  where the perturbation is modified accordingly. Because of the
  anti-Hermitian matrix, we see that these types of loop diagrams
  correspond to a decay of the field, i.e. the differential operator
  on the left side describes the propagation through a lossy medium.
  On the basis of this analysis and the analysis in Sec.
  \ref{sec:evolution-spin} we thus link the 
  loops in the Feynman diagrams with the decay associated with
  spontaneous emission.

  It remains to discuss the effect of light interacting with an atom
  that was previously subject to an interaction such that the atomic
  spin state has been changed. In terms of Feynman diagrams this is
  described as
      \begin{fmffile}{feynman_d_7_a}
  
\fmfcmd{%
    style_def wiggly_arrow expr p =
    cdraw (wiggly p);
    shrink (1);
       cfill (arrow p);
     endshrink;
     enddef;}

 \fmfcmd{%
    style_def DwigA expr p =
    draw_double (wiggly p);
    shrink (1.5);
       cfill (arrow p);
     endshrink;
     enddef;}

    \begin{align}
   \begin{fmfgraph*}(60,40)
     \fmfleftn{i}{4} \fmfright{o}
     \fmflabel{$t$}{o}
     \fmfpolyn{triagram,tension=.27,filled=shaded}{v}{3}
     \fmf{wiggly_arrow,tension=.1,left=.2}{i4,k}
     \fmf{wiggly_arrow,tension=.1,left=.3}{i3,v3}
     \fmf{wiggly_arrow,tension=.1,left=.3}{v3,i2}
       \fmf{fermion}{v2,k}
       \fmf{fermion}{i1,v1}
        \fmf{wiggly_arrow,tension=.1,left=.2}{k,o}
        \fmffreeze
        \fmfshift{.4w,.3h}{k}
        \fmfshift{.27w,.1h}{v1,v2,v3}
        \fmfblob{.1w}{k}
        \fmfdot{o}
   \end{fmfgraph*} \hspace{.5cm} \raisebox{.5cm}{.}
      \label{eq:second-order-feynman}
      \end{align}
  \end{fmffile}
  \noindent We shall postpone the analysis of this term and discuss it
  in connection with relating the fields to photon counting operators
  below.

  \subsection{Photon counting and Stokes operators}\label{sec:photon-counting}

  So far we have mainly been concerned with calculating the field
  $\tilde{\mathbf D}(\mathbf r,t)$. For experiments which eventually
  involves counting photons we are more interested in quantities like
  photon flux, and in particular the flux in some particular
  polarizational state.  We shall now discuss how to desribe such
  photon counting experiments within our theory.

  The general idea in this subsection is that we shall assume that we
  are able to measure the light-flux in a certain spatial mode by
  projecting the light field onto the mode and then integrating the
  flux of the light field at some detector plane, that we assume to be
  far away from the atomic ensemble. We will formulate such a
  measuring process in terms of an inner product,
 \begin{align}
   \label{eq:innerproduct_experiment}
   \ddleft \boldsymbol{\phi}(\mathbf r ,t) | \boldsymbol{\psi}(\mathbf
   r,t) \ddright \equiv \int_{-\infty} ^{\infty} dt \int_{\mathbb R^2}
   d^2r_{\perp} \boldsymbol{\phi}^{\dagger}(\mathbf r ,t) \cdot
   \boldsymbol{\psi}(\mathbf r,t).
 \end{align}
 We assume that the fields in general have some axis of propagation
 say $\mathbf r_{||}$. The spatial integral is then performed in some
 plane perpendicular to this axis at some point $r_{||}$ on this axis.
 This measuring process could be realized by
 e.g. 
 sending the light field through a single mode optical fibre prior to
 detection. 

 We are interested in the polarization of the field which is
 conveniently described by the so called Stokes operators defined
 below. These operators can be derived from a Stokes generator defined
 in a bra-ket-notation by
  \begin{align}
    \bar{\bar{S}} \equiv | \tilde{\mathbf D}^{(-)}(\mathbf r,t) \ddright \ddleft
    \tilde{\mathbf D}^{(-)}(\mathbf r,t) |,
  \end{align} 
  which we  represent as the following diagram
    \begin{fmffile}{feynman_d_9_a}
  
      \fmfcmd{%
    style_def wiggly_arrow expr p =
    cdraw (wiggly p);
    shrink (1);
       cfill (arrow p);
     endshrink;
     enddef;}

 \fmfcmd{%
    style_def DwigA expr p =
    draw_double (wiggly p);
    shrink (1.5);
       cfill (arrow p);
     endshrink;
     enddef;}
   \begin{align}
\begin{fmfgraph*}(40,30)
     \fmfleftn{i}{2} \fmfright{o}
     \fmflabel{$t$}{o}
     \fmf{DwigA,tension=.7,left=.3}{i2,o}
     \fmf{DwigA,tension=.7,left=.3}{o,i1}
     \fmfdot{o}
   \end{fmfgraph*} \qquad \raisebox{.3cm}{.} 
   \end{align}
\end{fmffile}
\noindent Measuring certain light-modes according to the inner product
in Eq.  (\ref{eq:innerproduct_experiment}), correspond to picking out
a certain matrix element of the Stokes generator. As an example we
assume that in some experiment we are able to measure the photon flux
of some linear polarization in some mode say $\tilde{\mathbf
  f}_{\mathbf q,x}(\mathbf r,t)$ after the interaction with the atoms.
The time dependence is here $\tilde{\mathbf f}_{\mathbf q,x}(\mathbf
r,t)=\mathbf f_{\mathbf q,x}(\mathbf r)e^{-i(\omega_{\mathbf
    q,x}-\laser)t}$. The integrated photon flux measured at the
detector plane, is then given by
\begin{align}
  \label{eq:phton_flux_general}
     \frac{2c^2}{\hbar \epsilon_0 \laser}  \ddleft
       \tilde{\mathbf f}^*_{\mathbf q,x} | \bar{\bar{\mathcal S}} | \tilde{\mathbf
       f}^*_{\mathbf q,x} \ddright,
\end{align}
where we normalize the outcome to count the number of photons. We have
here taken a spatial average of the Stokes generator as indicated by
the calligraphic font.

  Expanding this operator to second order, gives an
  additional term not covered by the analysis above. This extra term
  describes a process where both the negative frequency part
  and the positive frequency part of the displaced electric field
  interacts with the same atom.  This extra term comes from the
  following contribution to the Stokes generator
    \begin{fmffile}{feynman_d_7_c}
  
      \fmfcmd{%
    style_def wiggly_arrow expr p =
    cdraw (wiggly p);
    shrink (1);
       cfill (arrow p);
     endshrink;
     enddef;}

 \fmfcmd{%
    style_def DwigA expr p =
    draw_double (wiggly p);
    shrink (1.5);
       cfill (arrow p);
     endshrink;
     enddef;}
   \begin{align}
\begin{fmfgraph*}(40,30)
     \fmfleftn{i}{2} \fmfright{o}
     \fmflabel{$t$}{o}
     \fmf{DwigA,tension=.7,left=.3}{i2,o}
     \fmf{DwigA,tension=.7,left=.3}{o,i1}
     \fmfdot{o}
   \end{fmfgraph*}\hspace{.4cm} \raisebox{.5cm}{$= \: \cdots \:
     +\: \: $}
     \begin{fmfgraph*}(60,40)
     \fmfleftn{i}{4} \fmfright{o}
     \fmflabel{$t$}{o}
     \fmf{wiggly_arrow,tension=.1,left=.2}{i4,v1}
     \fmf{wiggly_arrow,tension=.1,left=.3}{v2,i1}
        \fmf{fermion}{i3,v1}
       \fmf{fermion}{i2,v2}
        \fmf{wiggly_arrow,tension=.1,left=.2}{v1,o}
        \fmf{wiggly_arrow,tension=.1,left=.2}{o,v2}
        \fmffreeze
        \fmfshift{.46w,.1h}{v1}
        \fmfshift{.46w,-.1h}{v2}
        \fmfblob{.1w}{v1}
        \fmfblob{.1w}{v2}
        \fmfdot{o}
   \end{fmfgraph*} \hspace{.4cm} \raisebox{.5cm}{$+\: \cdots$} \quad
   \raisebox{.4cm}{.} 
   \end{align}
\end{fmffile}
When taking the spatial average of this term we again generate a term
representing that the interaction happens at the same point. This
particular term would not have been there if we only considered the
spatial average of the displaced electric field. The generated term we
will illustrate as
    \begin{fmffile}{feynman_d_7_e}
  
      \fmfcmd{%
    style_def wiggly_arrow expr p =
    cdraw (wiggly p);
    shrink (1);
       cfill (arrow p);
     endshrink;
     enddef;}

 \fmfcmd{%
    style_def DwigA expr p =
    draw_double (wiggly p);
    shrink (1.5);
       cfill (arrow p);
     endshrink;
     enddef;}
   \begin{align}\label{eq:rais-.5cm-beginfmfgr}
\raisebox{-.5cm}{
     \begin{fmfgraph*}(60,40)
     \fmfleftn{i}{4} \fmfright{o}
     \fmflabel{$t$}{o}
     \fmf{wiggly_arrow,tension=.1,left=.2}{i4,v1}
     \fmf{wiggly_arrow,tension=.1,left=.3}{v2,i1}
        \fmf{fermion}{i3,v1}
       \fmf{fermion}{i2,v2}
        \fmf{wiggly_arrow,tension=.1,left=.2}{v1,o}
        \fmf{wiggly_arrow,tension=.1,left=.2}{o,v2}
        \fmffreeze
        \fmfshift{.46w,.1h}{v1}
        \fmfshift{.46w,-.1h}{v2}
        \fmfblob{.1w}{v1}
        \fmfblob{.1w}{v2}
        \fmfdot{o}
   \end{fmfgraph*}
  }
 \qquad \overrightarrow{\;\text{s.a.}\;}\;
 \begin{array}{c}
        \begin{fmfgraph*}(60,40)
     \fmfleftn{i}{4} \fmfright{o}
     \fmflabel{$t$}{o}
     \fmf{wiggly_arrow,tension=.1,left=.2}{i4,v1}
     \fmf{wiggly_arrow,tension=.1,left=.3}{v2,i1}
        \fmf{fermion}{i3,v1}
       \fmf{fermion}{i2,v2}
        \fmf{wiggly_arrow,tension=.1,left=.2}{v1,o}
        \fmf{wiggly_arrow,tension=.1,left=.2}{o,v2}
        \fmffreeze
        \fmfshift{.46w,.1h}{v1}
        \fmfshift{.46w,-.1h}{v2}
        \fmfv{decor.shape=circle,decor.size=.1w,decor.filled=30}{v1}
        \fmfv{decor.shape=circle,decor.size=.1w,decor.filled=30}{v2}
        \fmfdot{o}
   \end{fmfgraph*}
    \vspace{.2cm} \\ + \\ \vspace{.4cm}
\begin{fmfgraph*}(60,40)
     \fmfleftn{i}{3} \fmfright{o}
     \fmflabel{$t$}{o}
     \fmfpolyn{pentagon,pull=.9,tension=.7,filled=30,smooth}{v}{5}
      \fmf{wiggly_arrow,tension=.1,left=.2}{i3,v4}
      \fmf{wiggly_arrow,tension=.1,left=.2}{v1,i1}
      \fmf{fermion}{i2,v5}
       \fmf{wiggly_arrow,tension=.7,left=.6}{v3,o}
        \fmf{wiggly_arrow,tension=.1,left=.6}{o,v2}
        \fmfdot{o}
   \end{fmfgraph*}
 \end{array}
    \end{align}
 \end{fmffile}
 We constructed the interaction represented in the Feynman diagram as
 a gray circle, such that when taking the quantum mechanical average
 the term vanish. The new term generated when taking the spatial
 average, given as the lower right diagram of Eq.
 (\ref{eq:rais-.5cm-beginfmfgr}), describe the square of the
 fluctuations which is not vanishing. This was also the case for the
 terms containing the infinitely short propagator.  The new term
 however differs from the second order terms containing the infinitely
 short propagators because here we need to use the full macroscopic
 propagator. To calculate the effect of this term in detail, we
 therefore need to have an expression for the spatial modes describing
 the system. We will consider this term for a simplified system in Sec
 \ref{sec:appl-spin-meas}.

 To describe the experiments in Ref.  \cite{sherson} it is
 convenient to define a set of polarization dependent photon counting
 operators denoted as Stokes operators. 
 These are defined in accordance with Eq.
 (\ref{eq:phton_flux_general}) as
   \begin{subequations}\label{eq:begin-hat-s_xm}
     \begin{align}
       \hat s_1^{\;\mathbf q,\mathbf q'} =& \frac{K}{2}\Big[ \ddleft
       \tilde{\mathbf f}^*_{\mathbf q} | \bar{\bar{\mathcal S}} | \tilde{\mathbf
       f}^*_{\mathbf q} \ddright - \ddleft \tilde{\mathbf f}^*_{\mathbf q'} |
       \bar{\bar{\mathcal S}} |
       \tilde{\mathbf f}^*_{\mathbf q'} \ddright \Big] \\
       \hat s_2^{\;\mathbf q,\mathbf q'} =& \frac{K}{2}\Big[ \ddleft
       \tilde{\mathbf f}^*_{\mathbf q} | \bar{\bar{\mathcal S}} | \tilde{\mathbf
       f}^*_{\mathbf q'} \ddright + \ddleft \tilde{\mathbf f}^*_{\mathbf q'} |
       \bar{\bar{\mathcal S}} | \tilde{\mathbf f}^*_{\mathbf q} \ddright
       \Big] \\
       \hat s_3^{\; \mathbf q, \mathbf q'} =& \frac{K}{2i}\Big[ \ddleft
       \tilde{\mathbf f}^*_{\mathbf q} | \bar{\bar{\mathcal S}} | \tilde{\mathbf
       f}^*_{\mathbf q'} \ddright - \ddleft \tilde{\mathbf f}^*_{\mathbf q'} |
       \bar{\bar{\mathcal S}} | \tilde{\mathbf f}^*_{\mathbf q} \ddright \Big],
     \end{align}
   \end{subequations}
   where $K=\frac{2c^2}{\hbar \epsilon_0\laser}$.  Using
   commutation relations for the creation and annihilation operators
   these Stokes operators are seen to have the commutation relations
   for angular momentum operators.
  \begin{align}
    \big[ \hat s^{\mathbf q, \mathbf q'}_n ; \hat s^{\mathbf q,
      \mathbf q'}_m \big] = i\;\varepsilon_{nml} \hat s^{\mathbf q,
      \mathbf q'}_l,
  \end{align}
  We will calculate and discuss these Stokes operators to
  second order in the coupling coefficient $\beta$ in the following. 

  \subsection{Calculation of Stokes operators } \label{sec:appl-spin-meas} 
  
  In this section we shall calculate the Stokes operators to second
  order. In the experiments in Ref.
  \cite{juulsgaard1,juulsgaard2,sherson} the
  Stokes operators are measured by sending the light onto polarizing
  beamsplitters followed by a measurement of the difference in the
  intensity of the two outputs.  For instance if we take the indices
  ${\bf q}$ and ${\bf q'}$ to refer to the $x$ and $y$ polarizations
  of the light, the operator $\hat{s}_1^{x,y}$ in Eq.
  \eqref{eq:begin-hat-s_xm} can be measured by measuring the
  difference in the intensity of the $x$ and $y$ polarizations. The
  remaining operators $\hat{s}_2^{x,y}$ and $\hat{s}_3^{x,y}$ can
  respectively be related to the difference intensity with the
  polarizing beam splitter rotated by 45$^\circ$ and the difference
  intensity between the two circular polarizations.  For a general
  light beam, however, diffraction will cause the polarization of the
  light to depend on the spatial position and there is no well defined
  polarization. The simple measurement scheme is thus only applicable
  in the paraxial approximation, where we can separate out a position
  independent polarization vector. Far away from the ensemble we will
  therefore assume a paraxial approximation. That is, the
  mode-functions $\tilde{\mathbf f}_{\mathbf q}(\mathbf r,t)$ and
  $\tilde{\mathbf f}_{\mathbf q'}(\mathbf r,t)$ describing the Stokes
  operators far away from the atomic ensemble resemble plane waves
  with transverse profiles that change slowly compared to the
  wavelength.  The detector plane is placed far away from the atomic
  ensemble, and at this plane we will assume that the general set of
  basis-functions $\{ \mathbf f_{\mathbf q}\} $ can be approximated as
  \begin{align}\label{eq:par-axial-mode-fkt}
    \mathbf f_{\mathfb q}(\mathbf r)=\frac{1}{\sqrt{2\pi}}
    U_{n}(\mathbf r_{\perp})\mathbf e_j e^{ikz}.
  \end{align}
  We have here set the direction of propagation to be along the
  $z$-axis. The index $\mathbf q$ are now given as the set $\mathbf q =
  (k,n,j)$, where $k$ is some wavenumber, $n$ is an index referring to
  the transverse shape of the mode described by the scalar-field
  $U_n(\mathbf r)$, and $j$ describes the polarization
  of this mode, that can be either $x$- or $y$-polarized. The
  completeness relation Eq. (\ref{eq:barb-r-mathbf}), and
  orthonormality condition in this approximation thus gives
  {\allowdisplaybreaks
  \begin{subequations}
    \begin{align}
      \label{eq:completeness-relation}
       \sum_{n}U_n^*(\mathbf r_{\perp}) U_n(\mathbf r'_{\perp}) =
      \delta(\mathbf r_{\perp}-
      \mathbf r_{\perp}'), \\\label{eq:int-d2r_p-u_nm}
      \int d^2r_{\perp} U_n^*(\mathbf r_{\perp})
      U_{n'}(\mathbf r_{\perp}) = \delta_{nn'},
    \end{align}
  \end{subequations}}
  \hspace{-3pt}and the dispersion relation Eq.  (\ref{eq:boldsymb-barb-vhatm-1})
  at the detector plane is $\omega_{\mathbf q}^2= c^2 k^2$. 

  \begin{figure}
    \centering
    \vspace{.5cm}
    \includegraphics[width=.45\textwidth]{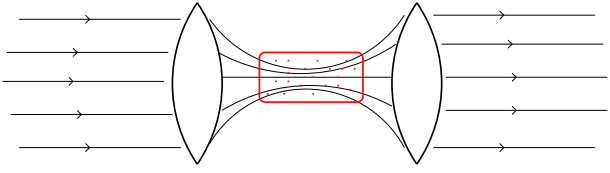}
    \caption{Schematic setup. We assume that away from the ensemble, the
      light-mode resembles a plane-wave with some transverse
      profile. A set of lenses focus the beam down into the ensemble.}
    \label{fig:focusing}
  \end{figure}
  The paraxial approximation above is convenient for expressing the
  measured observable in terms of the polarization of the field, but
  may not be sufficient to accurately describe experiments, where
  tightly focused beams are used. We shall therefore only assume this
  approximation to be applicable far away from the sample, and not
  necessarily inside the ensemble. Physically this could correspond to
  a situation, where an initially paraxial beam is focused onto the
  ensemble with a lens and converted back into a paraxial beam after
  the interaction by another lens, as shown in
  Fig. \ref{fig:focusing}. A similar treatment was used in
  Ref. \cite{van-enk}.

  Inside the
  ensemble we make the much weaker approximation that the set of
  spatial mode functions $U_{nq}({\bf r})$ is independent of the
  polarization of the field, so that the set ${\bf f_q}({\bf r})$ is
  given by
    \begin{align}
    \label{eq:modefunctions_ensemble}
    \mathbf f_{\mathbf q}(\mathbf r) = \frac{1}{\sqrt{2\pi}}
    U_{nq}(\mathbf r)\mathbf e_j(\mathbf r).
  \end{align}
  The mode $U_{nq}(\mathbf r)$ now takes into account that the spatial
  shape of the beam may change through the ensemble, and likewise the
  polarization vector $\mathbf e_j(\mathbf r)$, which we shall assume
  to be real-valued. The index $j$ will still be either $x$ or $y$,
  corresponding to the polarization of the mode far away from the
  sample, but the vector $\mathbf e_j(\mathbf r)$ will not necessarily
  be parallel to the $x$ or the $y$ axis. A more general description
  of the mode-functions would include a dependence of the polarization
  vector $\mathbf e_j$ on the polarization state $U_{mq}(\mathbf r)$,
  i.e., $\mathbf e_{mj}(\mathbf r)$. The correction this generalization
  gives to the Stokes operators, is presented in Appendix
  \ref{appendix-correction}, in relation to Sec.
  \ref{sec:weakly-coupl-single}.  When we make the relevant
  calculations to describe the Stokes operators defined in Eq.
  \eqref{eq:begin-hat-s_xm}, we will chose to consider modes
  corresponding to the index $\mathbf q=(k,m,x)$ and $\mathbf
  q'=(k,m',y)$.  We note that the set $\{ \mathbf f_{\mathbf q} \}$
  defined in this way is in general not complete, since, e.g, the
  assumption that the polarization vector is independent of the
  transverse mode number applies in the paraxial approximation but
  does not apply in general. When calculating the effect on the
  forward scattered field to first order we only get contributions
  from the near paraxial modes in the forward direction. When we go to
  second order there will, however, be effects of all the transverse
  modes, and in this case a correct treatment requires a more accurate
  treatment of the complete set of modes. Above we have already
  employed such a more general set of modes, when we discussed the
  effect of spontaneous emission, which involve all the transverse
  modes. In addition to this, a more accurate set of modes is also
  required for describing the effect of dipole-dipole interactions,
  which also involves all the transverse mode.


  We will in the following calculate the Stokes operators in the limit
  described above.  Diagrams containing a loop, we will not discuss,
  since these only leads to a decay of the light which we have
  discussed earlier. After taking spatial average the diagrams in
  consideration are
\begin{widetext}
        \begin{fmffile}{feynman_d_8}
  
      \fmfcmd{%
    style_def wiggly_arrow expr p =
    cdraw (wiggly p);
    shrink (1);
       cfill (arrow p);
     endshrink;
     enddef;}

 \fmfcmd{%
    style_def DwigA expr p =
    draw_double (wiggly p);
    shrink (1.5);
    cfill (arrow p);
     endshrink;
     enddef;}
   \begin{align}\label{eq:stokesgenerator-feynman}
     \begin{fmfgraph*}(40,30)
     \fmfleftn{i}{2} \fmfright{o}
     \fmflabel{$t$}{o}
     \fmf{DwigA,tension=.7,left=.3}{i2,o}
     \fmf{DwigA,tension=.7,left=.3}{o,i1}
     \fmfdot{o}
   \end{fmfgraph*}\hspace{.6cm} \raisebox{.5cm}{$\approx \: 
     $} \hspace{.2cm}&
        \begin{fmfgraph*}(40,30)
     \fmfleftn{i}{2} \fmfright{o}
     \fmflabel{$t$}{o}
     \fmf{wiggly_arrow,tension=.7,left=.3}{i2,o}
     \fmf{wiggly_arrow,tension=.7,left=.3}{o,i1}
     \fmfdot{o}
   \end{fmfgraph*} \hspace{.6cm} \raisebox{.5cm}{$+$} \hspace{.2cm}
     \begin{fmfgraph*}(60,40)
     \fmfleftn{i}{4} \fmfright{o}
     \fmflabel{$t$}{o}
     \fmf{wiggly_arrow,tension=.1,left=.2}{i4,v1}
     \fmf{wiggly_arrow,tension=.1,left=.3}{o,i1}
        \fmf{fermion}{i3,v1}
        \fmf{wiggly_arrow,tension=.1,left=.2}{v1,o}
        \fmffreeze
        \fmfshift{.46w,.1h}{v1}
        \fmfv{decor.shape=circle,decor.size=.1w,decor.filled=30}{v1}
        \fmfdot{o}
   \end{fmfgraph*} \hspace{.6cm} \raisebox{.5cm}{$+$} \hspace{.2cm}
   \begin{fmfgraph*}(60,40)
     \fmfleftn{i}{4} \fmfright{o}
     \fmflabel{$t$}{o}
     \fmf{wiggly_arrow,tension=.1,left=.2}{i4,o}
     \fmf{wiggly_arrow,tension=.1,left=.3}{v2,i1}
       \fmf{fermion}{i2,v2}
        \fmf{wiggly_arrow,tension=.1,left=.2}{o,v2}
        \fmffreeze
        \fmfshift{.46w,-.1h}{v2}
        \fmfv{decor.shape=circle,decor.size=.1w,decor.filled=30}{v2}
        \fmfdot{o}
   \end{fmfgraph*}  \hspace{.6cm} \raisebox{.5cm}{$+$} \hspace{.2cm}
  \begin{fmfgraph*}(60,40)
     \fmfleftn{i}{4} \fmfright{o}
     \fmflabel{$t$}{o}
     \fmf{wiggly_arrow,tension=.1,left=.2}{i4,v1}
     \fmf{wiggly_arrow,tension=.1,left=.3}{v2,i1}
        \fmf{fermion}{i3,v1}
       \fmf{fermion}{i2,v2}
        \fmf{wiggly_arrow,tension=.1,left=.2}{v1,o}
        \fmf{wiggly_arrow,tension=.1,left=.2}{o,v2}
        \fmffreeze
        \fmfshift{.46w,.1h}{v1}
        \fmfshift{.46w,-.1h}{v2}
        \fmfv{decor.shape=circle,decor.size=.1w,decor.filled=30}{v1}
        \fmfv{decor.shape=circle,decor.size=.1w,decor.filled=30}{v2}
        \fmfdot{o}
   \end{fmfgraph*}\hspace{.6cm} \raisebox{.5cm}{$+$} \hspace{.2cm}
   \notag \\ \vspace{4pt} &
        \begin{fmfgraph*}(60,40)
     \fmfleftn{i}{3} \fmfright{o}
     \fmflabel{$t$}{o}
     \fmfpolyn{pentagon,pull=.9,tension=.7,filled=30,smooth}{v}{5}
      \fmf{wiggly_arrow,tension=.1,left=.2}{i3,v4}
      \fmf{wiggly_arrow,tension=.1,left=.2}{v1,i1}
      \fmf{fermion}{i2,v5}
       \fmf{wiggly_arrow,tension=.7,left=.6}{v3,o}
        \fmf{wiggly_arrow,tension=.1,left=.6}{o,v2}
        \fmfdot{o}
   \end{fmfgraph*}  \hspace{.6cm} \raisebox{.5cm}{$+\quad \Bigg\{$}  \hspace{.2cm}
      \begin{fmfgraph*}(60,40)
     \fmfleftn{i}{5} \fmfright{o}
     \fmflabel{$t$}{o}
     \fmfpolyn{triagram,tension=.27,filled=30}{v}{3}
     \fmf{wiggly_arrow,tension=.1,left=.2}{i5,k}
     \fmf{wiggly_arrow,tension=.1,left=.3}{i4,v3}
     \fmf{wiggly_arrow,tension=.1,left=.3}{v3,i3}
       \fmf{fermion}{v2,k}
       \fmf{fermion}{i2,v1}
       \fmf{wiggly_arrow,tension=.1,left=.2}{o,i1}
        \fmf{wiggly_arrow,tension=.1,left=.2}{k,o}
        \fmffreeze
        \fmfshift{.4w,.3h}{k}
        \fmfshift{.27w,.1h}{v1,v2,v3}
        \fmfv{decor.shape=circle,decor.size=.1w,decor.filled=30}{k}
        \fmfdot{o}
   \end{fmfgraph*} \hspace{.6cm} \raisebox{.5cm}{$+$} \hspace{.2cm}
   \begin{fmfgraph*}(70,30) 
   \fmfleftn{i}{5} \fmfright{o}
   \fmflabel{$t$}{o} 
   \fmf{wiggly_arrow,tension=.1,left=.2}{i5,v1}
   \fmf{wiggly_arrow,tension=.1,left=.2}{v1,v2}
   \fmf{wiggly_arrow,tension=.1,left=.2}{v2,o}
   \fmf{fermion}{i4,v1}
   \fmf{fermion}{i3,v2}
   \fmf{wiggly_arrow,tension=.1,left=.2}{o,i1}
   \fmffreeze
   \fmfshift{.4w,.3h}{v1}
   \fmfshift{.65w,.4h}{v2}
   \fmfv{decor.shape=circle,decor.size=.1w,decor.filled=30}{v1}
   \fmfv{decor.shape=circle,decor.size=.1w,decor.filled=30}{v2}
   \fmfdot{o}
 \end{fmfgraph*}\hspace{.6cm} \raisebox{.5cm}{$+ \quad {\rm H.c.} \quad
   \Bigg\}. $} 
   \end{align}
 \end{fmffile}
  \end{widetext}
  Let us begin our discussion of this perturbation series by
  considering the first term on the right hand side of equation
  (\ref{eq:stokesgenerator-feynman}). This term is the zeroth-order
  term of the Stokes generator $\bar{\bar{\mathcal S}}^{(0)}$. In  the
  far-field limit $z\rightarrow \infty$ the matrix-element we need to
  calculate is
  \begin{align}
    \label{eq:calculatings_0}
    \ddleft \tilde{\mathbf f}_{kmj}^*(\mathbf
    r,t)|&\tilde{\boldsymbol{\mathcal D}}^{(-)}_0(\mathbf r,t)
    \ddright = \notag \\ & \iint_{-\infty}^{\infty}dtd^2r_{\perp}
    \frac{1}{\sqrt{2\pi}} U_m(\mathbf
    r_{\perp})e_je^{ikz-i(\omega_k-\laser)t} \cdot \notag \\ & \quad
    \sum_{qnl} \sqrt{\frac{\hbar \epsilon_0 \laser}{4\pi}}
    U^*_n(\mathbf r_{\perp})e_le^{-iqz-i(\omega_q-\laser)t} \hat
    a^{\dagger}_{qnl} \notag \\
    = & \sqrt{\frac{\hbar \epsilon_0 \laser}{2c^2}} \hat
    a^{\dagger}_{kmj},
  \end{align}
  and $\bar{\bar{\mathcal S}}^{(0)}$ thus gives us
  \begin{align}
    \label{eq:stokesgenerator_zeroth}
    K \ddleft \tilde{\mathbf f}_{kmj}^*(\mathbf
    r,t)|\bar{\bar{\mathcal S}}^{(0)}|\tilde{\mathbf f}_{km'j'}^*(\mathbf
    r,t) \ddright =  \hat a^{\dagger}_{kmj} \hat a_{km'j'}.
  \end{align}
  The zeroth order Stokes
operator $\hat s_1^{\mathbf q \mathbf q'}$ for $\mathbf q=(k,m,x)$ and
$\mathbf q'=(k,m',y)$ gives
\begin{subequations}
  \label{eq:s1-zeroth}
  \begin{align}
    \hat s_1^{\mathbf q \mathbf q'} \equiv \hat s_1^{mm'}
    =&\frac{1}{2}\big( \hat a^{\dagger}_{kmx}\hat a_{kmx} - \hat
    a^{\dagger}_{km'y}\hat a_{km'y} \big).  \intertext{The two
      remaining zeroth order Stokes operators are found accordingly,}
    \hat s_2^{mm'}
    =&\frac{1}{2}\big( \hat a^{\dagger}_{kmx}\hat a_{km'y} + \hat
    a^{\dagger}_{km'y}\hat a_{kmx} \big), \\
   \hat s_3^{mm'}
    =&\frac{1}{2i}\big( \hat a^{\dagger}_{kmx}\hat a_{km'y} - \hat
    a^{\dagger}_{km'y}\hat a_{kmx} \big).
  \end{align}
\end{subequations}

In the following we will calculate the first-order
components of the Stokes operators. We assume the quantum mechanical
average of the atomic spin $\mathbf J$ to be parallel the $x$-axis.
The relevant interaction matrix can in this case be written
\begin{align}
  \label{eq:relevant-matrix-interaction}
  \bar{\bar{ m}}[\hat{\mathbf J}]=ic_1\beta \left[
    \begin{array}{ccc}
      0 & \hat J_z(\mathbf r) & -\hat J_y(\mathbf r) \\
      - \hat J_z(\mathbf r) & 0 & 0 \\
      \hat J_y (\mathbf r) & 0& 0 
    \end{array}\right],
\end{align}
and after spatial averaging we simply write
\begin{align}
  \label{eq:after_spatial_average}
  \<\bar{\bar{ m}}[\hat{\mathbf J}]\>_{\mathrm{sa.}}\equiv
  \bar{\bar{\mathscr M}}[\bar{\mathbf J}]=-ic_1\beta \rho(\mathbf
  r)\left(
    \begin{array}{c}
      0 \\ \bar J_y(\mathbf r) \\ \bar J_z(\mathbf r)
    \end{array} \right) \times.
\end{align}
The second and the third term on the right hand side of
Eq. \eqref{eq:stokesgenerator-feynman} are the first order terms
of the Stokes generator, $\bar{\bar{\mathcal S}}^{(1)}$. To calculate
the contribution to the Stokes operators from these terms we have to
evaluate the expression
\begin{align}
  \label{eq:calcul_first_order_stokes}
 \ddleft \tilde{\mathbf f}^*_{kmj}(\mathbf r,t) | \:c^2 \hspace{-2pt}
 \iint_{t_0}^{t}dtd^3r'\: \bar{\bar P}^{(-)}(\mathbf r,t&|\mathbf
 r',t')\cdot \notag \\ &\bar{\bar{\mathscr M}}^t[\bar{\mathbf
   J}]\tilde{\boldsymbol{\mathcal D}}_0^{(-)}(\mathbf r',t') \ddright .
\end{align}
The initial time $t_0$ we will set to $-\infty$, and because we assume
our detector plane to be infinitely far away from the atomic ensemble,
we can take $t\rightarrow \infty$.  Using the expression for the set
$\{ \mathbf f_{\mathbf q} \}$ given by Eq.
\eqref{eq:par-axial-mode-fkt} for the detector plane and Eq.
\eqref{eq:modefunctions_ensemble} inside the ensemble, Eq.
\eqref{eq:calcul_first_order_stokes} reduces to
\begin{align}
  \label{eq:first_order_result}
  \left( -i \sqrt{\frac{\hbar \laser \epsilon_0}{2c^2}} \right)
  \frac{\klaser c_1 \beta }{2} \int d^3r' \sum_{nl} \rho (\mathbf r')
  {\Theta^{mn}_{jl} (\mathbf r')}^* \hat a_{knl}^{\dagger}, 
\end{align}
where
\begin{align}
   \Theta^{mn}_{jl} (\mathbf r')& \equiv U_{km}(\mathbf
  r')^*U_{kn}(\mathbf r')\mathbf e_j(\mathbf r)\cdot \big[ \left(
    \begin{array}{c}
      0 \\ \bar J_y(\mathbf r') \\ \bar J_z(\mathbf r')
    \end{array} \right) \times \mathbf e_l(\mathbf r') \big] \notag
  \\ &=\Psi^{mn}_k(\mathbf r')\big[
  \delta_{lx} \delta_{jy} - \delta_{jx} \delta_{ly} \big] \big[ \left(
    \begin{array}{c}
      0 \\ \bar J_y(\mathbf r') \\ \bar J_z(\mathbf r')
    \end{array} \right) \cdot \mathbf e_z(\mathbf r') \big],
  \intertext{with}
  & \Psi^{mn}_k(\mathbf r') = U_{km}(\mathbf
  r')^*U_{kn}(\mathbf r').\label{eq:bar-j_zmathbf-r}
\end{align}
In the final equality we have introduced the local basis vector
$\mathbf e_z(\mathbf r) = \mathbf e_x(\mathbf r) \times \mathbf
e_y(\mathbf r)$.  The effect of the first-order term of the Stokes
generator $\bar{\bar{\mathcal S}}^{(1)}$ to the Stokes operators thus
reads
\begin{align}
  \label{eq:firstorder_stokes_operator_general}
  K \ddleft &\tilde{\mathbf f}_{kmj}^*(\mathbf r,t)|\bar{\bar{\mathcal
      S}}^{(1)}|\tilde{\mathbf f}_{km'j'}^*(\mathbf
  r,t) \ddright = \notag \\
  &\klaser c_1 \beta \int d^3r' \sum_{nl} \rho (\mathbf r') \:
  \frac{1}{2}\big\{ {\Theta^{mn}_{jl} (\mathbf r')}^* \hat
  a_{knl}^{\dagger} \hat a_{km'j'} \notag \\ & \hspace{3cm}+
  {\Theta^{m'n}_{j'l} (\mathbf r')} \hat a_{kmj}^{\dagger} \hat
  a_{knl} \big\}.
\end{align}

The remaining terms of the right hand side of Eq.
(\ref{eq:stokesgenerator-feynman}), that is the second-order terms,
can be calculated in a similar way. The results may be found 
in Appendix \ref{sec:calul-second-order}.  The calculations given in
Eq.  (\ref{eq:firstorder_stokes_operator_general}),
(\ref{eq:second_order_calc_stokes_terms_1}),
(\ref{eq:second_order_calc_stokes-2}) and
(\ref{eq:stokes_term_second_4}) is the starting-point for a discussion
of the dynamics of the system subject to a general light field of many
modes. 

The description that we have used here, where we define
the Stokes operators in term of expectation value between different
orthogonal modes, is very convenient for a theoretical description of
the process. It does, however, not directly correspond to the
experimentally measured observables unless one, e.g., separates out
particular modes with single mode optical fibers.  We shall therefore
defer the discussion of the consequences of these results to the next
section, where we use these result to calculate the evolution of
observables more relevant to experiments.


We will now give the equation for the atomic spin.  The incoherent
terms describing decay due to spontaneous emission have already been
discussed. Here we will consider the coherent interaction up to second
order in the perturbation series.  Below we show the
diagrammatic representation of the coherent perturbation series for
the atomic spin up to second order.
  \begin{widetext}
\begin{fmffile}{feynman_d_10a}
  \fmfcmd{%
    style_def wiggly_arrow expr p =
    cdraw (wiggly p);
    shrink (1);
       cfill (arrow p);
     endshrink;
     enddef;}

 \fmfcmd{%
    style_def DwigA expr p =
    draw_double (wiggly p);
    shrink (1.5);
       cfill (arrow p);
     endshrink;
     enddef;}
\begin{align} \label{eq:diagram-spin-coherent-second-order} 
\begin{fmfgraph*}(60,30) \fmfleft{i} \fmfright{o}
         \fmflabel{$t$}{o}
    \fmf{dbl_plain_arrow}{i,o} 
    \fmfdot{o}
    \end{fmfgraph*} 
    \hspace{.6cm} &\raisebox{.4cm}{=} \hspace{.1cm}
  \begin{fmfgraph*}(35,30) \fmfleft{i} \fmfright{o}
         \fmflabel{$t$}{o}
    \fmf{fermion}{i,o} 
    \fmfdot{o}
  \end{fmfgraph*}\qquad
  \raisebox{.4cm}{+}\:
      \begin{fmfgraph*}(70,30) \fmfleftn{i}{3} \fmfright{o}
        \fmflabel{$t$}{o}
        \fmfpolyn{triagram,tension=.7,filled=30}{v}{3}
    \fmf{wiggly_arrow,tension=.5,left=.5}{i3,v1} 
    \fmf{wiggly_arrow,tension=.5,left=.5}{v1,i2}
    \fmf{fermion}{i1,v2}
    \fmf{fermion}{v3,o}
    \fmffreeze
    \fmfshift{0.2w,0h}{v1,v2,v3}
    \fmfshift{.09w,.1h}{i2}
    \fmfdot{o} 
    \end{fmfgraph*} \qquad  \raisebox{.4cm}{+} \quad
    \begin{fmfgraph*}(60,50) \fmfleftn{i}{4} \fmfright{o}
        \fmflabel{$t$}{o}
        \fmfpolyn{triagram,tension=.8,filled=30}{v}{3}
    \fmf{wiggly_arrow,tension=.5,left=.5}{i3,k,v1} 
    \fmf{wiggly_arrow,tension=.5,left=.5}{v1,i2}
    \fmf{fermion}{i1,v2}
    \fmf{fermion}{v3,o}
    \fmf{fermion}{i4,k}
    \fmffreeze
    \fmfshift{0.25w,0h}{v1,v2,v3}
    \fmfshift{.15w,0h}{k}
    \fmfshift{-.1w,0h}{i4}
    \fmfshift{0w,.06h}{i2}
    \fmfshift{0w,-.06h}{i3}
    \fmfdot{o} \fmfv{decor.shape=circle,decor.size=.1w,decor.filled=30}{k}
    \end{fmfgraph*} \qquad \raisebox{.4cm}{+} \quad
    \begin{fmfgraph*}(60,50) \fmfleftn{i}{4} \fmfright{o}
        \fmflabel{$t$}{o}
        \fmfpolyn{triagram,tension=.8,filled=30}{v}{3}
    \fmf{wiggly_arrow,tension=.5,left=.5}{i4,v1} 
    \fmf{wiggly_arrow,tension=.5,left=.5}{v1,k,i3}
    \fmf{fermion}{i1,v2}
    \fmf{fermion}{v3,o}
    \fmf{fermion}{i2,k}
    \fmffreeze
    \fmfshift{0.25w,0h}{v1,v2,v3}
    \fmfshift{.15w,0h}{k}
    \fmfshift{-.1w,0h}{i3}
    \fmfshift{0w,-.1h}{i2}
    \fmfshift{-.18w,-.11h}{i4}
    \fmfdot{o} \fmfv{decor.shape=circle,decor.size=.1w,decor.filled=30}{k}
    \end{fmfgraph*} \notag \\ &\hspace{8cm}
        \raisebox{.4cm}{+} \quad
          \raisebox{.6cm}{{\large{\bf :}} } 
      \begin{fmfgraph*}(60,50) \fmfleftn{i}{5} \fmfright{o}
        \fmflabel{$t$}{o}
        \fmfpolyn{triagram,tension=.8,filled=30}{v}{3}
    \fmf{wiggly_arrow,tension=.5,left=.5}{i5,v1} 
    \fmf{wiggly_arrow,tension=.5,left=.5}{v1,i4}
    \fmfpolyn{triagram,tension=.7,filled=30}{k}{3}
    \fmf{fermion}{k1,v2}
    \fmf{fermion}{v3,o}
    \fmf{wiggly_arrow,tension=.5,left=.5}{i3,k2} 
    \fmf{wiggly_arrow,tension=.5,left=.5}{k2,i2}
    \fmf{fermion}{i1,k3}
    \fmffreeze
    \fmfshift{0.2w,0h}{v1,v2,v3}
    \fmfshift{.2w,-.1h}{k1,k2,k3}
    \fmfshift{-.01w,.1h}{i2}
    \fmfdot{o} 
    \end{fmfgraph*} \hspace{.6cm}\raisebox{.6cm}{{\large{\bf :}} }
    \end{align}
   \end{fmffile}

 \end{widetext}
\vspace{1cm} 

We will denote the first order term in the expansion,
Eq. (\ref{eq:diagram-spin-coherent-second-order}),  as $\mathcal
J^{(1)}$. Employing  again the approximations done in the previous
calculations, that is, using the set of light modes $\{ \mathbf
f_{\mathbf q}\}$ given in Eq. (\ref{eq:modefunctions_ensemble}) and
setting the initial time to $-\infty$ and the final time to $\infty$,
the term can be written 
\begin{align}
  \label{eq:firstorder_spin_result}
  \mathcal J^{(1)} = -&\beta c_1 \klaser \sum_{kmm'}
  \Psi_k^{mm'}(\mathbf r) \Big( \bar{\mathbf J}(\mathbf r) \times
  \mathbf e_z(\mathbf r) \Big) \notag \\ &\frac{1}{2i} \Big[ \hat
  a^{\dagger}_{kmx}\hat a_{km'y}- \hat a^{\dagger}_{kmy}\hat a_{km'x}
  \Big]\notag \\ = -&\beta c_1 \klaser \sum_{kmm'}
   \Big( \bar{\mathbf J}(\mathbf r) \times
  \mathbf e_z(\mathbf r) \Big) \notag \\ &
  \Big\{ {\rm Re}[ \Psi_k^{mm'}(\mathbf r) ] \hat s_3^{mm'} + {\rm Im} [
  \Psi_k^{mm'}(\mathbf r) ] \hat s_2^{mm'} \Big\}.
\end{align}
We notice that compared to the simple theory in Ref.
\cite{brian_thesis} there is an additional term proportional to the
imaginary part of the function $\Psi^{mm'}(\mathbf r)$. A similar
correction can also be found for the Stokes operators for the light.
Also notice that the dynamics of the spin to first order happens in a
plane orthogonal to the vector $\mathbf e_z(\mathbf r)$. This is the
reason why the term in Eq. (\ref{eq:stokes_ters_calc_3}) vanish, since
there we are considering the effect of the dynamics of the atomic spin
on an axis parallel to the $\mathbf e_z(\mathbf r)$-vector. The
calculation of the second-order terms is presented in Appendix
\ref{sec:calc-second-order}.  In the following section we will examine
the effect of these calculations under conditions attainable in
experiments.

  \section{Experimental application  and
    validity}\label{sec:exper-appl-valid} 
  In this section we shall consider different limits where we can
  reduce our general theory to a theory resembling the simple
  description obtained in one dimensional theories
  \cite{brian_thesis,duan}. Furthermore we discuss the validity of the
  approximations made to arrive at these simple limits as well as the
  validity of our perturbative treatment of the interaction.
  \bigskip
  
  \subsection{Measurement procedure} 
  In the previous section we discussed how our theory could be used to
  calculate Stokes operators corresponding to specific transverse
  modes of the field. While such a treatment is appealing from a
  theoretically perspective, it is less desirable experimentally,
  since the isolation of single transverse modes is complicated
  (although it could be done by passing the light through single mode
  optical fibers). Here we shall therefore express our result in terms
  of a simpler experimental procedure. Suppose that the detections is
  performed by sending the light onto a polarizing beamsplitter and
  recording the intensity of the two output port with two cameras. The
  difference between the intensities can now be used to define
  position dependent Stokes operators $\hat{s}_i({\bf r_\perp})$,
  i.e., $\hat{s}_1({\bf r_\perp})$ corresponds to the difference in
  intensity between $x$ and $y$ polarization at position ${\bf
    r_\perp}$ in the detector plane. Similarly $\hat{s}_2({\bf
    r_\perp})$ and $\hat{s}_3({\bf r_\perp})$ can, respectively, be
  related to the difference intensity with the polarizer rotated by
  $45^\circ$ and the difference intensity between the two circular
  polarizations.  These operators may in general be determined by
  \begin{widetext}
    \begin{subequations}     \label{eq:stokes_multimode_operator}
      \begin{align}
           \hat s_1(\mathbf r_\perp) =& \sum_{kmm'} \frac{1}{2}\Big(
        U_{m}^*(\mathbf r_\perp)\hat a^{\dagger}_{kmx} \hat a_{km'x}
        U_{m'}(\mathbf r_\perp) - U_m^*(\mathbf r_\perp)\hat
        a^{\dagger}_{kmy} \hat a_{km'y} U_{m'}(\mathbf r_\perp)
        \Big)\\
          \hat s_2(\mathbf r_\perp) =& \sum_{kmm'} \frac{1}{2}\Big(
        U_m^*(\mathbf r_\perp)\hat a^{\dagger}_{kmx} \hat a_{km'y}
        U_{m'}(\mathbf r_\perp) + U_m^*(\mathbf r_\perp)\hat
        a^{\dagger}_{kmy} \hat a_{km'x} U_{m'}(\mathbf r_\perp)
        \Big)\\
        \hat s_3(\mathbf r_\perp) =& \sum_{kmm'} \frac{1}{2i}\Big(
        U_m^*(\mathbf r_\perp)\hat a^{\dagger}_{kmx} \hat a_{km'y}
        U_{m'}(\mathbf r_\perp) - U_m^*(\mathbf r_\perp)\hat
        a^{\dagger}_{kmy} \hat a_{km'x} U_{m'}(\mathbf r_\perp)
        \Big).
      \end{align}
    \end{subequations}
  \end{widetext}

  Below we shall derive expressions for the operators
  (\ref{eq:stokes_multimode_operator}) and discuss how to implement a
  light-matter quantum interface based on these operators.  In subsec.
  \ref{sec:strongly-coupl-multi} we for simplicity first consider an
  extreme paraxial limit, where we assume that essentially no
  diffraction occurs during the propagation. In this limit the
  dynamics becomes extremely simple. In subsec.
  \ref{sec:weakly-coupl-single} we consider a more interesting limit,
  where we may have multiple modes which may experience diffraction.
  Here we show that measurement of the operators $\hat{s}_i({\bf
    r_\perp})$ still allows us to simplify the dynamics of the system.
  In a suitable limit we find a simple two mode transformation between
  transverse modes of the light field and single modes of the atomic
  ensembles.
 

\subsection{Extreme paraxial approximation}
\label{sec:strongly-coupl-multi}

In the extreme paraxial approximation, we completely ignore any
dynamics transverse to the propagation direction of the light modes
and approximate the set of modes $\{\mathbf f_{\mathbf q} \}$ with Eq.
(\ref{eq:par-axial-mode-fkt}) throughout the ensemble.  Since the
typical distance for diffraction is given by $l_d\sim A/\lambda$, the
condition for the validity of this approximation is $L \ll l_d$, or
expressed in terms of the Fresnel number $\mathcal F\gg 1$.
  
The full expressions for the Stokes operators are quite involved, and
we therefore leave the incoherent part of the evolution to Appendix
\ref{sec:calc-spont-emiss}.  Keeping only the coherent part of the
interaction, we find the Stokes operators to second order in the
interaction to be
  \begin{widetext}
    \begin{subequations}        \label{eq:full-eq-stoke-par-multi}
      \begin{align}
        \hat s_{1,out}(\mathbf r_{\perp})=& \hat s_{1,in}(\mathbf
        r_{\perp}) - \klaser c_1 \beta \int dz' \rho(z',\mathbf
        r_{\perp})\bar J_z(z',\mathbf r_{\perp}) \hat s_{2,in}(\mathbf
        r_{\perp}) \notag \\ & \hspace{3cm}- \frac{1}{2}(\klaser\beta
        c_1)^2\iint dz'dz'' \rho(z',\mathbf r_{\perp})\rho(z'',\mathbf
        r_{\perp})\bar J_z(z',\mathbf r_{\perp})\bar J_z(z'',\mathbf
        r_{\perp}) \hat s_{1,in}(\mathbf
        r_{\perp}),\\
        \hat s_{2,out}(\mathbf r_{\perp})=& \hat s_{2,in}(\mathbf
        r_{\perp}) + \klaser c_1 \beta \int dz' \rho(z',\mathbf
        r_{\perp})\bar J_z(z',\mathbf r_{\perp}) \hat s_{1,in}(\mathbf
        r_{\perp}) \notag \\ & \hspace{3cm} - \frac{1}{2}(\klaser\beta
        c_1)^2\iint dz'dz'' \rho(z',\mathbf r_{\perp})\rho(z'',\mathbf
        r_{\perp})\bar J_z(z',\mathbf r_{\perp})\bar J_z(z'',\mathbf
        r_{\perp}) \hat s_{2,in}(\mathbf
        r_{\perp}), \\
        \hat s_{3,out}(\mathbf r_{\perp})=& \hat s_{3,in}(\mathbf
        r_{\perp}).
      \end{align}
    \end{subequations}
  \end{widetext}
  In this limit we see that the Stokes operator $\hat s_3$ is
  decoupled from the coherent dynamics of the system, and only evolves
  due to spontaneous emission [derived in Eq.
  (\ref{eq:spon-corr-emmis})].
  
  Similarly we may find the coherent dynamics of the atomic spin.
  Leaving again the incoherent part to Appendix
  \ref{sec:calc-spont-emiss}, we find
  \begin{widetext}
    \begin{subequations}        \label{eq:coherent-spin-dynamic}
      \begin{align}
        \bar J_{x,out}(\mathbf r) =& \bar J_{x,in}(\mathbf r) - \beta
        c_1\klaser \sum_k \bar J_{y,in}(\mathbf r)\hat
        s_{3,in}^k(\mathbf r_{\perp}) - \frac{1}{2}(\beta c_1 \klaser
        )^2 \sum_{kk'} \bar J_{x,in}(\mathbf r) \hat
        s^k_{3,in}(\mathbf r_{\perp}) \hat s^{k'}_{3,in}(\mathbf
        r_{\perp})\\
        \bar J_{y,out}(\mathbf r) =& \bar J_{y,in}(\mathbf r) + \beta
        c_1\klaser \sum_k \bar J_{x,in}(\mathbf r)\hat
        s_{3,in}^k(\mathbf r_{\perp}) - \frac{1}{2}(\beta c_1 \klaser
        )^2 \sum_{kk'} \bar J_{y,in}(\mathbf r) \hat
        s^k_{3,in}(\mathbf r_{\perp}) \hat s^{k'}_{3,in}(\mathbf
        r_{\perp}) \\
        \bar J_{z,out}(\mathbf r) =& \bar J_{z,in}(\mathbf r).
      \end{align}
    \end{subequations}
  \end{widetext}
  Analogous to what we found for $\hat{s}_3$, we see that the operator
  $\bar J_z$ is decoupled from the coherent dynamics of the system.
  This result can directly be associated to the conservation of
  angular momentum along the $z$-axis.  In the extreme paraxial
  approximation this is true to all orders in the coherent
  interaction.

  The results in Eq.  (\ref{eq:full-eq-stoke-par-multi}) and
  (\ref{eq:coherent-spin-dynamic}) is essentially equivalent to the
  simplified one-dimensional description of the system given in Refs.
  \cite{duan,brian_thesis}. The only difference is that the
  expressions derived here now apply for each value of ${\bf
    r}_\perp$, whereas the previous treatments assumed the system was
  transversely homogeneous and only considered the variables
  integrated over ${\bf r}_\perp$.
  
  A further simplification of Eq. (\ref{eq:coherent-spin-dynamic}) can
  be obtained if we introduce the rotation vector
    \begin{align}
    \label{eq:kappa-et-def}
    \boldsymbol{\Omega}=\beta c_1\klaser \sum_k \hat s_{3,in}^k(\mathbf
    r_{\perp}) \mathbf e_z.
  \end{align}
  With this definition we find that Eq. (\ref{eq:coherent-spin-dynamic}) describes nothing
  but a rotation of the spin around the $\mathbf e_z$-axis
  \begin{align}
    \label{eq:rotation-ez-spin}
    \bar{\mathbf J}_{out} = \bar{\mathbf J}_{in} + \bar{\mathbf J}_{in} \times
    \boldsymbol{\Omega} + \frac{1}{2} \big( \bar{\mathbf J}_{in} \times
    \boldsymbol{\Omega} \big) \times \boldsymbol{\Omega} .
  \end{align}

  \subsection{Multi-mode coupling}\label{sec:weakly-coupl-single}
  In the previous subsection we basically ignored all the dynamics
  transverse to the propagation direction. Now we turn to a more
  interesting situation, where we may describe effects associated with
  diffraction of the light beams. Our goal in this section is to find
  a set of conditions under which we can have a simple dynamics, where
  the individual transverse modes of the light field talks to a single
  mode of the atomic ensemble. Such an interaction would enable the
  storage of information from several light modes into spatial modes
  of the ensemble, e.g., using the protocol in \cite{juulsgaard2}.
  The realization of this interaction would thus expand the
  information storage capacity of the atomic ensembles. A similar
  problem is considered in Ref. \cite{denise}. In related work such
  storage of multimode memory has recently been achieved in atomic
  ensembles using electromagnetically induced transparency \cite{eit-memory}.
  
  To achieve simple results in the
  end, we will here consider a situation, where we have a strong
  classical beam polarized in the $x$-direction in a single transverse
  mode $U_{ok}(\mathbf r)$ (denoted by the index $o$). For the
  $y$-polarization we, however, include a complete set of modes, which
  may or may not include a term identical to the mode of the
  $x$-polarization. 
  For the strong mode we will approximate $\hat a^{\dagger}_{kox}=\hat
  a_{kox}=\sqrt{N^o_x}\gg 1$ where $N_x^o$ is the number of photons in
  this particular mode. Since the Stokes operators are dominated by
  the terms involving the classical component, the only important
  contributions in the Stokes operator
  (\ref{eq:stokes_multimode_operator}) are the terms containing
  the strong classical mode. 
   Eq. (\ref{eq:stokes_multimode_operator}) are thus approximated by
  \begin{subequations}      \label{eq:simplified-stokes-xp}
    \begin{align}
      \hat s^{(in)}_1(\mathbf r_\perp)\approx \:&  \frac{1}{2} |U_{o\klaser}(\mathbf r_\perp)|^2 N_x^o, \\
      \hat s^{(in)}_2(\mathbf r_\perp)\approx\: & \frac{\sqrt{ N_x^o}}{2}\sum_{km}
      \Big( {\rm Re}[ U_{ok}^*(\mathbf r_\perp)U_{mk}(\mathbf r_\perp)] \hat X^m_P
      \notag \\ & \hspace{2cm} - {\rm
        Im}[ U_{ok}^*(\mathbf r_\perp)U_{mk}(\mathbf r_\perp)] \hat P^m_P \Big),\\
      \hat s^{(in)}_3(\mathbf r_\perp)\approx\: & \frac{\sqrt{ N_x^o}}{2}\sum_{km}
      \Big( {\rm Re}[ U_{ok}^*(\mathbf r_\perp)U_{mk}(\mathbf r_\perp)] \hat P^m_P
      \notag \\ & \hspace{2cm} + {\rm Im}[ U_{ok}^*(\mathbf
      r_\perp )U_{mk}(\mathbf r_\perp )] \hat X^m_P \Big),
    \end{align}
  \end{subequations}
  where
  \begin{subequations} \label{eq:modefunkewhgirew}
    \begin{align}
           \hat X^m_P =& \frac{1}{\sqrt 2}\big(\hat
      a^{\dagger}_{kmy} + \hat a_{kmy} \big), \\
      \hat P^m_P =& \frac{1}{i \sqrt 2}\big(\hat a^{\dagger}_{kmy} -
      \hat a_{kmy} \big).
    \end{align}
  \end{subequations}

  In order to obtain simple result in the measurement process, let us
  assume that we can choose the mode functions $U_{mk}(\mathbf r)$ to
  be real in the detection plane. This could, e.g., be achieved by
  sending the light through a lens which converts the incoming modes
  into extreme paraxial beams as shown in Fig. \ref{fig:focusing} (note that
  since we only make this assumption in the detection plane, this
  assumption does not restrict the shape inside the ensemble).
  Experimentally the operators $\hat X^m$ and $\hat P^m$ defined here
  can then be measured by simply integrating the measured $ \hat
  s_i(\mathbf r_{\perp}) $ with a suitable weight function, e.g.,
  \begin{equation}\label{eq:sqrtfr-dmathbf-r_per}
    \sqrt{\frac{2}{N_x^0}}\int d\mathbf r_\perp  \frac{U_m(\mathbf
      r_\perp)} {U_o(\mathbf r_\perp)}\hat s_2(\mathbf r_{\perp})  =\hat X^m_P,
  \end{equation}
  where we have used the expansion in
  (\ref{eq:stokes_multimode_operator}) as well as the orthogonality
  relation of the transverse mode functions
  \eqref{eq:int-d2r_p-u_nm}.
 
  In our equations of motions we for simplicity  only keep
  terms to first order in $\beta$ and $\sqrt{N^o_x}$, and neglect all
  other terms. The equations of motion for the Stokes operators give
  in this limit
  \begin{widetext}
    \begin{subequations}        \label{eq:single-mode-weak-coupl-stokes}
      \begin{align}
        \hat X^m_{out} =& \hat X^m_{in} + \klaser \beta c_1
        \sqrt{\frac{N^o_x}{2}} \int d^3r' \: \rho(\mathbf r') \left(
          \begin{array}{c}
            0\\ \bar J_y(\mathbf r) \\ \bar J_z(\mathbf r)
          \end{array}\right) \cdot \mathbf e_{z}(\mathbf
        r){\rm Re}[\Psi^{mo}_ k(\mathbf r)]  \\
        \hat P^m_{out} =& \hat P^m_{in} + \klaser \beta c_1
        \sqrt{\frac{N^o_x}{2}} \int d^3r' \: \rho(\mathbf r') \left(
          \begin{array}{c}
            0\\ \bar J_y(\mathbf r) \\ \bar J_z(\mathbf r)
          \end{array}\right) \cdot \mathbf e_{z}(\mathbf
        r){\rm Im}[\Psi^{mo}_ k(\mathbf r)] ,
      \end{align}
    \end{subequations}
  \end{widetext}
  where $\Psi^{mo}$ is defined in terms of the mode functions $U_m$ in
  Eq. \eqref{eq:bar-j_zmathbf-r}.  Employing the same set of
  approximations for the spin equation we find
  \begin{widetext}
    \begin{align}
      \label{eq:spin-eq-weak-multi-mode}
      \bar{\mathbf J}_{out}(\mathbf r) \approx \bar{\mathbf
        J}_{in}(\mathbf r) + \klaser \beta c_1 
      \sqrt{\frac{N^o_x}{2}} \sum_n \Big[ {\rm Re}[\Psi^{no}_ k(\mathbf
      r)]\hat P^n_{in} - {\rm Im}[\Psi^{mo}_ k(\mathbf r)]\hat X_{in}^n
      \Big] \Big(\bar{\mathbf J}_{in}(\mathbf r) \times \mathbf
      e_{z}(\mathbf r) \Big).
    \end{align}
  \end{widetext}
  Note, that the expressions we have derived here, allow for a general
  set of transverse modes which may experience diffraction, and thus
  go beyond the extreme paraxial approximation made in the previous
  section. In the expressions above we do, however, still use the
  paraxial approximation in Eq.  \eqref{eq:modefunctions_ensemble},
  where we ignore the dependence of the polarization vector on the
  mode number. In Appendix \ref{appendix-correction} we relax this
  approximation.

  The expressions in Eqs. \eqref{eq:single-mode-weak-coupl-stokes} and
  \eqref{eq:spin-eq-weak-multi-mode} differ from the simple results of
  the last section because of the extra terms proportional to ${\rm
    Im}[\Psi^{mo}_ k(\mathbf r)]$.  These terms complicate the
  dynamics and, e.g., means that one cannot use the protocol in Ref.
  \cite{juulsgaard2} to store information in the ensemble.  There are,
  however, certain limits where the extra terms in Eq.
  (\ref{eq:spin-eq-weak-multi-mode}) disappear. One situation is when
  the mode we are considering in the $y$-polarization is identical to
  the classical mode in the $x$-polarization (except from the
  different orientation of the polarization). This situation
  corresponds to the experimental situation, where the weight factor
  $U_m/U_o$ in Eq. (\ref{eq:sqrtfr-dmathbf-r_per}) is unity, such that
  the final result is obtained by integrating the intensity over the
  transverse plane.  This case therefore corresponds the experimental
  situation where the light is detected by photo detectors instead of
  cameras.  In this case ${\rm Im}[\Psi^{mo}_ k(\mathbf r)]$ vanish
  identically and the evolution of the light operators again resemble
  the result of the last section, where, e.g., the $\hat s_2$
  component was conserved, which translates into
  $\hat{P}^m_{out}=\hat{P}^m_{in}$. Note, however, that unlike the
  situation considered below, the atomic operators in this situation
  gets an admixture of several different input light modes, and will
  not in general reduce to the dynamics considered in Ref.
  \cite{juulsgaard2}.

  Let us now consider a different limit ideally suited for a
  multi-mode memory. We assume that we are in the paraxial
  approximation, where we can ignore the spatial dependence of the
  polarization vectors.  For simplicity we also assume that the
  classical mode $U_o(\mathbf r)$ has a uniform intensity and
  that the density is constant over the region, where $U_m$ is non-zero
  in the atomic ensemble.  
  We furthermore assume that the macroscopic polarization is constant and
  along the $x$-axis, $\bar J_x$, and finally  we assume that
  $\Psi^{mo}$ is real (for a discussion of the validity of this
  approximation we refer to the next subsection).
  In the spin equation (\ref{eq:spin-eq-weak-multi-mode}) we will only
  keep terms proportional to the macroscopic spin component $\bar
  J_x$. In this situation the relevant equations reads
  \begin{widetext}
    \begin{subequations}        \label{eq:relevant-equations}
      \begin{align}
        \hat X^m_{P,out} =& \hat X^m_{P,in} + \klaser \beta c_1 U_{o}
        \sqrt{\frac{N^o_x}{2}} \rho \int d^3r' \: \bar J_z(\mathbf r)
        U_{m}(\mathbf r) e^{-ikz}  \\
        \hat P^m_{P,out} =& \hat P^m_{P,in} \\
        \bar{J}_{y,out}(\mathbf r) =& \bar{ J}_{y,in}(\mathbf r) +
        \klaser \beta c_1 U_{o} \sqrt{\frac{N^o_x}{2}} \sum_n
        U_{n}(\mathbf r)e^{-ikz} \hat P^n_{in} \bar J_{x,in}
        \\
        \bar J_{z,out}(\mathbf r) =& \bar J_{z,in}(\mathbf r).
      \end{align}
    \end{subequations}
  \end{widetext}
  Here the factor $\exp(-ikz)$ comes from the classical field and
  cancels the $\exp(ikz)$ dependence of the mode function $U_m$, since
  $U_m\exp(-ikz)$ should be real according to the assumption of $\Psi$
  being real.  This set of equations can be symmetrized and simplified
  by introducing a set of collective operators
  \begin{subequations} \label{eq:collective-operators}
    \begin{align}
      \tilde X_A^m =& \sqrt{\frac{\rho}{J_x L}}\int d^3r \bar
      J_y(\mathbf r) U_m(\mathbf r)e^{-ikz}, \\
      \tilde P_A^m =&\sqrt{\frac{\rho}{J_x L}}\int d^3r \bar
      J_z(\mathbf r) U_m(\mathbf r)e^{-ikz},
    \end{align}
  \end{subequations}
  where $L$ is the length of the ensemble. The coefficients here are
  chosen such that the operators $\hat X_A^m$ and $\hat P_A^m$ fulfil
  the standard commutation relation for position and momentum
  \begin{align}
    \label{eq:commutation-relation-XaPa}
    \big[ \tilde X_A^m , \tilde P_A^{m'} \big] = i\delta_{mm'} .
  \end{align}
  With these definitions Eqs. (\ref{eq:relevant-equations}) reduce to
  \begin{subequations}      \label{eq:reduce-intre-equa}
    \begin{align}
      \tilde X_{P,out}^m =&  \tilde X_{P,in}^m + \kappa  \tilde
      P_{A,in}^m,\\
       \tilde P_{P,out}^m =&  \tilde P_{P,in}^m, \\
       \tilde X_{A,out}^m =&  \tilde X_{A,in}^m + \kappa  \tilde
      P_{P,in}^m,\\
       \tilde P_{A,out}^m =&  \tilde P_{A,in}^m,
       \intertext{where}
       \kappa =& \klaser \beta c_1 U_o \sqrt{\frac{N^o_x \rho \bar J_x
         L}{2}}.
    \end{align}
  \end{subequations}
  These equations describe a system where one transverse light-mode
  couples to a single mode of the atomic ensemble, which in term
  couple back to the same light mode. This two-mode mode dynamics is
  exactly identical to the dynamics derived in Ref. \cite{duan} for a
  single transverse mode. The dynamics can thus, e.g., be used to realize a multi-mode
  version of the memory protocol implemented Ref. \cite{juulsgaard2}.
  In this protocol $\hat P^m_{P,in}$ is stored in the atomic mode
  $\hat X_{a,out}^m$, while at the same time the atomic mode $\hat
  P^m_{A,in}$ is transferred to the light-mode $\hat X_{P,out}^m$, as described by Eq. \eqref{eq:reduce-intre-equa}.
  After detection of the light operator $\hat X_{P,out}^m$ one can
  then realize a quantum memory by feeding back the measurement result
  to the atoms as it was shown in Ref. \cite{juulsgaard2}.
  
  \subsection{Validity}
  \subsubsection{Validity of the simple multi-mode dynamics}
  In the previous subsection we derived a simple multi-mode dynamics
  useful for making a multi-mode light matter quantum interface.  For
  experimental implementation of these idea an important question is
  the validity of the approximations leading to Eq.
  (\ref{eq:reduce-intre-equa}).
  First of all we need that the imaginary part of $\Psi^{om}(\mathbf
  r)$ in Eq.  (\ref{eq:single-mode-weak-coupl-stokes}) should vanish.
  Furthermore, in order to define orthogonal spin-modes that do not
  couple different transverse modes, we need $|U_o(\mathbf r)|$ to be
  uniform. Taking the classical mode to be given by
  $U_o(\mathbf{r})=U_o e^{ikz}$, where $U_o$ is real, we also need the
  quantum mode $U_m(\mathbf r)$ to be real-valued apart from the
  $e^{ikz}$ dependence. Let us now take the modes $U_m(\mathbf r)$ to
  be Hermite-Gaussian beams \cite{lasers}. Such modes can be
  represented by
  \begin{subequations}
    \begin{align}
      \label{eq:gaussian-beam-mandel}
      U_{mn}(\mathbf r) =& \frac{B w_0}{w(z)}H_m\left(\sqrt 2
        \frac{x}{w(z)}\right) H_n\left(\sqrt 2 \frac{y}{w(z)}\right)
      \notag \\ & \qquad \times e^{i[kz-(m+n+1)\tanh z/z_0]}\notag \\
      & \qquad \times
      e^{ik(x^2+y^2)/2R(z)}e^{-(x^2+y^2)/w^2(z)}, 
      \intertext{where}
      w(z) =& w_0\sqrt{1+z^2/z_0^2}, \\
      R(z) =& z+\frac{z_0^2}{z}, \\
      z_0=& \frac{\pi w_0^2}{\lambda}.
    \end{align}
  \end{subequations}
  Here $w_0$ is the minimum waist of the beam, $k$ is the wave-number,
  $\lambda$ is the wavelength, $B\in\mathbb R$ is a normalization
  coefficient, and $H_n$ is the set of Hermite polynomials. The
  condition that $U_{mn}(\mathbf r)$ must be real-valued gives the
  conditions
  \begin{align}
    \label{eq:conditions}
    \lambda R(z) \gg w^2(z) \quad |(1+m+n)\frac{z}{z_0}|\ll 1
  \end{align}
  These are in fact equivalent conditions, and introducing the Fresnel
  number $\mathcal F \equiv w^2(z)/\lambda L$ we find the condition
  \begin{align}
    \label{eq:fresnel-number-large}
    \mathcal F \gg 1+m+n.
  \end{align}
  
\subsubsection{Validity of perturbation theory}  
The theory we have developed in this paper is based on perturbation
theory in the interaction between light and atoms.  In this subsection
we discuss the limits of validity of this perturbative treatment. We
will be considering worst case scenarios to find the limit, where our
perturbation series Eq.  (\ref{eq:stokesgenerator-feynman}) and
(\ref{eq:diagram-spin-coherent-second-order}) converge. An important
parameter for these estimates will be the effective coupling constant
for the collective operators $\kappa$ defined in Eq.
(\ref{eq:reduce-intre-equa}). For applications to light-matter quantum
interfaces this parameter should be of order unity. As we shall see
below, this is still possible without violating the applicability of
perturbation theory.  Another important parameter is the optical
depth, $OD$, defined by $OD \sim \rho \lambda^2
L$. 
The optical depth plays an important factor when describing the effect
of the incoherent interaction, e.g., the spontaneous emission.

Throughout this work, we have assumed that the atomic ensemble is
polarized along the $x$-axis, so that the atomic spin components $\rho
\bar J_y$, $\rho \bar J_z$ only carries quantum noise. Also we have
assumed that the classical component of the light is linearly
polarized so that, e.g., circular components are governed by quantum
noise. These assumptions will be important for estimating the terms
below.

We first consider the expansion of the light field
(\ref{eq:stokesgenerator-feynman}), and in particular the coherent
part of the interaction. The effective perturbation coefficient for
the first order term is found to scale at most as $(\beta \klaser
\sqrt{ N_A})/A \sim \kappa / \sqrt{N_P}$ (may be found by estimating
Eq.  \eqref{eq:firstorder_stokes_operator_general}).  Here $A$ is the
transverse area of the atomic ensemble, and $N_P$ is the total number
of photons in a pulse. Going to second order an
important term is described in Eq. (\ref{eq:stokes_ters_calc_3}).
Since we are not including the the time evolution of the macroscopic
polarization in the average interaction, this term has a potential
scaling as large as $\kappa^2$.  We showed, however, that in the
paraxial approximation the term vanish. Going beyond the paraxial
approximation as done in Appendix \ref{appendix-correction}, we find
that for linearly polarized light the scaling is $\kappa^2 /
\sqrt{N_P}$. The last contribution to Eq.
(\ref{eq:stokesgenerator-feynman}) is the incoherent interaction
considered in Appendix \ref{sec:calc-spont-emiss}. The scaling of this
effect $\kappa^2\cdot(N_A/N_P)/OD$ .

Now we consider the spin series
(\ref{eq:diagram-spin-coherent-second-order}) for a single atom. The
incoherent part of the evolution of the spin is described in Eq.
(\ref{eq:fracb-c_12-rhom}), and scales as $\kappa^2 / OD$, it can be
ignored for sufficiently large $OD$.  The first
order term scale as $\kappa /\sqrt{N_A}$ for linearly polarized light.
To increase this coefficient we need circularly polarized light, which
makes it interesting to examine the second order term describing the
change of the polarization of the due to the interactions with atoms.
This process is described in Eq.
\eqref{eq:secondordertermspin_vanish}, which represent the optically
induced dipole-dipole interaction. This particular term vanish when we
take quantum mechanical averages, because we have subtracted the only
non-vanishing component, but we can still calculate the root mean
square contribution. The effect can then be separated into a short
range part and a long-range contribution. The long range contribution
can be estimated to give a contribution of order
$\kappa^2\sqrt{d/(L\cdot OD)}$, where $d$ is the smallest dimension of
the setup, i.e., the smaller of the length and the transverse sizes of
the beam and the ensemble.  The short range part actually diverges
within our present approximations.  If, however, we regularize the
integral by excluding the volume, where the dipole-dipole interaction
of an excited and a ground state atom $V\sim \gamma \lambda^3/r^3$ is
of the same order as the detuning $\Delta$, we find a contribution
$\kappa^2\sqrt{\Delta/\gamma}\sqrt{\lambda/(L\cdot OD)}$.  The
justification for this regularization is that when we made the
adiabatic elimination we assumed a constant detuning $\Delta$. This
approximation breaks down when two atoms are sufficiently close that
the dipole-dipole interaction is the strongest effect in the problem,
in which case it is more appropriate to describe the atoms in terms of
molecular states. Both the short and long range part of the
interaction are thus small for sufficiently large optical depth $OD$
and for sufficiently long ensembles (large $L$). It should, however,
be noted that here we have only performed a very rough treatment of
the dipole-dipole interaction, and it would be desirable to make a
more accurate treatment of the effects of these terms. Also it should
be noted that the estimates we have performed here apply to non-moving
atoms, i.e., cold atoms. If we include the motion of the atoms, i.e.,
warm atoms as in Refs.  \cite{juulsgaard1,juulsgaard2,sherson}, there
will be a reduction of these terms because the sign of the interaction
will change in time.


In summary, sufficient requirements for the convergence of the series
for the light fields are
\begin{equation}
  \label{eq:requirements_convergence}
  \frac{\kappa}{\sqrt{N_P}} \ll 1, \quad
  \frac{\kappa^2}{\sqrt{N_P}}\ll 1, \quad  \frac{\kappa^2}{OD}\cdot \frac{N_A}{N_P}\ll 1 ,
\end{equation}
and for the spin equation sufficient requirements are
\begin{equation}
  \begin{split}  \frac{\kappa}{\sqrt{N_A}} \ll 1, \quad &  \frac{\kappa^2}{OD}
    \ll 1 , \quad
    \kappa^2 \sqrt{\frac{d}{L\cdot OD}}\ll 1, \\
    &  \kappa^2 \sqrt{\frac{\Delta}{\gamma}}\sqrt{\frac{\lambda}{L\cdot OD}}\ll 1.
\end{split}
\end{equation}
By having many atoms and photons as well as a large optical depth, it
is thus possible to achieve $\kappa\sim 1$ without violating the
applicability of perturbation theory.

The main idea in this work is to develop a perturbation series, where
we explicitly take into account the reshaping of the light modes
caused by the mean effect of the interaction.  Let us for comparison
compare with the series, if the mean effect of the interaction had not
been subtracted. For the Stokes operators the perturbative series is
given in Eq. \eqref{eq:stokesgenerator-feynman}. If we do not subtract
the average effect of the interaction, the scalar part of the
interaction [the $c_0$ component in Eq.
\eqref{eq:barb-vmathbf-j_j=b-1}] will give first order corrections to
the field of order $\kappa \sqrt{N_A/N_P}$ times the incoming field.
With $N_A\sim N_P$ as it is suggested in Ref. \cite{duan}, this term
will give a factor of order unity for $\kappa\sim 1$, and this
therefore cannot be considered a small term.  For the calculation of
the Stokes operators, however, the two large components in the first
order terms in Eq. \eqref{eq:stokesgenerator-feynman} cancel out.  The
calculation may thus yield reasonable result even without performing
the more involved procedures described in this article, but the
validity of the procedure would be questionable. (Some experiments
actually uses $N_p\gg N_A$ \cite{juulsgaard1}, where this problem may
be of minor concern).  Furthermore, one of the major limiting factors
identified above, is the dipole-dipole interactions.
The effect of this term is much more complicated to evaluate if we had
not subtracted the average interaction, but the term certainly will be
larger, because the interactions in Eq.
\eqref{eq:secondordertermspin_vanish} would include a non-vanishing
term, and not just the quantum fluctuations. Again this term would
thus seriously question the applicability of perturbation theory. In
contrast the present approach allows us to rigorously apply
perturbation theory in experimentally relevant regimes.

\section{Conclusion}  \label{sec:conclusion}
In quantum optics the propagation of light through an atomic medium is
often described in a one-dimensional approximation, where one
completely ignores the transverse structure of the beam and only
considers the longitudinal propagation. In this paper we have
investigated the validity of this approximation by developing a full
three-dimensional theory describing the interaction. The challenge in
this work has been to develop a theory capable of describing the
microscopic interaction with a single atoms as well as macroscopic
effects such as the diffraction of the laser beam caused by the
refractive index of the gas. In essence the theory we have developed
here includes both the micro- and macroscopic effect by separating the
interaction into an average part and the fluctuation from the average.
In this formulation macroscopic effects such as diffraction are
naturally associated with the average part whereas the microscopic
fluctuations describe processes such as the mapping of quantum
fluctuations between light and atoms. Furthermore we have shown that
spontaneous emission from the atoms naturally appear as an effect
caused by the fluctuations associated with the point particle nature
and the random positions of the atoms.

Based on our separation into the average and the fluctuations we have
developed a perturbative expansion in the fluctuations. The advantage
of this procedure is that it has a wider region of applicability than
a direct perturbative treatment. For instance in an experimental setup
an index of refraction of the gas just change of the beam profile
which often only has a minor effect on the experiment. On the other
hand, such 'trivial' effects may have a large influence on the
theoretical calculation. If one considers perturbation theory based on the
vacuum solutions to the wave equation, the perturbative expansion will
include all the terms responsible for the reshaping of the beam, and
this may break the validity of perturbation theory. On the other hand
our theory performs perturbation theory on modes which are solutions
to the wave equation including the index of refraction of the gas. Our
theory is thus applicable even for situations where the beam is
considerably distorted by the refractive index of the gas.

A major motivation for this work has been to investigate the validity
of the one-dimensional approximation in the description of the
experiments in Refs. \cite{juulsgaard1,juulsgaard2,sherson,mabuchi}. In Sec.
\ref{sec:exper-appl-valid} we explicitly considered some situations
where we could reduce our general theory to a theory resembling the
one used to describe these experiments in the one dimensional
approximation \cite{duan,brian_thesis}. To achieve a simple description
resembling the previous theories, an essential requirement is that we
are in the paraxial approximation. If we are not in this limit, the
polarization of the light change as its propagate through the
ensemble, which complicates the interaction with the atoms.
Furthermore, for the particular interaction considered here, we also
find it to be desirable to be in a regime where the Fresnel number is
much larger than unity $\mathcal{F}\gg1$. In these limits our theory
essentially reproduce the results of the simple theory. The only
difference is that instead of the vacuum mode functions, the mode functions
appearing in the theory should represent the modes, which are solutions
to the diffraction problem including the index of refraction of the
gas.

In the present paper we have mainly focused on developing the theory
and deriving how the usual approximations arise from our more
complicated approach. The theory is, however, fully consistent and
thus capable of including any higher order corrections not previously
included in the theoretical description. In particular it could be
interesting to study the effect of light induced dipole-dipole
interactions. While such processes may not be relevant for
understanding the current experiments, they may play an important role
in future experiments, e.g., with Bose-Einstein condensates, where the
density may be fairly high. Another interesting extension of our
theory could be to study different types of interactions such as for
instance electromagnetically induced transparency \cite{lukin-eit}.

\acknowledgments AS is grateful to the hospitality of the university
of Innsbruck, where early stages of this works were initiated in the
spring 2000. We are grateful to J. Cirac, L.-M. Duan, J.-H.
M{\"u}ller, K. M\o lmer, E. Polzik, J. Sherson, and P. Zoller for useful
discussions. This work was supported by the Danish Natural Science
Research Council.

\newpage
\appendix
\section{Adiabatic elimination}
\label{sec:adiab-elim}
In this appendix we derive an effective Hamiltonian involving only the
atomic ground state.  The Hamiltonian (\ref{eq:dipole_hamilton}) can
be expanded on the complete set of states describing the atom. Let
such a set be comprised of a set of exited states $\{ |e_j \> \}$ and
a set of ground states $\{ |g_i \> \}$ so that the Hamiltonian reads
\begin{align}
  \mathcal H= \sum_j ( \omega_j +\omega _0) |e_j\>\< e_j| + \sum_{i} \omega_0 |g_i\>\< g_i|  +  \mathcal H_{\text{int}}.
\end{align}
For convenience we have here set $\hbar =1$ and only consider a single
atom. The set of ground states are assumed to have the same energy,
$\omega _0$ and $\omega_j$ is the transition frequency from the ground
state to the exited state $|e_j\>$. The interaction Hamiltonian is
given in Eq. (\ref{eq:hatm-h_text-=}), and when expanded on the
set of internal atomic states it reads
\begin{align}\label{eq:mathc-h_text-frac1}
  \mathcal H_{\text{int}}=-\frac{1}{\epsilon_0} \sum_{ij} &\hat{\mathbf
    D}^{(-)}(t) \cdot \< g_i|\hat{\mathbf P}|e_j\>
  |g_i\>\<e_j| \notag \\ &+ \< e_j|\hat{\mathbf P}|g_i\>
  |e_j\>\<g_i| \cdot \hat{\mathbf D}^{(+)}(t) ,
\end{align}
where we have used the rotating wave approximation as well as the fact
that the matrix elements $\< e_j|\hat{\mathbf P}|e_{j'}\>$
and $\< g_i|\hat{\mathbf P}|g_{i'}\>$ vanish. To shorten the
notation we suppress the spatial dependence. We will use
that the displaced electric field  primarily oscillate at the laser
frequency, and change to the interaction picture
\begin{align}
  \hat{\mathbf D}^{(-)}(t) \propto e^{i\laser t}. 
\end{align}
Using Heisenberg's equations of motion we may derive an equation of
motion for $|g_i\>\<e_j|$
\begin{align}\label{eq:fracddt-g_ie_j-=}
  \frac{d}{dt} |g_i\>\<e_j| = -i \Delta_j  &|g_i\>\<e_j| -
  \frac{i}{\epsilon_0} \sum_{j'} \Big\{  \< e_j|\hat{\mathbf P}|g_{i}\>
  |e_{j'}\>\<e_j| \notag \\ &- \< e_j|\hat{\mathbf P}|g_{i'}\>
  |g_{i}\>\<g_{i'}| \Big\} \cdot \tilde{\mathbf D}^{(+)}(t)    ,
\end{align}
where $\tilde{\mathbf D}^{(\pm)}$ is slowly varying. In the limit of weak
driving we may set $\frac{d}{dt} |g_i\>\<e_j| =0$, and obtain an approximate solution
\begin{align}\label{eq:g_ie_j-appr-frac1}
  |g_i\>\<e_j| \approx \frac{1}{\epsilon_0 \Delta_j} \sum_{i'} \<
  e_j|\hat{\mathbf P}|g_{i'}\> |g_{i}\>\<g_{i'}| \cdot \hat{\mathbf
    D}^{(+)}(t),
\end{align}
where we have neglected the exited state population. The atomic part of the
Hamiltonian can be written
\begin{align}\label{eq:mathcal-h_0=-sum_j}
  \mathcal H_0= \sum_j \Delta_j |e_j\>\<&g_0|g_0\>\< e_j| +\sum_i
  \laser |g_i\>\< g_i| \notag \\ & + \sum_{ij} (\omega_0-\laser) (|e_j\>\<
  e_j| + |g_i\>\< g_i| ),
\end{align}
where $|g_0\>$ is any ground state. By inserting expression
(\ref{eq:g_ie_j-appr-frac1}) and the Hermitian conjugate into
Eq. (\ref{eq:mathc-h_text-frac1}) and
(\ref{eq:mathcal-h_0=-sum_j}) we find the simple result
\begin{align}
  \mathcal H = -\frac{1}{\epsilon_0} \Big( \hat{\mathbf D}^{(-)}(t)& \cdot
  \sum_{jii'} \frac{1}{\epsilon_0 \Delta_j}  \<
  g_i|\hat{\mathbf P}|e_{j}\>\Big) \notag \\ &|g_{i}\>\<g_{i'}| \Big( \<
  e_j|\hat{\mathbf P}|g_{i'}\> \cdot \hat{\mathbf D}^{(+)}(t)\Big) .
\end{align}
(neglecting a zero-point energy term in the Hamiltonian). We may now
identify the matrix operator $\bar{\bar V}[\hat J]$
\begin{align}\label{eq:barbar-vhat-j}
  \bar{\bar V}[\hat J] = \sum_{jii'} \frac{1}{\epsilon_0 \Delta_j}  \<
  g_i|\hat{\mathbf P}|e_{j}\>  \<
  e_j|\hat{\mathbf P}|g_{i'}\> |g_{i}\>\<g_{i'}| \: \cdot , 
\end{align}
and we immediately get the result stated in equation
(\ref{eq:hatm-h_text-=}). The notation ``$\cdot$''  in this expression means usual vector product with
the vector to the right. Furthermore we may also find the relation
between the polarization and the displaced electric field
\begin{align}\label{eq:hatmathbf-p+t-=}
  \hat{\mathbf P}^{(-)}(t) =& \sum_{ij}  |e_{j}\> \<
  e_j|\hat{\mathbf P}|g_{i}\> \<g_{i}| \notag \\ 
  =&\: \sum_{jii'} \frac{1}{\epsilon_0 \Delta_j} \<
  e_j|\hat{\mathbf P}|g_{i}\> \<
  g_{i'}|\hat{\mathbf P}|e_{j}\> |g_{i'}\>\<g_{i}| \cdot \hat{\mathbf
    D}^{(-)}(t)  \notag \\
  =& \: \bar{\bar V}^t[\hat{\mathbf J}] \hat{\mathbf D}^{(-)}(t).
\end{align}
We have here only written the positively oscillating component, the
negatively oscillating component is found by Hermitian conjugation,
which from equation (\ref{eq:barbar-vhat-j}) is the same as
transposition of the matrix.

 \section{Calculation of infinitely short propagator}
 \label{sec:calc-infin-short}
 In this appendix we calculate the infinitely short propagator in the
 local density approximation. We will
 for simplicity only consider the simple interaction given by
 \begin{align}
   \bar{\bar{\mathcal V}}[\mathbf J] = \beta \rho(\mathbf r)
   \Big( c_0 {\mathbf J}(\mathbf r)^2 - ic_1 {\mathbf
     J}(\mathbf r) \times  \Big). 
 \end{align}
 We further shorten the notation by introducing the coefficients
 $a_0=1-\beta \rho(\mathbf r) c_0 {\mathbf J}(\mathbf r)^2 $ and
 $a_1=\beta \rho(\mathbf r) c_1 |{\mathbf J}(\mathbf r)|$. 

 If we Fourier-transform equation (\ref{eq:boldsymb-barb-vhatm}), the
 equation we wish to solve is 
 \begin{subequations}
   \begin{align}
     \hat{\mathbf k} \times \hat{\mathbf k} \times ( a_0 + i a_1
     \hat{\mathbf j} \times ) \boldsymbol{\varepsilon}^{\mathbf k} =& -
     \frac{\omega_{\mathbf k}^2}{c^2 k^2} \label{eq:hatmathbf-k-times}\\
     \hat{\mathbf k} \cdot \boldsymbol{\varepsilon}^{\mathbf k} =& 0,
   \end{align}
 \end{subequations}
 where the vectors $\hat{\mathbf k}$ and $\hat{\mathbf j}$ are unit
 vectors representing respectively the direction of the plane wave
 solution and the orientation of the atomic spin.  The solutions to
 the above equations is the following set of polarization-vectors
 \begin{align}
   \boldsymbol{\varepsilon}^{\mathbf k}_{\pm} = N^{\mathbf k}_{\pm}\Big(
   \frac{\hat{\mathbf j}\times \hat{\mathbf k}}{|\hat{\mathbf j}\times
     \hat{\mathbf k}|} \pm i \frac{\hat{\mathbf k} \times
     (\hat{\mathbf j}\times \hat{\mathbf k})}{|\hat{\mathbf k} \times
     ( \hat{\mathbf j}\times
     \hat{\mathbf k})|} \Big) \equiv N^{\mathbf k}_{\pm} \big(
   \hat{\mathbf v}_1 \pm i \hat{\mathbf v}_2 \big) ,
 \end{align}
 where $\hat{\mathbf v}_1$ and $\hat{\mathbf v}_2$ are unit vectors
 given by the first and second fraction respectively.  The
 normalization constant $N^{\mathbf k}_{\pm}$ is determined by using
 the inner product in Eq. (\ref{eq:-boldsymb-r}). In this way we find the
 real space representation of the basis-functions $\mathbf f_{\mathbf
   k}(\mathbf r)$
 \begin{align}
   \mathbf f^{\mathbf k}_{\pm}(\mathbf r) =
   \frac{1}{\sqrt{2(2\pi)^3(a_0 \pm a_1(\hat{\mathbf j} \cdot \hat{\mathbf
         k}))}} \big( \hat{\mathbf v}_1 \pm i \hat{\mathbf v}_2 \big)
   e^{i \mathbf k \cdot \mathbf r} .
 \end{align} 
 The dispersion relation is then derived from (\ref{eq:hatmathbf-k-times})
  \begin{align}\label{eq:omega_mathbf-kpm2-=}
   \omega_{\mathbf k^{\pm}}^2 = c^2 k^2 ( a_0 \pm a_1(\hat{\mathbf
     j} \cdot \hat{\mathbf k} )).
 \end{align}

 The infinitely short propagator can then be calculated to be the
 following
 \begin{align}
   \bar{\bar P}^{(-)}(\mathbf r,t-t') = \frac{-i}{2\laser c^2} \sum_{s\in
     \{ +,- \} } \int d^3 k \; \omega_{\mathbf k^s}^2 &(\mathbf
   f^{\mathbf k}_{s}(\mathbf r))^* \mathbf f^{\mathbf k}_{s}(\mathbf
   r)\cdot  \notag \\ 
   &e^{\frac{i(t-t')}{2\laser}(\omega_{\mathbf ks}^2 - \laser^2)}.
 \end{align}
 We introduce the matrix given by the following juxtaposition:
 \begin{align}
   \bar{\bar M}(\hat{\mathbf k},\hat{\mathbf j},s) = \big( \hat{\mathbf v}_1- is
   \hat{\mathbf v}_2 \big) \big( \hat{\mathbf v}_1 + i s
   \hat{\mathbf v}_2 \big) .
 \end{align}
 Changing to spherical coordinates and making the substitutions $x=\cos
 \theta$ and $k'= k\sqrt{1-a_0 + sa_1 x}$ as well as using the
 dispersion relations given in equation (\ref{eq:omega_mathbf-kpm2-=})
 the integral reduce to
 \begin{widetext}
   \begin{align}
     \bar{\bar P}^{(-)}(\mathbf r,t-t') = \frac{-i}{2\laser c^2}\sum_{s\in
       \{ +,- \} } \int_0^{\infty} dk' \int_{-1}^1 dx \int_0^{2\pi}
     d\phi \; \frac{c^2 k'^4}{2(2\pi)^3(a_0+sa_1x)^{5/2}} \bar{\bar
       M}(x,\phi,s) e^{ic(t-t')(k'^2-\klaser^2)/(2\klaser)}.
   \end{align}
 \end{widetext}
 Neglecting the denpendence of $k'$ outside the exponential and using
 that the difference ${k'}^2 -\klaser^2$ for large $\klaser$ runs from $-\infty$ to
 $\infty$, the $k'$ integral gives a delta-function in time. Including
 the $\phi$ integration in a matrix $\bar{\bar M}$ we finally get
 \begin{align}
   \bar{\bar P}^{(-)}(\mathbf r,t-t')=\frac{-i \klaser^3\delta (t-t') }{16
     \pi^2 c^2}\hspace{-.2cm} \sum_{s\in \{ +,- \} } \int_{-1}^1 dx
   \frac{\bar{\bar M}(x,s)}{(a_0+sa_1x)^{\frac{5}{2}}},
 \end{align}
with  the matrix $\bar{\bar M}$ given by
 \begin{align}
   \bar{\bar M}(x,s) = \pi \left[
     \begin{array}{ccc}
       2(1-x^2) & 0 & 0 \\
       0& 1+x^2 & 2isx \\ 
       0& -2isx & 1+x^2
     \end{array} \right] . 
 \end{align}
 The $s$-sum is evaluated by substitution in the integral and the final
 expression for the infinitely short propagator is
 \begin{align}
   \bar{\bar P}^{(-)}(\mathbf r,t-t')=\frac{-i \klaser^3\delta(t-t')}{8 \pi c^2}
   \int_{-1}^1 dx \frac{\bar{\bar M}(x,+)}{\pi(a_0+a_1x)^{5/2}}.
 \end{align}
 These integral  may be evaluated, and we will express the infinitely
 short propagator as
 \begin{align}
   \bar{\bar P}^{(-)}(\mathbf r,t-t')= \frac{-i \delta (t-t')}{c^2} \left[ 
     \begin{array}{ccc}
       \varrho_{||} & 0 & 0 \\
       0& \varrho_{\perp} & - i \varrho_{\Gamma} \\ 
       0& i \varrho_{\Gamma} & \varrho_{\perp}
     \end{array} \right] .
 \end{align}
 The coefficients are for $a_0 - a_1 > 0$, given by
 \begin{subequations}\label{eq:beginalign-varrho_-=}
   \begin{align}
     \varrho_{||} =& \frac{-\klaser ^3}{3\pi a_1^3} \Big\{
     \frac{-4a_0 + 2a_1}{\sqrt{a_0 - a_1}} + \frac{4a_0 +
       2a_1}{\sqrt{a_0+a_1}}
     \Big\} \\
     \varrho_{\perp} =& \frac{-\klaser ^3}{3\pi a_1^3} \Big\{
     \frac{2a_0^2 -3a_0a_1 + \frac{1}{2} a_1^2}{ (a_0 - a_1)^{3/2}} -
     \frac{2a_0^2 +3a_0a_1 + \frac{1}{2} a_1^2 }{ (a_0+a_1)^{3/2}}
     \Big\} \\
     \varrho_{\Gamma} =& \frac{\klaser ^3}{6 \pi a_1^2} \Big\{
     \frac{2a_0 -3a_1 }{ (a_0 - a_1)^{3/2}} -
     \frac{2a_0 +3a_1  }{ (a_0+a_1)^{3/2}}
     \Big\}.  
   \end{align}
 \end{subequations}

\section{Reciprocal equation for Green's function}
  \label{sec:recipr-equat-greens}
  In this appendix we derive the reciprocal equation for the Green's
  function. Before doing so we will need some
  results concerning the representation of the Green's function. Let
  us define the following inner product:
  \begin{align}
    \< \boldsymbol{\phi}|\boldsymbol{\psi}\> = \int d^3rdt\;\bar{\bar{\mathcal
      M}}(\mathbf r)\boldsymbol{\phi}(\mathbf r,t) \cdot
    \boldsymbol{\psi}^{\dagger}(\mathbf r,t).
  \end{align}
  We will generally work in the $L^2$-space equipped with this inner
  product. Using that the matrix operator $\bar{\bar{\mathcal M}}$ is
  Hermitian, one finds the differential operator $\mathcal D$ given in
  equation (\ref{eq:mathcal-d-=}) to be
  Hermitian in our inner product space
  \begin{align}
    \< \boldsymbol{\phi}|\mathcal D\boldsymbol{\psi}\> =& \< \mathcal
    D \boldsymbol{\phi}|\boldsymbol{\psi}\>.
  \end{align}
  That $\mathcal D$ is Hermitian means that the eigenfunctions $F_k$ to
  $\mathcal D$
  \begin{align}\label{eq:mathcal-d-}
    \mathcal D \; \mathbf F_{\mathbf k}(\mathbf r,t) =
    \lambda_{\mathbf k} \mathbf F_{\mathbf k}(\mathbf r,t),
  \end{align}
  define a complete basis of our inner product space $\big\{ \mathbf
  F_{\mathbf k} \big\}$.  A representation of the identity functional
  given in equation (\ref{eq:mathc-dbarb-gmathbf}) may therefore be
  \begin{align}\label{eq:sum_mathbf-k-mathbf}
    \sum_{\mathbf k} \mathbf F^{\dagger}_{\mathbf k}(\mathbf
    r,t)\mathbf F_{\mathbf k}(\mathbf r_0,t_0).
  \end{align}
  It can be checked that this is exactly a functional identity
  representation in our inner product space by expanding any function
  on the basis $\big\{ \mathbf F_{\mathbf k} \big\}$. 

 To get a formal
  expression of the Green's function defined in equation
  (\ref{eq:mathc-dbarb-gmathbf}) we expand the Green's function in
  this basis, and using equation (\ref{eq:mathcal-d-}) and
  (\ref{eq:sum_mathbf-k-mathbf}) we find
  \begin{align} \label{eq:barbar-gmathbf-r} \bar{\bar G}(\mathbf
    r,t|\mathbf r_0,t_0) = \sum_{\mathbf k} \frac{1}{\lambda_{\mathbf
        k}} \mathbf F^{\dagger}_{\mathbf k}(\mathbf r,t)\mathbf
    F_{\mathbf k}(\mathbf r_0,t_0) .
  \end{align}
  Starting from equation (\ref{eq:mathc-dbarb-gmathbf}) we make the
  substitution $t\rightarrow -t$, $t_0 \rightarrow -t_1$ and $\mathbf
  r_0 \rightarrow \mathbf r_1$ 
  and we write:
  \begin{align}\label{eq:mathcal-d-barbar}
    \mathcal D^* \bar{\bar G}(\mathbf r,-t|\mathfb r_1,-t_1) =
    \bar{\bar I}\delta(\mathbf r,\mathbf r_1)\delta(t,t_1).
  \end{align}
  In the next step we take inner product with equation
  (\ref{eq:mathc-dbarb-gmathbf}) and $\bar{\bar G}(\mathbf
  r,-t|\mathfb r_1,-t_1)$ from the left with respect to unprimed
  coordinates, and equation (\ref{eq:mathcal-d-barbar}) and $\bar{\bar
    G}(\mathbf r,t|\mathfb r_0,t_0) $ also from the left with respect
  to unprimed coordinates. The resulting two equations are then subtracted.
  The term containing $\laser^2$ vanish trivially, and using rules for
  differentiating a product, the resulting equation may be written as
    \begin{align}\label{eq:2ilaser-iint-d3r}
    &2i\laser \iint d^3r dt \; \bar{\bar{\mathcal M}}(\mathbf r)
    \frac{\partial}{\partial t} \Big[ \bar{\bar G}(\mathbf
    r,-t|\mathfb r_1,-t_1) \cdot \bar{\bar G}(\mathbf r,t|\mathfb
    r_0,t_0) \Big] \notag \\ &+ c^2 \iint d^3 dt \Big[ \bar{\bar{\mathcal
      M}}(\mathbf r) \bar{\bar G}(\mathbf r,-t|\mathfb r_1,-t_1) \cdot
    \boldsymbol{\nabla}\times \boldsymbol{\nabla}\times \bar{\bar{\mathcal
      M}}(\mathbf r) \notag \\ & \bar{\bar G}(\mathbf r,t|\mathfb
    r_0,t_0) - \bar{\bar{\mathcal M}}(\mathbf r) \bar{\bar G}(\mathbf
    r,t|\mathfb r_0,t_0) \cdot \boldsymbol{\nabla}\times
    \boldsymbol{\nabla}\times \bar{\bar{\mathcal M}}(\mathbf r) \notag \\
    &\bar{\bar G}(\mathbf r,-t|\mathfb r_1,-t_1) \Big] = \bar{\bar
      G}(\mathbf r_1,t_1|\mathfb r_0,t_0) - \bar{\bar G}(\mathbf
    r_0,-t_0|\mathfb r_1,-t_1).
  \end{align}
  Using the cut-off property of the Green's function, the first term
  on the left hand side is
  seen to vanish. Using the explicit expression for the Green's
  function (\ref{eq:barbar-gmathbf-r}) along with Gauss' theorem, one
  may show that the second term also vanish.  The final result is
  therefore
  \begin{align}\label{eq:barbar-gmathbf-r_1}
    \bar{\bar G}(\mathbf r_1,t_1|\mathfb r_0,t_0) = \bar{\bar
      G}(\mathbf r_0,-t_0|\mathfb r_1,-t_1).
  \end{align}
  From Eq. (\ref{eq:mathc-dbarb-gmathbf}),
  (\ref{eq:barbar-gmathbf-r_1}) and using the substitutions
  $t\rightarrow -t'$, $t_0\rightarrow t$, $\mathbf r \rightarrow
  \mathbf r'$ and $\mathbf r_0 \rightarrow \mathbf r$ we end up with
  the \emph{reciprocal equation}
  \begin{align}\label{eq:big-2ilas-t-1}
    \Big( -2i\laser\frac{\partial}{\partial t'} - \laser^2 + c^2
    \boldsymbol{\nabla}' \times \boldsymbol{\nabla}' \times&
    \bar{\bar{\mathcal M}}(\mathbf r') \Big) \bar{\bar G}(\mathbf
    r,t|\mathbf r',t')\notag \\ =&\: \bar{\bar I} \delta(\mathbf
    r,\mathbf r')\delta(t,t').
  \end{align}

   In the following we derive the general solution to the equation 
     \begin{align}\label{eq:big-2ilas-t}
    \Big( 2i\laser\frac{\partial}{\partial t} -\laser^2 + c^2
    \boldsymbol{\nabla}\times \boldsymbol{\nabla}\times \bar{\bar{\mathcal
      M}}(\mathbf r) \Big) \boldsymbol{\psi}(\mathbf r,t)=
    \boldsymbol{\rho}(\mathbf r,t),
  \end{align}
  where $\boldsymbol{\psi}(\mathbf r,t)$ is an unknown field,
  $\boldsymbol{\rho}(\mathbf r,t)$ is a source term effecting the
  solution, and $\bar{\bar{\mathcal M}}$ is some Hermitian matrix
  operator, which may depend on position.
   We  make an inner product of equation
  (\ref{eq:big-2ilas-t}) with $\bar{\bar G}(\mathbf r,t|\mathbf
  r',t')$ from the left and an inner product of equation
  (\ref{eq:big-2ilas-t-1}) with $\boldsymbol{\psi}(\mathbf r,t)$ from
  the right and subtract these two equations.
  In this calculation we are integrating over the time interval $t'\in \big] t_0 ,t^+
  \big[$,  where we understand $t^+ =\lim_{\epsilon \rightarrow 0}
  [t+\epsilon]$. Again we find that terms containing $\laser^2$
  vanish. Similar to above we will use rules for
  differentiation a product, and we eventually end up with 
  \begin{align}\label{eq:boldsymb-r-t}
    \boldsymbol{\psi}&(\mathbf r,t) - \iint_{t_0}^{t^+} d^3r'dt'
    \bar{\bar{\mathcal M}}(\mathbf r')\bar{\bar G}(\mathbf r,t|\mathbf
    r',t') \cdot \boldsymbol{\rho}(\mathbf r',t') = \notag \\ &
    -2i\laser \iint_{t_0}^{t^+} d^3r'dt' \bar{\bar{\mathcal
        M}}(\mathbf r') \frac{\partial}{\partial t} \Big[ \bar{\bar
      G}(\mathbf r,t|\mathbf r',t') \cdot \boldsymbol{\psi}(\mathbf
    r',t') \Big] \notag \\ & +c^2\iint_{t_0}^{t^+} d^3r'dt'
    \bar{\bar{\mathcal M}}(\mathbf r') \Big\{ \notag \\ &
    \hspace{1.8cm} \boldsymbol{\psi}(\mathbf r',t') \cdot
    \boldsymbol{\nabla}' \times \boldsymbol{\nabla}' \times
    \bar{\bar{\mathcal M}}(\mathbf r') \bar{\bar G}(\mathbf
    r,t|\mathbf r',t') \notag \\ & \hspace{2.5cm} - \bar{\bar
      G}(\mathbf r,t|\mathbf r',t') \cdot \boldsymbol{\nabla}' \times
    \boldsymbol{\nabla}' \times \boldsymbol{\psi}(\mathbf r',t')
    \Big\}.
  \end{align}

  Using the same boundary conditions as was done in the calculation
  leading to the reciprocal equation we conclude that the last term in
  equation (\ref{eq:boldsymb-r-t}) vanish. The right hand side of the
  equation thus reduce to
  \begin{align}
    -2i\laser \int d^3r' \bar{\bar{\mathcal M}}(\mathbf r') \Big[
    \bar{\bar G}(\mathbf r,t|\mathbf r',t') \cdot
    \boldsymbol{\psi}(\mathbf r',t') \Big]_{t_0}^{t^+} = \notag \\
    2i\laser \int d^3r' \bar{\bar{\mathcal M}}(\mathbf r')
    \bar{\bar G}(\mathbf r,t|\mathbf r',t_0) \cdot
    \boldsymbol{\psi}(\mathbf r',t_0).
  \end{align}
Here we have used that the upper time limit
  vanish due to the cut-off in the Green's function.  Rearranging
  terms we finally arrive at the general solution to the diffusion
  equation
  \begin{align}
    \boldsymbol{\psi}(\mathbf r,t) =& 2i\laser \int d^3 r'
    \;\bar{\bar{\mathcal M}}(\mathbf r') \bar{\bar G}(\mathbf
    r,t|\mathbf r',t_0) \cdot \boldsymbol{\psi}(\mathbf r',t_0) \notag
    \\ + \iint_{t_0}^t& d^3r'dt'\;\bar{\bar{\mathcal M}}(\mathbf r')
    \bar{\bar G}(\mathbf r,t|\mathbf r',t') \cdot
    \boldsymbol{\rho}(\mathbf r',t').
  \end{align}

 \section{Lorentz-Lorenz relation} 
\label{sec:lorenz-lorenz-corr}
In the main text we mainly consider lowest order corrections to the
index of refraction. To verify that our theory can also correctly
reproduce higher order corrections, we shall in this appendix show how
to derive the so called Lorentz-Lorenz or Clausius-Mossotti relation
for the electric permittivity within our theoretical framework
\cite{jackson}.  To lowest order the permittivity is given by Eq.
(\ref{eq:-=c2int-d3r})
\begin{align}\label{eq:barb-r}
  \bar{\bar{\epsilon}}(\mathbf r)^{-1}=1-\bar{\bar{\mathcal V}}^t[\mathbf J].
  \end{align}
  To calculate the higher order correction it is convenient to first
  Fourier transform the Dyson equation (\ref{eq:tild-d+mathbf-r-1})
  describing the light field with respect to time
\begin{align}
  \label{eq:dyson_light_fouriertrans} 
    \tilde{\mathbf D}^{(-)}&(\mathbf r,\omega) =\tilde{\mathbf
      D}_0^{(-)}(\mathbf r,\omega) \notag \\ &+ c^2\int d^3r'\;
    \bar{\bar P}^{(-)}(\mathbf r,\mathbf r',\omega) \cdot \bar{\bar
      m}[\hat{ \mathbf J}]^t \tilde{\mathbf D}^{(-)}(\mathbf r',\omega). 
\end{align}
This equation is the starting point for the analysis. ( The Fourier
transformation is here defined as 
\begin{align}
  \label{eq:fouriertransformation_definition}
  f(\omega)=\int_{0}^{\infty} dt e^{(i\omega - \eta)t} f(t),
\end{align}
where $\eta$ is an infinitely small convergence factor.) 

From
Eq. (\ref{eq:barbar-p+mathbf-r-2}) we find the Fourier transformed
propagator $\bar{\bar P}^{(-)}$ to read
\begin{align}
  \label{eq:fouriertrans_propagator}
  \bar{\bar P}^{(-)}(\mathbf r,\mathbf r',\omega ) = \frac{1}{c^2}
  \sum_{\mathbf k} \frac{\omega_{\mathbf k}^2 \mathbf f^*_{\mathbf
      k}(\mathbf r) \mathbf f_{\mathbf k}(\mathbf r')}{\omega_{\mathbf
      k}^2 - \laser^2 + 2\laser(\omega + i \eta)} .
\end{align}
The real space representation of this propagator is in general
difficult to calculate, however, for a scalar interaction the
calculation simplify considerably. For $\omega \approx 0$ which is
reasonable in our case, since we are dealing with slowly varying
operators, the propagator reads
  \begin{widetext}
    \begin{align}\label{eq:g_nmmathbf-r-=}
      \bar{\bar P}^{(-)}(\mathbf n) =&  \int \frac{d^3k}{c^2(2\pi)^3}
      \sum_{\boldsymbol{ \varepsilon} \perp \mathbf k} \boldsymbol{\varepsilon}
      \boldsymbol{\varepsilon} \frac{\mathbf k^2 e^{i\mathbf k \cdot \mathbf
          n}}{  \mathbf k^2 - \klaser^2 } \notag \\
      =& -\frac{\klaser^3}{c^24\pi} \frac{e^{i\klaser n}}{\klaser n} \Big[ \left( 1+
        \frac{3i}{\klaser n} - \frac{3}{(\klaser n)^2}\right)
      \frac{\mathbf n \mathbf n}{\mathbf n^2} -
      \left( 1+ \frac{i}{\klaser n} - \frac{1}{(\klaser
          n)^2}\right)\bar{\bar I}
      \Big] + \frac{2}{3}\bar{\bar I} \delta(\mathbf n),
    \end{align}
  \end{widetext}
  where $\mathbf n=\mathbf r- \mathbf r'$, $n=|\mathbf n|$, and
  $\bar{\bar I}$ is the identity matrix. 
  We notice that the propagator gives us the well known result
  for the radiated field of an oscillating dipole. In addition
  we have a term describing a self-interaction. This propagator  is also 
  discussed in Ref. \cite{olivier95:_refrac_bose}. 
  In the following we
  shall only be considering the self interaction part of the
  propagator. 
  
  When considering the density correlation function to second order
  $\< \rho(\mathbf r_1) \rho(\mathbf r_2) \>$ we have so far used the
  ideal gas approximation in Eq. (\ref{eq:densitydensity}), where
  there are no correlations between different atoms. In reality we can
  never have two atoms at the same position and this give a small
  correction to $\< \rho(\mathbf r_1) \rho(\mathbf r_2) \>$, which
  must vanish for $\mathbf r_1 = \mathbf r_2$ (apart from the delta
  function, which represent the single atom contribution).  This can
  formally be described by introducing so called irreducible
  correlation functions $h_2$ such that
  \begin{align}
    \label{eq:density_irre}
    \< \rho(\mathbf r_1) \rho(\mathbf r_2) \> = \< \rho(\mathbf r_1)\>
    \< \rho(\mathbf r_2) \> + h_2(\mathbf r_1, \mathbf r_2),
  \end{align} 
  where $h_2$ now takes care of the core-repulsion of the atoms (here
  we exclude the delta function). For $\mathbf r_1 = \mathbf r_2 $ we
  thus finds that $h_2(\mathbf r_1,\mathbf r_1)=-\< \rho(\mathbf
  r_1)\>^2$ .

  The above can be used along with the real space representation of
  the propagator to give the second order correction to the
  permittivity. We will not consider terms that vanish when we take
  quantum mechanical mean. The relevant part of the second order term
  thus gives in shorthand notation $- \int \bar{\bar P}^{(-)}
  (2/3)(\bar{\bar{\mathcal V}}^t[\mathbf J])^2 \tilde{\mathbf
    D}^{(-)}$. When we introduce this interaction to the differential
  equation (\ref{eq:-=c2int-d3r}) we find the permittivity to second
  order
  \begin{align}
    \label{eq:secondorderpermea}
    \bar{\bar{\epsilon}}(\mathbf r)^{-1}=1-\bar{\bar{\mathcal
        V}}^t[\mathbf J]+\frac{2}{3}(\bar{\bar{\mathcal
      V}}^t[\mathbf J])^2.
  \end{align}
  The calculation can be continued to infinite order
  \cite{lorenz-lorenz}, and the result reads
  \begin{align}
    \label{eq:infinite-order-permeabilty}
    \bar{\bar{\epsilon}}(\mathbf r)^{-1}=&1-\bar{\bar{\mathcal
        V}}^t[\mathbf J] - \bar{\bar{\mathcal V}}^t[\mathbf J]
    \sum_{n=1}^{\infty} \left( \frac{-2}{3}\bar{\bar{\mathcal
          V}}^t[\mathbf J]\right)^n \notag \\
    =& \frac{1-\frac{1}{3}\bar{\bar{\mathcal V}}^t[\mathbf J] }{ 1 +
      \frac{2}{3}\bar{\bar{\mathcal V}}^t[\mathbf J]}.
  \end{align}
  This is the Lorenz-Lorenz relation, and we thus see that the effect can
  be included in the theory by dressing the spatial mode functions
  according to the result above. 
  
  \section{Calculations of second-order Stokes generator}
  \label{sec:calul-second-order}
  In this appendix we present detailed calculations of the second-order
  terms of Eq.  (\ref{eq:stokesgenerator-feynman}).  We will denote
  the fourth term of the right hand side of Eq.
  \eqref{eq:stokesgenerator-feynman} as $\bar{\bar{\mathcal
      S}}^{(2)}_A$, and one finds
\begin{align}
  \label{eq:second_order_calc_stokes_terms_1}
  K& \ddleft \tilde{\mathbf f}_{kmj}^*(\mathbf r,t)|\bar{\bar{\mathcal
      S}}^{(2)}_A|\tilde{\mathbf f}_{km'j'}^*(\mathbf
  r,t) \ddright = \left(\frac{1}{2}\right)^2(\klaser \beta c_1)^2\notag \\
  & \iint d^3r d^3r' \sum_{\substack{ln\\l'n'}} \rho(\mathbf r)\rho (\mathbf
  r')
  \Theta^{mn}_{jl}(\mathbf r)^* \Theta ^{m'n'}_{j'l'}(\mathbf r') \hat
  a^{\dagger}_{knl} \hat a_{kn'l'}.
\end{align}

The seventh term of the right hand side of Eq.
\eqref{eq:stokesgenerator-feynman} plus its complex conjugate we will
denote as $\bar{\bar{\mathcal S}}^{(2)}_B$. To calculate this
term we extend the limits of the time integration
from minus to plus infinity. This we can do by introducing a factor of
one half, and approximating the imaginary term $i\int_{-\infty}^0dt \sin
(\omega t)$ to be zero. This corresponds to the usual treatment of
such terms in the Markov approximation to spontaneous emission when
one ignores the Lamb shift. We then find the following contribution to the
Stokes operators
\begin{align}
  \label{eq:second_order_calc_stokes-2}
  &K \ddleft \tilde{\mathbf f}_{kmj}^*(\mathbf r,t)|\bar{\bar{\mathcal
      S}}^{(2)}_B|\tilde{\mathbf f}_{km'j'}^*(\mathbf r,t) \ddright =
  \left(\frac{1}{2}\right)^3(\klaser \beta c_1)^2\notag \\ & \iint
  d^3r d^3r' \sum_{\substack{ln\\l'n'}}\rho(\mathbf r)\rho (\mathbf
  r')\Big\{ \Theta^{mn}_{jl}(\mathbf r)^* \Theta ^{nn'}_{ll'}(\mathbf
  r')^* \hat a^{\dagger}_{kn'l'} \hat a_{km'j'} \notag \\ &
  \hspace{3.7cm} +\Theta^{m'n}_{j'l}(\mathbf r) \Theta
  ^{nn'}_{ll'}(\mathbf r') \hat a^{\dagger}_{kmj} \hat a_{kn'l'}
  \Big\}.
\end{align}
One notice that the factors of $1/2$ in
Eq. (\ref{eq:second_order_calc_stokes_terms_1}) and
(\ref{eq:second_order_calc_stokes-2}) exactly add up to give one half
of the square of the first-order term, as is shown in
Eq. (\ref{eq:full-eq-stoke-par-multi})  

The sixth term on the right hand side of Eq.
(\ref{eq:stokesgenerator-feynman}), plus its Hermitian conjugate, we
will denote as $\bar{\bar{\mathcal S}}^{(2)}_C$, and we find 
\begin{align}
  \label{eq:stokes_ters_calc_3}
  K &\ddleft \tilde{\mathbf f}_{kmj}^*(\mathbf r,t)|\bar{\bar{\mathcal
      S}}^{(2)}_C|\tilde{\mathbf f}_{km'j'}^*(\mathbf r,t) \ddright =
  \left(\frac{1}{2}\right)^3(\klaser \beta c_1)^2\notag \\ & \int
   d^3r' \sum_{\substack{ln\\ql'n'\\n''l''}}\rho (\mathbf
  r') \Big\{ \mathcal
  C_{jl}^{l'l''}(\mathbf r') \Psi^{mn}_k(\mathbf r')^*
  \Psi^{n'n''}_q(\mathbf r') \notag \\ & \hspace{1cm}\hat a^{\dagger}_{qn'l'} \hat
  a_{qn''l''} \hat a^{\dagger}_{knl} \hat a_{km'j'} + \hat
  a^{\dagger}_{kmj} \hat a^{\dagger}_{qn'l'} \hat a_{qn''l''} \hat
  a_{knl} \notag \\ & \hspace{3.7cm} \mathcal
  C_{j'l}^{l'l''}(\mathbf r') \Psi^{m'n}_k(\mathbf r')
  \Psi^{n'n''}_q(\mathbf r')^* \Big\},
\end{align}
where we have introduced the coefficients
\begin{subequations}
  \begin{align}
    \label{eq:coefficients_stokes_terms}
    \mathcal C_{jl}^{l'l''}(\mathbf r)=& \mathbf e_j(\mathbf r)\cdot
    \big\{ \big( \bar{\mathbf J}(\mathbf r) \times \big[ \mathbf
    e_{l'}(\mathbf r) \times \mathbf e_{l''}(\mathbf r) \big] \big)
    \times \mathbf e_{l}(\mathbf r) \big\}. 
  \end{align}
\end{subequations}
This term cam be shown to vanish by expanding the spin-operator
$\bar{\mathbf J}$ on the basis defined by the polarization vectors
$\mathbf e_x(\mathbf r)$, $\mathbf e_y(\mathbf r)$ and $\mathbf
e_z(\mathbf r)$ and using that the indices $j,l,l'$ and
$l''$ only run over $x$ and $y$.  

Finally we will calculate the effect of the fifth term on the right
hand side of Eq.  (\ref{eq:stokesgenerator-feynman}), which we will
denote $\bar{\bar {\mathcal S}}^{(2)}_D$. In this calculation it is
important to remember that the term will scale as $\beta^2 \rho$, and
reads
\begin{align}
  \label{eq:stokes_term_second_4}
  K\ddleft & \tilde{\mathbf f}_{kmj}^*(\mathbf r,t)|\bar{\bar{\mathcal
      S}}^{(2)}_D|\tilde{\mathbf f}_{km'j'}^*(\mathbf
  r,t) \ddright = \left(\frac{1}{2}\right)^2(\klaser \beta)^2\notag \\
  & \int d^3r \sum_{\substack{ln\\l'n'}} \rho(\mathbf r)
  \Psi^{nm}_k(\mathbf r) \Psi^{m'n'}_k(\mathbf r) \hat
  a^{\dagger}_{knl} \hat a_{kn'l'} \Bigg\{ \notag \\ &c_1^2
  \Big(\bar{\mathbf J}(\mathbf r) \cdot \mathbf e_z(\mathbf r)
  \Big)^2\big( \delta_{jy}\delta_{lx}-\delta_{jx}\delta_{ly}\big)\big(
  \delta_{j'y}\delta_{l'x}-\delta_{j'x}\delta_{l'y}\big) 
  \notag \\ &\hspace{3cm} +c_0^2 \mathbf
  J(\mathbf r)^4 \delta_{jl}\delta_{j'l'}
  \Bigg\}.
\end{align}

\section{Calculation of second-order Spin-terms}
\label{sec:calc-second-order}
In this section we calculate the second order terms for the atomic
spin, represented as the  third and fourth
term of the right-hand side of Eq.
(\ref{eq:diagram-spin-coherent-second-order}). These terms we will denote
$\mathcal J^{(2)}_A$, and using the previous notation one finds
\begin{widetext}
  \begin{align}
    \label{eq:secondordertermspin_vanish}
     \mathcal J^{(2)}_A =&\frac{-i}{2}\left( \frac{\beta c_1 \klaser }{2}
    \right)^2 \sum_k \int d^3 r' \sum_{\substack{m'm''\\l}}
    \big( \bar{\mathbf J}(\mathbf r) \times \mathbf e_z(\mathbf r)
    \big)  \left( \left(
      \begin{array}{c}
        0 \\ \bar{J_y}(\mathbf r') \\ \bar J_z(\mathbf r')
      \end{array} \right) \cdot \mathbf e_z(\mathbf r')
    \right)   \rho(\mathbf r')\hat a^{\dagger}_{km'l} \hat a_{km''l}
     \notag \\ &\hspace{8cm} \sum_m \Big\{ \Psi^{mm''}_k(\mathbf
    r)\Psi^{m'm}_{k}(\mathbf r')  - \Psi^{m'm}_k(\mathbf
    r)\Psi^{mm''}_{k}(\mathbf r')   \Big\}.
  \end{align}
We can examine this term by assuming that the only photon carrying
modes of the light are the two modes $\mathbf f_{kox}$ and $\mathbf
f_{ko'y}$ and neglect all other modes. In this case the term reduce
to 
 \begin{align}
    \label{eq:secondordertermspin_vanish_single_mode}
     \mathcal J^{(2)}_A =&\left( \frac{\beta c_1 \klaser }{2}
    \right)^2 \sum_k \int d^3 r' \sum_{\substack{m\\(n,l)\in\{(o,x),(o',y)\}}}
    \big( \bar{\mathbf J}(\mathbf r) \times \mathbf e_z(\mathbf r)
    \big)  \big( \left(
      \begin{array}{c}
        0 \\ \bar{J_y}(\mathbf r') \\ \bar J_z(\mathbf r')
      \end{array} \right) \cdot \mathbf e_z(\mathbf r')
    \big)   \rho(\mathbf r')\hat a^{\dagger}_{knl} \hat a_{knl}
      {\rm Im} [ \Psi^{mn}_k(\mathbf
    r)\Psi^{nm}_{k}(\mathbf r')].
  \end{align}
\end{widetext}
This term represents an atom at position $\mathbf{r'}$ interacting
with the light field and emitting a photon into mode $m$, which
propagates to the position $\mathbf{r}$, where it is absorbed by an
atom followed by stimulated emission into the classical beam. This
process is also known as optically induced dipole-dipole interaction,
and indeed the sum over all modes $m$ can be used to introduce the
dipole propagator in (\ref{eq:g_nmmathbf-r-=}). Note, however, that
above we have written the term in the paraxial approximation, where we
ignore the dependence of the polarization vector on the mode number.
Since the sum over $m$ involves all modes, and not just the paraxial
modes, an accurate treatment requires a more complicated expression
involving the polarization vectors along the lines of Appendix
\ref{appendix-correction} (we use this more complicated expression in
our estimates of the size of the effect).

The last term we will consider is the term describing an atom
interacting with the light field at two different times. This term is
represented as the fifth term on the right hand side of Eq.
(\ref{eq:diagram-spin-coherent-second-order}), and is given on vector
component form in Eq.  \eqref{eq:fracb-c_12-rhom-1}. We will denote
this term with $\mathcal J^{(2)}_B$. A short calculation gives
\begin{align}
  \label{eq:long_calculation_term}
  \mathcal J^{(2)}_B= -\frac{1}{2}&\left( \frac{\beta c_1 \klaser}{2}
  \right) ^2 \sum_{kk'} \sum_{\substack{mm' \\ nn'}}
  \sum_{\substack{jj' \\ l}} \mathbf e_l\left( \bar{\mathbf
      J} \cdot \mathbf e_{j'} \right)\Bigg\{ \notag \\
  &\Psi^{mn}_k \Psi^{m'n'}_{k'} \Big[ \hat
  a^{\dagger}_{kml} \hat a^{\dagger}_{k'm'j} \hat a_{knj} \hat
  a_{k'n'j'} \notag \\ & \hspace{1.7cm} - \hat a^{\dagger}_{kmj} \hat
  a^{\dagger}_{k'm'j} \hat a_{knj'} \hat a_{k'n'l} \Big] + {\rm H.c.}
  \Bigg\},
\end{align}
where we have suppressed the spatial dependence to shorten the
notation. Doing the sum over $j$,$j'$ and $l$ we obtain
  \begin{align}
    \label{eq:resultant_sum_over_l}
    \mathcal J^{(2)}_B&= -\frac{1}{2}\left( \frac{\beta c_1
        \klaser}{2} \right) ^2 \sum_{kk'} \sum_{\substack{mm' \\ nn'}}
    \left( \bar{\mathbf J} - \mathbf e_z ( \bar{\mathbf J} \cdot
      \mathbf e_z) \right) \cdot \notag \\ & \hspace{0.7cm}\Psi^{mn}_k
    \Psi^{m'n'}_{k'} \Big\{ \; 2 \hat a^{\dagger}_{kmx} \hat
    a^{\dagger}_{k'm'y} \hat a_{kny} \hat a_{k'n'x} \notag \\ & - \hat
    a^{\dagger}_{kmy} \hat a^{\dagger}_{k'm'y} \hat a_{knx} \hat
    a_{k'n'x} - \hat a^{\dagger}_{kmx} \hat a^{\dagger}_{k'm'x} \hat
    a_{kny} \hat a_{k'n'y} \Big\}.
  \end{align}
  The first order term in Eq. (\ref{eq:firstorder_spin_result})
  describe the first order effect of rotation of the spin around the
  $e_z({\bf r})$ axis. The second order term in
  (\ref{eq:resultant_sum_over_l}) describe the second order term of
  this rotation. From the rotation frequency in the first order term
  $\propto s_3$ (assuming $\Psi$ to be real), one would thus expect this
  term to scale as $\beta^2(s_3)^2$ which is different from the term
  in (\ref{eq:resultant_sum_over_l}).  This difference arises because
  we have separated the term into normal ordered components such that
  the second order term in (\ref{eq:resultant_sum_over_l}) only
  contributes when at least two photons are present. When we did the
  normal ordering in the diagram we introduced an additional term,
  which we described by the third term in Eq.
  \eqref{eq:beginfmfgr-30-fmfl}


  \section{Calculation of spontaneous
    emission}\label{sec:calc-spont-emiss}

  In this section we calculate the corrections to Eq.
  (\ref{eq:full-eq-stoke-par-multi}) and Eq.
  \eqref{eq:coherent-spin-dynamic}, due to the incoherent interaction.
  To do this we  need a result for the infinitely short
  propagator. From the definition of the propagator
  (\ref{eq:barbar-p+mathbf-r-2}) and the calculation of in it (\ref{eq:barbar-p+mathbf-r-3}), we find the relation
  \begin{align}
    \label{eq:infinitely-short-and-modes}
    \sum_n |U_n(\mathbf r_{\perp})|^2 =
    \frac{2}{\klaser}\varrho(\mathbf r_{\perp})
  \end{align}
  where $\varrho(\mathbf r_{\perp})=\klaser^3/(16\pi^2)$ is the zeroth
  order term of the expansion of $\varrho_{||}(\mathbf r)$ in $\beta$
  given in Eq.  (\ref{eq:beginalign-varrho_-=}). This result is
  important when calculating $\bar{\bar{\mathcal S}}^{(2)}_d$ and for
  relating this term with the incoherent interactions, responsible for
  spontaneous emission. When including this term and the decay
  described in Sec. \ref{sec:evolution-light}, the incoherent
  interaction reduce to
    \begin{widetext}
    \begin{subequations}        \label{eq:spon-corr-emmis}
      \begin{align}
        \hat s_{1,out}(\mathbf r_{\perp})=& \ldots - \frac{\beta^2
          \klaser\varrho(\mathbf r_{\perp})}{2}\int dz' \rho(z')
        \Big\{ c_1^2( \bar J^2_y(z')- \bar J^2_z(z')) \hat
        s_{0,in}(\mathbf r_{\perp}) +( c_0^2\mathbf J^4(z') + c_1^2[
        4J_z^2(z') + J_y^2(z') ])\hat
        s_{1,in}(\mathbf r_{\perp})\Big\}, \\
        \hat s_{2,out}(\mathbf r_{\perp})=& \ldots - \frac{\beta^2
          \klaser\varrho(\mathbf r_{\perp})}{2}\int dz' \rho(z')
        \Big\{ c_0^2\mathbf J^4(z') + c_1^2[ 3J_z^2(z') + J_y^2(z')
        +J_x^2(z') ]\Big\} \hat
        s_{2,in}(\mathbf r_{\perp}), \\
        \hat s_{3,out}(\mathbf r_{\perp})=& \ldots - \frac{\beta^2
          \klaser\varrho(\mathbf r_{\perp})}{2}\int dz' \rho(z')\Big\{
        c_0^2\mathbf J^4(z') + c_1^2[ J_z^2(z') + J_y^2(z') +J_x^2(z')
        ]\Big\} \hat s_{3,in}(\mathbf r_{\perp}),
      \end{align}
    \end{subequations}
  \end{widetext}
  where we have only kept terms that are nonvanishing after taking
  quantum mechanical average of the atomic spin. The operators $\hat
  s_{0,in}(\mathbf r_{\perp})$ measures the total photon flux, and is
  given as
  \begin{align}
    \hat s_0(\mathbf r_\perp) = \sum_{kmm'} \frac{1}{2}&\Big(
    U_{m}^*(\mathbf r_\perp)\hat a^{\dagger}_{kmx} \hat a_{km'x}
    U_{m'}(\mathbf r_\perp) \notag \\ &+ U_m^*(\mathbf r_\perp)\hat
    a^{\dagger}_{kmy} \hat a_{km'y} U_{m'}(\mathbf r_\perp) \Big)
  \end{align}
  It is important to
  note that in a discussion of the various contributions to decay
  one should include all terms in the perturbative expansion,
  including the loop diagrams (\ref{eq:beginfmfgr-35-fmfl}).  If
  these are not included one finds the contribution from the term in Eq.
  (\ref{eq:stokes_term_second_4}) to increase the the operator $\hat
  s_3$. 

  Similarly we find the effect of
  spontaneous emission on the spin equation to read
  \begin{widetext}
    \begin{subequations}        \label{eq:spon-emission-spin-eq}
      \begin{align}
        \bar J_{x,out}(z) =& \ldots - \beta^2 c_1^2\klaser
        \varrho(\mathbf r_{\perp}) \sum_k \big\{ \bar J_{x,in}(z)[
        \hat s_{0,in}^k(\mathbf r_{\perp}) + \frac{1}{2}\hat
        s_{1,in}^k(\mathbf r_{\perp}) ] + \frac{1}{2}\bar J_{y,in}(z)
        \hat s_{2,in}^k(\mathbf r_{\perp})
        \big\} \\
        \bar J_{y,out}(z) =& \ldots - \beta^2 c_1^2\klaser
        \varrho(\mathbf r_{\perp}) \sum_k \big\{ \bar J_{y,in}(z)[
        \hat s_{0,in}^k(\mathbf r_{\perp}) + \frac{1}{2}\hat
        s_{1,in}^k(\mathbf r_{\perp}) ] + \frac{1}{2}\bar J_{x,in}(z)
        \hat s_{2,in}^k(\mathbf r_{\perp})
        \big\} \\
        \bar J_{z,out}(z) =& \ldots - \beta^2 c_1^2\klaser
        \varrho(\mathbf r_{\perp}) \sum_k \bar J_{z,in}(z) \hat
        s_{0,in}^k(\mathbf r_{\perp}).
      \end{align}
    \end{subequations}
  \end{widetext}
  The above result is derived  from Eq. (\ref{eq:fracb-c_12-rhom}) by
  using the paraxial approximation and only keeping  terms of order
  $\beta^2$. A minor correction is introduced since we in
  Eq. (\ref{eq:coherent-spin-dynamic}) chose a representation that was
  in fact not normal ordered. 

  \section{Beyond paraxial approximation}
  \label{appendix-correction}

  In this section we will go slightly beyond the approximation made in
  Eq. \eqref{eq:modefunctions_ensemble}, and consider the set
  \begin{align}
    \label{eq:ceneralized-ensemble-mode}
    \mathbf f_{\mathbf q}(\mathbf r) = \frac{1}{\sqrt{2\pi}}
    U_{nq}(\mathbf r)\mathbf e_{nj}(\mathbf r).
  \end{align}
  We will  consider the correction this generalization makes to the
  result given i Eq. \eqref{eq:single-mode-weak-coupl-stokes}, and
  therefore define spin-components  in the local basis
  given by the set $\mathbf e_{mx}(\mathbf r), \mathbf e_{my}(\mathbf
  r)$ and $\mathbf e_{mz}(\mathbf r)$ 
  \begin{align}
    \label{eq:local-spin-components}
    \bar{\mathcal J}_{\mathbf e_{mi}}(\mathbf r) = \left(
      \begin{array}{c}
        0 \\ \bar{ J}_y(\mathbf r) \\ \bar J_z(\mathbf r)
      \end{array}\right) \cdot \mathbf e_{mi}(\mathbf r)
  \end{align}
  for $i\in \{ x,y,z\}$. These vectors are defined by the fact that,
  e.g., $\mathbf e_{oy}(\mathbf r)$ should be transverse and
  perpendicular to the polarization vector arising from the mode
  function $U_{ok}(\mathbf r) \mathbf e_{ox}(\mathbf r)$. $\mathbf
  e_{oz}$ is then defined by $\mathbf e_{oz}=\mathbf e_{ox}\times
  \mathbf e_{oy}$. Similarly for the quantum modes $m$ the definition
  of $\mathbf e_{mx}$ follow from the fact that it should be
  perpendicular to the polarization vector from the mode
  $U_{mk}(\mathbf r) \mathbf e_{my}(\mathbf r)$.

  With these definitions Eq. \eqref{eq:single-mode-weak-coupl-stokes}
  gives
  \begin{widetext}
    \begin{subequations}
      \begin{align}
        \label{eq:genera-eq-stokes-multi}
        \hat X^m_{out} =& \hat X^m_{in} + \klaser \beta c_1
        \sqrt{\frac{N^o_x}{2}} \int d^3r' \: \rho(\mathbf r') {\rm
          Re}[\Psi^{mo}_ k(\mathbf r)] \Big\{ \bar{\mathcal
          J}_{\mathbf e_{oz}}(\mathbf r)[\mathbf e_{ox}(\mathbf r)
        \cdot \mathbf e_{mx}(\mathbf r)]- \bar{\mathcal J}_{\mathbf
          e_{ox}}(\mathbf r)[\mathbf e_{ox}(\mathbf r) \cdot \mathbf
        e_{mz}(\mathbf r)] \Big\}
        \\
        \hat P^m_{out} =& \hat P^m_{in} + \klaser \beta c_1
        \sqrt{\frac{N^o_x}{2}} \int d^3r' \: \rho(\mathbf r') {\rm
          Im}[\Psi^{mo}_ k(\mathbf r)]\Big\{ \bar{\mathcal J}_{\mathbf
          e_{oz}}(\mathbf r)[\mathbf e_{ox}(\mathbf r) \cdot \mathbf
        e_{mx}(\mathbf r)]- \bar{\mathcal J}_{\mathbf e_{ox}}(\mathbf
        r)[\mathbf e_{ox}(\mathbf r) \cdot \mathbf e_{mz}(\mathbf r)]
        \Big\}.
      \end{align}
    \end{subequations}
  \end{widetext}
  Similarly we find the correction to Eq.
  \eqref{eq:spin-eq-weak-multi-mode} to give
  \begin{widetext}
    \begin{align}
      \label{eq:correction-spin-equations-beyond}
      \bar{\mathbf J}_{out}(\mathbf r) \approx \bar{\mathbf
        J}_{in}(\mathbf r) + \klaser \beta c_1 \sqrt{\frac{N^o_x}{2}}
      \sum_n \Big[ {\rm Re}[\Psi^{no}_ k(\mathbf r)]\hat P^n_{in} -
      {\rm Im}[\Psi^{no}_ k(\mathbf r)]\hat X_{in}^n \Big] \Big\{
      \bar{\mathbf J}_{in}(\mathbf r) \times \Big( \mathbf
      e_{ox}(\mathbf r) \times \mathbf e_{ny}(\mathbf r) \Big) \Big\}.
    \end{align}
  \end{widetext}


\begin{thebibliography}{3}
  \bibitem{briegel} H. J. Briegel {\it et al.}, Phys. Rev. Lett. {\bf
      81}, 5932 (1998).
\bibitem{kimblecavity} L.-M. Duan, A. Kuzmich, and H. J. Kimble,
  Phys. Rev. A {\bf 67}, 032305 (2003).
\bibitem{rempe} P. Maunz, T. Puppe, I. Schuster, N. Syassen, P. W. H.
  Pinkse, and G. Rempe, Nature {\bf 428}, 50-52 (2004).
\bibitem{blinov} B. B. Blinov, D. L. Moehring, L.-M. Duan, and
  C. Monroe, Nature {\bf 428}, 153-157 (2004). 
\bibitem{yamamoto} J. Vu\v ckovi\'c, M. Pelton, A. Scherer, and
  Y. Yamamoto, Phys. Rev. A {\bf 66}, 023808 (2002).
\bibitem{imamoglu} A. Badolato, K. Hennessy, M. Atatüre, J. Dreiser,
  E. Hu, P.M. Petroff, and A. Imamo\u glu, Science {\bf 308},
  pp. 1158-1161 (2005).
\bibitem{duan} L.-M. Duan, J. I. Cirac, P. Zoller, and E. S. Polzik,
  Phys. Rev. Lett. {\bf 85}, 005643 (2000).
\bibitem{juulsgaard1} B. Julsgaard, A. Kozhekin, and E. S. Polzik,
  Nature {\bf 413}, 400 (2001).
\bibitem{juulsgaard2} B. Julsgaard, J. Sherson, J. I. Cirac, J. Fiurasek, and
E. S. Polzik, Nature {\bf 432}, 482 (2004).
\bibitem{sherson} J. F. Sherson, H. Krauter, R. K. Olsson, B. Julsgaard, K. Hammerer,
J. I. Cirac, and E. S. Polzik, Nature {\bf 443}, 557 (2006).
\bibitem{mabuchi} J. M. Geremia, J. K. Stockton, and H. Mabuchi, Science
  {\bf 304},  270 (2004). 
\bibitem{kuzmichtheory} A. Kuzmich, N. P. Bigelow and L. Mandel,
  Europhys. Lett. {\bf 42},481 (1998).
\bibitem{DLCZ} L.-M. Duan, M. D. Lukin, J. I. Cirac and P. Zoller,
  Nature {\bf 414}, 413 (2001).
\bibitem{kuzmich} D. N. Matsukevich, and A. Kuzmich, Science {\bf 306},
  663 (2004).
\bibitem{lukin} C.H. van der Wal {\it et al.}, Science {\bf 301}, 196 (2003).
\bibitem{kimbleensemble} C.W. Chou {\it et al.}, Nature {\bf 438},
  828 (2005).
\bibitem{schmiedmayer} S. Chen {\it et al.}, Phys. Rev. Lett. {\bf 97},
  173004 (2006).
\bibitem{vuletic} H.W. Chan, A.T. Black, and V. Vuleti\'c,
  Phys. Rev. Lett. {\bf 90}, 063003 (2003).  
  \bibitem{lukin-eit} M. D. Lukin, Rev. Mod. Phys. {\bf 75}, 457 (2003).
  \bibitem{kraus} B. Kraus W. Tittel, N. Gisin, M. Nilsson, S. Kr{\"o}ll, and J. I. Cirac, Phys. Rev. A {\bf 73}, 020302 (2006).
\bibitem{oxana1} D. V. Kupriyanov, O. S. Mishina, I. M. Sokolov, B. Julsgaard, and E. S. Polzik 
Phys. Rev. A {\bf 71}, 032348 (2005).
\bibitem{oxana2} O. S. Mishina, D. V. Kupriyanov, J. H. M{\"u}ller, and E. S. Polzik,
Phys. Rev. A {\bf 75}, 042326 (2007).
\bibitem{raymer} M. G. Raymer and J. Mostowski, Phys. Rev. A {\bf 24}, 1980 (1981).
\bibitem{mostowski83} J. Mostowski and B. Sobolewska, Phys. Rev. A {\bf 28}, 2573 (1983).
\bibitem{mostowski84} J. Mostowski and B. Sobolewska, Phys. Rev. A {\bf 30}, 610 (1984).
\bibitem{duan02} L. -M. Duan, J. I. Cirac, and P. Zoller, Phys. Rev. A
{\bf 66}, 023818 (2002).
\bibitem{claude97:photons_and_atoms} C. Cohen-Tannoudji, J.
  Dupont-Roc, and G. Grynberg, {\it Photons and Atoms, Introduction
    to Quantum Electrodynamics} (Wiley, New York, 1997).
\bibitem{footnote1} We have here a formally divergent term, the dipole
    self-energy. One can, however, show that this term has no effect on
    the dynamics of the system. 
\bibitem{brian_thesis} Brian Julsgaard, {\it Entanglement and Quantum
    Interactions with Macroscopic Gas Samples} (Ph.D. Thesis,
  University of \AA rhus, 2003).
\bibitem{glauber91:_quant} R. J. Glauber and M. Lewinstein, Phys.
  Rev. A {\bf 43}, 467 (1991).
\bibitem{van-enk} S. J. van Enk and H. J. Kimble, Phys. Rev. A {\bf 63},
  023809 (2000).
\bibitem{denise} D. V. Vasilyev, I. V. Sokolov, and E. S. Polzik,
  arXiv:0704.1737v1 [quant-ph].
\bibitem{eit-memory} R. M. Camacho, C. J. Broadbent, L. Ali-Khan, and
  J. C. Howell, Phys. Rev. Lett. {\bf 98}, 043902 (2007)
\bibitem{lasers} P. W. Milonni and J. H. Eberly {\it Lasers} (Wiley, 1988).
\bibitem{olivier95:_refrac_bose} O. Morice, Y. Castin, and J. Dalibard,
Phys. Rev. A {\bf 51}, 3896 (1995). 
\bibitem{lorenz-lorenz} A. Lagendijk, B. Nienhuis, B.A. van Tiggelen,
  and P. de Vries, Phys. Rev. Lett. {\bf 79}, 657 (1997).
\bibitem{jackson} A. D. Jackson, {\it Classical Electrodynamics, Third
    Edition} (Wiley, New York, 1998).
\end{thebibliography}
\end{document}